\documentclass[12pt]{article}

\usepackage{latexsym}
\usepackage{epsf}

\begin{document}
\ifx\href\undefined\else\hypersetup{linktocpage=true}\fi
\raggedbottom
\title{\bf Models of the Pseudogap State in High -- Temperature
Superconductors}
\author{M.V.Sadovskii\\ 
Institute for Electrophysics, Russian Academy of Sciences,\\ 
Ekaterinburg,\ 620016,\ Russia}

\date{}

\maketitle

\begin{abstract}

We present a short review of our basic understanding of the physics of copper --
oxide superconductors and formulate the list of ``solved'' and ``unsolved''
problems. The main problem remains theoretical description of the properties
of the normal state, requiring clarification of the nature of the
so called pseudogap state.

We review simplified models of the pseudogap state, based on the scenario
of strong electron scattering by (pseudogap) fluctuations of ``dielectric''
(AFM, CDW) short -- range order and the concept of ``hot'' spots (patches)
on the Fermi surface. Pseudogap fluctuations are described as appropriate
static Gaussian random field scattering electrons.

We derive the system of recurrence equations for the one -- particle Green's
function and vertex parts, taking into account all Feynman diagrams for
electron scattering by pseudogap fluctuations. Results of calculations of 
spectral density, density of states and optical conductivity are presented, 
demonstrating both pseudogap and localization effects.

We analyze the anomalies of superconducting state (both $s-$ and $d$ -- wave
pairing) forming on the ``background'' of these pseudogap fluctuations. 
Microscopic derivation of Ginzburg -- Landau expansion allows calculations of
critical temperature $T_c$ and other basic characteristics of a superconductor,
depending on the parameters of the pseudogap. We also analyze the role of
``normal'' (nonmagnetic) impurity scattering. It is shown that our simplified
model allows semiquantitative modelling of the typical phase diagram of
superconducting cuprates\footnote{Extended version of the talk given by the 
author on the seminar of I.E.Tamm Theoretical Department of P.N.Lebedev Physical
Institute, Russian Academy of Sciences, Moscow, October 7, 2003.}.

\end{abstract}

\newpage

\tableofcontents

\newpage

\section{Basic problems of the physics of high -- temperature superconductors.}

We shall start with brief review of the present day situation in the physics
of high -- temperature superconductors, which may be useful for the reader
not involved in this field.

\subsection{What is really KNOWN about copper oxides:}

High -- temperature superconducting (HTSC) copper oxides are intensively 
studied for more than 15 years now. In these years, considerable progress 
has been achieved in our understanding of the nature and basic physical 
properties of these systems, despite their complicated phase diagram, 
containing almost all the main phenomena studied by solid state physics 
(Fig. \ref{ph_diag}).

It is apparent that the number of facts are now well established and are not
due to any revision in the future
\footnote{Surely, any such list is subjective enough and is obviously based
on some prejudices and ``tastes'' of the author. It is well known, that in
HTSC physics quite opposite views are often ``peacefully coexistent'' with
each other. However, such list may hopefully be of some interest.}
If we talk about superconducting state, the list of ``solved'' problems may be
formulated as follows:

\begin{figure} 
\epsfxsize=7cm 
\epsfbox{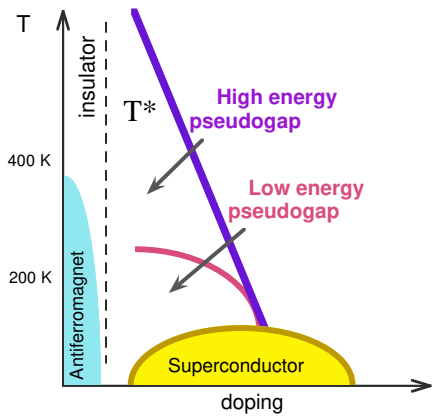}
\caption{Typical phase diagram of copper oxide high -- temperature 
superconductor. Different phases and anomalous regions are shown
schematically. Carrier concentration, corresponding to a maximum of
transition temperature $T_c$ is called ``optimal''. Systems with lower
concentrations are called ``underdoped'', those with larger --- ``overdoped''.}
\label{ph_diag} 
\end{figure}

\begin{itemize}

\item{\sl Nature of superconductivity $\rightleftharpoons$ Cooper pairing.}

It is definitely established that in these system we are dealing with

\begin{enumerate}

\item{$d$ -- wave pairing,}

and in the spectrum of elementary excitations we observe

\item{Energy gap $\Delta\cos 2\phi$,}

where the polar angle $\phi$ determines the direction of electronic momentum
in two -- dimensional inverse space, corresponding to $CuO_2$ plane.
Thus, energy gap becomes zero and changes sign on the diagonals of two --
dimensional Brillouin zone.

\item{The size of pairs is relatively small: $\xi_0\sim 5-10 a$,}

where $a$ is the lattice constant. This means that high -- temperature
superconductors belong to a crossover region between 
``large'' pairs of BCS theory and ``compact'' Bosons picture of very strongly
coupled electrons.

\noindent

Superconducting state always appears as

\item{Second order phase transition

with more or less usual thermodynamics.}

\end{enumerate}

All these facts were established during approximately first five years of
studies of HTSC copper oxides. We do not give any references to original
works here as the list will be to long, and quote only two review articles 
\cite{Legg, Kirt}.

Situation becomes much more complicated when we consider the properties of
the normal state. Almost all anomalous properties of these unusual systems
appear first of all in the normal state. However, here we also can list some 
definitely established facts:

\item{\sl Existence of the Fermi surface.}

In this sense these systems are clearly {\em metals}, though rather unusual
(``bad''). Most of the data on Fermi surfaces were obtained from ARPES 
\cite{ZX1,ZX2}, and the progress here in recent years is quite spectacular
due to the great increase (more than order of magnitude!) in resolution both
in energy and momentum. As an illustration of these advances in Fig.
\ref{f_surf} we show {\em experimentally} determined
Fermi surface of  $La_{2-x}Sr_xCuO_4$ with $x=0.063$ \cite{Zhou}. 
Note the existence of characteristic ``flat'' parts of this Fermi surface.
This form of the Fermi surface is rather typical for the majority of HTSC
cuprates in superconducting region of the phase diagram.

\begin{figure} 
\epsfxsize=10cm 
\epsfbox{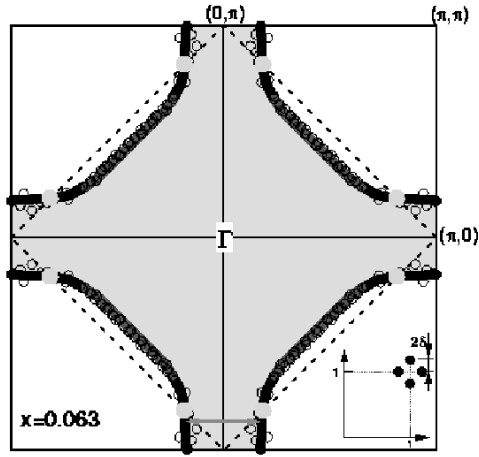}
\caption{Experimentally determined Fermi surface of
$La_{2-x}Sr_xCuO_4$ with $x=0.063$. At the insert -- schematic intensity of
neutron scattering with four incommensurate peaks close to the scattering
vector $(\pi, \pi)$. Dashed lines show the borders of antiferromagnetic 
Brillouin zone, which appear after the establishment of antiferromagnetic
long -- range order (period doubling). Points, where these borders intersect
the Fermi surface are called ``hot'' spots.} 
\label{f_surf} 
\end{figure}

\item{\sl Metal --- Insulator transition,}

which takes place with the change of chemical composition (as the number of
carriers drops). Stochiometric $La_2CuO_4$ is an antiferromagnetic insulator
(apparently of Mott type) with Neel temperature of the order of 400K. 
There exists well defined optical (energy) gap, and antiferromagnetism is due
to the ordering of {\em localized} spins on $Cu$ ions and is well described
by two -- dimensional Heisenberg model. This is more or less typical also for
other HTSC oxides, which thus belong to a wide class of {\em strongly
correlated} electronic systems. At the same time, this dielectric state is
rapidly destroyed by introduction of few percents of doping impurities.
Note that the Fermi surface shown in Fig. \ref{f_surf}, is observed, in fact,
in the system which is on the edge of metal -- insulator transition, which
takes place at $x\sim 0.05$.

\item{\sl Strong anisotropy of all properties 
(quasi -- two -- dimensional nature!).}

Current carriers (in most cases --- holes) are propagating more or less freely
along $CuO_2$ planes ($a$ and $b$ directions in orthorombic crystal), while
transverse motion in orthogonal direction ($c$ axis) is strongly suppressed.
Conductivity anisotropy is usually of the order of $10^2$ --- $10^5$. 
This fact is seriously damaging from the point of view of practical 
applications. At the same time it is still unclear whether 
two -- dimensionality is a necessary condition for realization of 
high -- temperature supeconductivity.

\end{itemize}

\subsection{What is still UNKNOWN about copper oxides:}

Let us now list the main unsolved problems, remaining in the center of
rather sharp discussions with participants sometimes not hearing each other.
On the first place stays of course the

\begin{itemize}

\item{{\sl Mechanism of Cooper pairing.}

Most researchers (not all!) do not have any doubt, that in HTSC -- oxides
we are dealing with Cooper pairing within more or less standard BCS
``scenario''. However, the question is --- what kind of interaction
leads to pair formation? A number of microscopic mechanisms are under
consideration:

\begin{enumerate}

\item{Electron -- phonon \cite{Maks,Kulic}.}
The main unsolved problem here is an explanation of $d$ -- wave symmetry of
Cooper pairing. Usually, electron -- phonon coupling leads to
$s$ -- wave pairing. Recently there was some progress in possible ways to 
solve this problem \cite{Kulic2}.

\item{Spin -- fluctuation \cite{Scal,Mor,CPS,YY}.}

Historically, within this approach the $d$ -- wave nature of pairing in
oxides was {\em predicted}. Further experimental confirmation, as well as
the possibility of semiquantitative description of many properties of these
systems made this approach probably most favorable (from my point of view!) 
among many other mechanisms.

\item{Exchange -- $RVB$, $SO(5)$, ...?}

These models \cite{PWA, Zhang}, as well as a number of more ``exotic'', were
formulated in early days of HTSC research. Enormous intellectual resources
of leading theorists were and are spent on their development. 
However (again it is only my personal view), the completeness and ``stability'' 
of results here is incomparably less than in more traditional approaches, while
real connection with experiments is relatively slight, so that all these
models still remain a kind of ``brain gymnastics''. Obviously, I do not deny
usefulness of these models from purely theoretical point of view. For example,
the discussion of symmetry connections of antiferromagnetism and 
superconductivity \cite{Zhang} is of great interest. At the same time, the most
natural and useful among phenomenological approaches remains the description of
superconductivity in oxides based on anisotropic version of the standard
Ginzburg -- Landau theory \footnote{This fact is of special significance
as this seminar has taken place approximately an hour after the news 
came on the award of 2003 Nobel Prize in physics to V.L.Ginzburg.}.

\end{enumerate}

}

At this time it is unclear, which of these mechanisms is realized or is
dominant in real HTSC -- oxides. And the reason for this lies in the fact,
that the properties of superconducting state are relatively independent of
microscopic mechanism of pairing. It is rather difficult to propose some
``crucial'' experiment, which will definitely confirm one of these mechanisms.
In some sense, situation here is somehow similar to those existing, actually
for decades, in the theory of magnetism. Since the middle of the thirties,
it is well known, that the nature of magnetism is connected with exchange
interaction. There exist plenty microscopic mechanisms of exchange interaction
between spins (e.g. direct exchange, superexchange, $s-d$ interaction,
RKKY etc.). However, it is not always possible to tell, which of these 
mechanisms acts in some real magnetic system. Classical example here is the
problem of magnetism of iron!

\item{{\sl Nature of the normal state.}

Here any consensus among researchers is almost absent. Usually, recognizing
the metallic nature of these systems, theorist discuss several alternative
possibilities:

\begin{enumerate}

\item{Fermi -- liquid (Landau).}

\item{``Marginal'' or ``bad'' Fermi -- liquid \cite{Var}.}

\item{Luttinger liquid \cite{PWA}.}

All the talking on the possible breaking of Fermi -- liquid behavior
(absence of ``well defined'' quasiparticles) originates, from theoretical
point of view, from  strong electronic correlations and quasi -- two --
dimensionality of electronic properties of these systems, while from
experimental side this is mainly due to practical impossibility  of performing
studies of electronic properties of the {\em normal} state at low enough
temperatures (when these properties are in fact ``shunted'' by
superconductivity, which is impossible to suppress without significant
changes of the system under study). One must remember, that Fermi -- liquid
behavior, by definition, appears only as limiting property at low enough
temperature. Thus, it is not surprising at all that it is ``absent'' in
experiments performed at temperatures of the order of $10^2$K! 
At the same time, as we shall see below, recently there were an important
experimental developments, possibly clarifying the whole problem.

\item{Disorder and local inhomogeneities.}

Practically all high -- temperature superconductors are internally disordered
due to chemical composition (presence of doping impurity). Thus, any 
understanding of their nature is impossible without the detailed studies of
the role of an internal disorder in the formation of electronic properties
of strongly correlated systems with low dimensionality. In particular,
localization effects are quite important in these systems, and studies of these
has already a long history \cite{SC_Loc}. Situation with the role of internal
disorder has complicated recently with the arrival of the new data, obtained
in by scanning tunneling microscopy (STM), which clearly demonstrated 
inhomogeneous nature of the local density of states and superconducting
energy gap on microscopic scale, even in practically ideal single -- crystals
of copper oxides. Of many recent papers devoted to these studies, we quote
only two \cite{Pan,Davis}, where further references can be found. These
results significantly  modify previous ideas on microscopic phase separation,
``stripes'' etc. On the other hand, the presence of such inhomogeneities
makes these systems a kind of ``nightmare'' for theorists, though the 
picture of inhomogeneous superconductivity due to fluctuations in the local
density of states was analyzed rather long time ago \cite{SC_Loc,BPS}.

\item{Pseudogap.}

Now it is clear that most of the anomalies of the normal state of copper
oxides is related to the formation of the so called pseudogap state.
This state is realized in a wide region of the phase diagram as shown in
Fig. \ref{ph_diag}, corresponding mainly to ``underdoped'' compositions and
temperatures $T<T^*$. It is important to stress, that the line $T^*$ on 
Fig. \ref{ph_diag} is rather approximate and apparently do not correspond to
any phase transition and signify only a kind of crossover to the region of
well developed pseudogap anomalies\footnote{Sometimes the notion of
``high energy pseudogap'' is introduced and defined by $T^*$, while the
``low energy pseudogap'' is defined by another crossover line in the region
of $T<T^*$ closer to superconducting ``dome'', as shown in 
Fig. \ref{ph_diag}.}. In short, numerous experiments \cite{Tim,MS} show that
in the region of $T<T^*$ the systems, which are in the normal (non 
superconducting) state, do possess some kind of the gap -- like feature in
the energy spectrum. However, this is not a real gap, but some kind of 
precursor of its appearance in the spectrum (which is the reason for
the term ``pseudogap'', which appeared first in qualitative theory of
amorphous and liquid semiconductors \cite{Mott}.). Some examples of the
relevant experiments will be given below.

Crudely speaking, the gap in the spectrum can be either of superconducting or
insulating nature. Accordingly, there exist two possible theoretical
``scenarios'' to explain pseudogap anomalies in copper oxides. The first one
anticipates formation of Cooper pairs already at temperatures higher, than
the temperature of superconducting transition, with phase coherence appearing
only in superconducting state at $T<T_c$. The second assumes, that the origin
of the pseudogap state is due to fluctuations of some kind of short range
order of ``dielectric'' type, developing in the underdoped region. 
Most popular here is the picture of antiferromagnetic (AFM) fluctuations,
though fluctuating charge density waves (CDW) cannot be excluded, as well as
structural deformations or phase separation at microscopic scales. 
In my opinion, most of the recent experiments provide evidence for this
second scenario. So, in further discussion we shall deal only with this type
of models, mainly speaking about antiferromagnetic fluctuations.

\end{enumerate}

}

\end{itemize}

It is common view, that the final understanding of the nature of high --
temperature superconductivity is impossible without clarification of the 
nature of the normal state. Generally speaking, we have to explain all
characteristic features of the phase diagram, shown in Fig. \ref{ph_diag}. 
This is the main task of the theory and it is still rather far from the
complete solution. In the following we shall concentrate on the discussion of
some simple models of the pseudogap state and attempts to illustrate possible
ways to find a solution of this main problem.

\section{Basic experimental facts on the pseudogap behavior in
high -- temperature superconductors.}

Phase diagram of Fig. \ref{ph_diag} shows that, depending on concentration
of current carriers in highly conducting $CuO_2$ plane, a number of phases and
regions with anomalous properties can be observed in copper oxides. For small
concentrations, all the known HTSC -- cuprates are antiferromagnetic 
insulators. With the growth of carrier concentration Neel temperature $T_N$ 
rapidly drops from the values of the order of hundreds of $K$, vanishing
at concentration of holes $p$ less than or of the order of $0.05$, and system
becomes metallic. As concentration of holes is increasing further, the system 
becomes superconducting, and temperature of superconducting transition grows
with the growth of carrier concentration, passing through characteristic
maximum at $p_0\approx 0.15-0.17$ (optimal doping), and then drops and vanish
at $p\approx 0.25-0.30$, though metallic behavior in this (overdoped) region
remains. In the overdoped region with  $p>p_0$ metallic properties are more
or less traditional (Fermi -- liquid), while for $p<p_0$ the system is a kind
of anomalous metal, not described (in the opinion of majority of authors)
by the usual theory of Fermi -- liquid.

Anomalies of physical properties attributed to the pseudogap formation
are observed in metallic phase for $p<p_x$ and temperatures  $T<T^*$, 
where $T^*$ drops from the values of the order of $T_N$ at $p\sim 0.05$
vanishing at some ``critical'' concentration of carriers $p_c$, slightly
greater than $p_0$, so that $T^*$ -- line in Fig.\ref{ph_diag} is actually
continuing under the superconducting ``dome''. According to the data
presented in Ref. \cite{TL} $T^*$ vanishes at $p=p_c\approx 0.19$. In the 
opinion of some other authors (proponents of superconducting scenario for the
pseudogap) $T^*$ -- line just passes to $T_c$ -- line of superconducting
transition somewhere close to the optimal doping $p_0$. In my opinion, most
recent data, apparently, confirm the first variant of the phase 
diagram (more details can be found in Ref. \cite{TL}).

Pseudogap anomalies, in general, are interpreted as due to suppression
(in this region) of the density of single -- particle excitations close to
the Fermi level, which corresponds to the general concept of the pseudogap
\cite{Mott}. The value of $T^*$ then determines characteristic scale of
the observed anomalies and is proportional to effective energy width of the
pseudogap. Let us now consider typical experimental manifestations of
pseudogap behavior.

\begin {figure} 
\epsfxsize=6cm 
\epsfbox{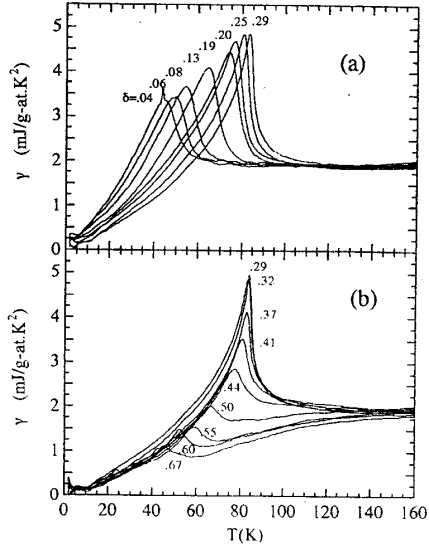}
\caption {Coefficient $\gamma$ of electronic specific heat in overdoped
(a) and underdoped (b) $Y_{0.8}Ca_{0.2}Ba_2Cu_3O_{7-\delta}$ .}
\label{Cybco}
\end {figure}

Consider first experimental data  on electronic specific heat
of cuprates. In metals this contribution is usually written as: $C=\gamma(T)T$, 
so that in the normal state ($T>T_c$) $\gamma\sim N(0)$, where $N(0)$ -- 
is the density of states at the Fermi level. At $T=T_c$ we have the well known 
anomaly due to second order phase transition, so that $\gamma(T)$ demonstrates
characteristic peak (discontinuity). As a typical example, in Fig. \ref{Cybco} 
we show experimental data for $Y_{0.8}Ca_{0.2}Ba_2Cu_3O_{7-\delta}$ for 
different values of $\delta$ \cite{Loram}.  
We can see, that in optimally doped and overdoped samples $\gamma(T)$ is
practically constant for all $T>T_c$, while for underdoped samples considerable
drop of $\gamma(T)$ appears for $T<150-200K$. This is a direct evidence of the
appropriate drop of the density of states at the Fermi level due to pseudogap
formation for $T<T^*$.

Note also that the value of specific heat discontinuity at superconducting
$T_c$ is significantly suppressed as we move to the underdoped region.
More detailed analysis shows \cite{TL} that the drop of this discontinuity 
$\Delta\gamma_c$ actually starts at some ``critical'' carrier concentration
$p_c\approx 0.19$, which is connected with the ``opening'' of the pseudogap.

Pseudogap formation in the density of states is clearly seen also in
experiments on single particle tunneling. Thus, in highly cited Ref. \cite{Renn}
tunneling experiments were performed on single crystals of
$Bi_2Sr_2CaCu_2O_{8+\delta}$ ($Bi-2212$) with different oxygen content. For
underdoped samples pseudogap formation in the density of states was clearly 
observed at temperatures significantly higher than superconducting $T_c$.
This pseudogap smoothly transformed into superconducting gap at $T<T_c$, 
which is often seen as an evidence of its superconducting nature. However,
in Refs. \cite{Kras1,Kras2}, where tunneling experiments were performed on
the same system, it was directly shown, that superconducting gap exists on
the ``background'' of the wider pseudogap and vanishes at $T=T_c$, while
pseudogap persists at higher temperatures.

Pseudogap also shows itself in transport properties of HTSC -- systems in
normal state, Knight shift and NMR relaxation. In particular, changes from the
usual for optimally doped linear temperature dependence of resistivity for
underdoped samples and $T<T^*$ are often attributed to its formation.
Also the value of Knight shift in such samples for $T<T^*$ becomes temperature
dependent and drops as temperature lowers. Similar behavior is observed in
underdoped samples for $(TT_1)^{-1}$, where $T_1$ -- is NMR relaxation time.
Let us remind, that in usual metals Knight shift is just proportional to the
density of states at the Fermi level $N(0)$, while $(TT_1)^{-1}\sim N^2(0)$ 
(Korringa behavior), and resistivity  $\rho$ is proportional to the scattering
rate (inverse mean free time) $\gamma\sim N(0)$. Thus, significant lowering of
these characteristics is naturally attributed to the drop in the density of
states $N(0)$ at the Fermi level. Note that these arguments are, of course, 
oversimplified, particularly when we are dealing with temperature dependences.
E.g. in case of resistivity this dependence is determined by inelastic 
scattering and physics of these processes in HTSC is still unclear.
Thus, the decrease in the density of states (due to partial dielectrization of
the spectrum) can also lead, in fact, to the growth of resistivity.

\begin{figure}
\epsfxsize=14cm
\epsfbox{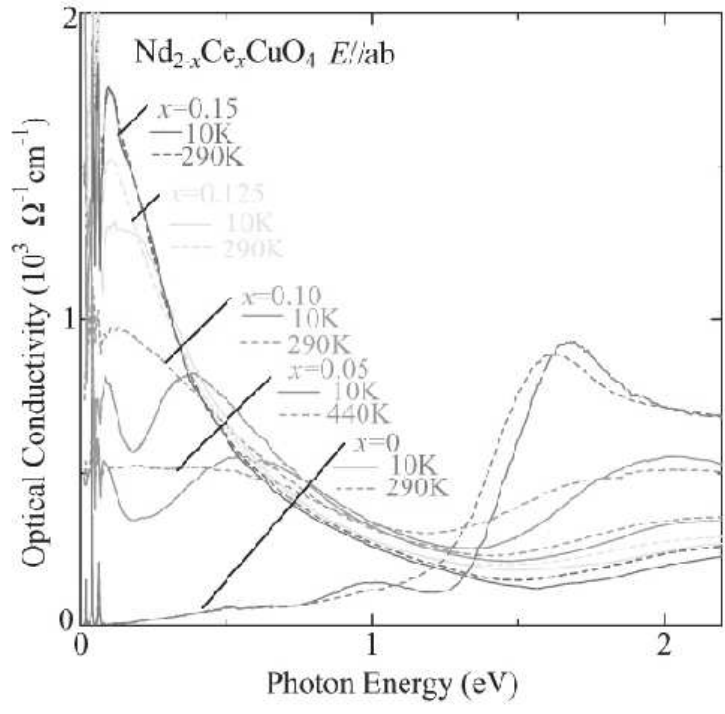}
\caption{Real part of optical conductivity in the $ab$ -- plane of  
$Nd_{2-x}Ce_{x}CuO_4$ for different temperatures and compositions $x$.}
\label{optcond}
\end{figure}

Pseudogap in underdoped cuprates is also observed in 
experiments on optical conductivity, both for electric field polarization
along highly -- conducting $CuO_2$ plane and also along orthogonal direction
($c$ -- axis). These experiments are reviewed in detail in Ref. \cite{Tim}. 
As a typical example in Fig. \ref{optcond} we show the data of Ref. 
\cite{Onose} on optical conductivity in the $CuO_2$ plane for different
compositions of electronically conducting oxide $Nd_{2-x}Ce_{x}CuO_4$.
Characteristic feature here is the appearance (most clearly seen for underdoped
samples with $x=0.05$ and $x=0.10$), of the shallow minimum at frequencies
of the order of $0.25\ eV$ and of smooth maximum due to absorption through
unusually wide pseudogap around $\omega\sim 0.5\ eV$. The nature of an
additional absorption maximum seen at $\omega\sim 0.1\ eV$ can, apparently,
bw attributed to localization (cf. below).

In underdoped cuprates with hole conductivity, optical conductivity is usually
characterized by a narrow ``Drude like'' absorption peak at small frequencies,
followed by a shallow minimum and smooth maximum due to pseudogap absorption
through pseudogap of the order of $0.1 eV$ \cite{Tim,MS}. Additional peak due
to localization is observed rarely, usually after the introduction of an
additional disorder \cite{B1,B2,B3}.

Most spectacular effects due to pseudogap formation are seen in experiments
on angle resolved photoemission (ARPES). ARPES intensity (energy and momentum
distribution of photoelectrons) is determined by \cite{ZX2}:
\begin{equation}
I({\bf k}\omega)=I_0({\bf k})f(\omega)A({\bf k}\omega)
\label{IARP}
\end{equation}
where $\bf k$ -- is the momentum in the Brillouin zone, $\omega$ -- energy of
initial state, measured with respect to the Fermi level (chemical 
potential)\footnote{In real experiments $\omega$ is measured with respect to
the Fermi level of some good metal, e.g. $Pt$ or $Ag$, placed in electric 
contact with a sample.}, $I_0({\bf k})$ includes some kinematic factors and the
square of the matrix elements of electron -- photon interaction and in crude
approximation is considered to be some constant,
\begin{equation}
A({\bf k},\omega)=-\frac{1}{\pi}ImG({\bf k},\omega+i\delta)
\label{sdens}
\end{equation}
where $G({\bf k},\omega)$ -- is Green's function of an electron, determines
the spectral density. The presence of Fermi distribution
$f(\omega)=[exp(\omega/T)+1]^{-1}$ reflects the fact, that only occupied states
can produce photoemission. Thus, in such a crude approximation it can be said,
that ARPES experiments just measure the product $f(\omega)A({\bf k}\omega)$, 
and we get direct information on the spectral properties of single -- particle 
excitations.

Consider qualitative changes in single -- electron spectral density
(\ref{sdens}) due to pseudogap formation. In the standard Fermi -- liquid
theory, single -- electron Green' function can be written as:
\begin{equation}
G(\omega,{\bf k})=\frac{Z_{\bf k}}{\omega-\xi_{\bf k}-i\gamma_{\bf k}}+
G_{incoh}
\label{FLGr}
\end{equation}
where $\xi_{\bf k}=\varepsilon_{\bf k}-\mu$ -- is quasiparticle energy with
respect to the Fermi level (chemical potential) $\mu$, $\gamma_{\bf k}$ -- 
quasiparticle damping. The residue in the pole $0<Z_{\bf k}<1$, $G_{incoh}$ -- 
is some non singular contribution due to many particle excitations.
Then the spectral density is:
\begin{equation}
A(\omega,{\bf k})=\frac{1}{\pi}Z_{\bf k}\frac{\gamma_{\bf k}}
{(\omega-\xi_{\bf k})^2+\gamma^2_{\bf k}}+...
\label{FLsdens}
\end{equation}
where dots denote more or less smooth contribution from $G_{incoh}$, while
quasiparticle spectrum determines a narrow (if damping $\gamma_{\bf k}$ is
small in comparison to $\xi_{\bf k}$) Lorentzian peak. 
In the usual Fermi -- liquid $\gamma\sim\omega^2\sim |{\bf k-k}_F|^2$ and
quasiparticles are well defined in some (narrow enough) vicinity of the
Fermi surface. In the model of ``marginal'' Fermi -- liquid
$\gamma\sim\omega\sim|{\bf k-k}_F|$ and quasiparticles are
``marginally'' defined due to $\gamma\sim\xi_{\bf k}$. In the presence of
static scattering (e.g. due to impurities) a constant (independent of $\omega$) 
contribution $\gamma_0$ appears in damping.

\begin{figure}
\epsfxsize=12cm
\epsfysize=16cm
\epsfbox{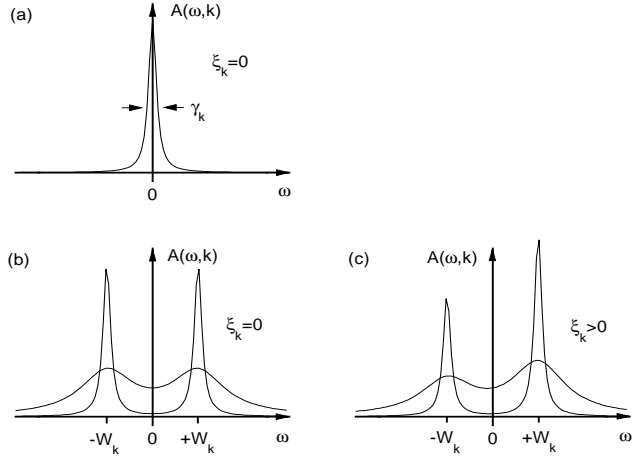}
\caption{Qualitative evolution of spectral density.
(a) -- normal metal (Fermi -- liquid), $\xi_{\bf k}=0$ -- at the Fermi surface.
(b) -- two narrow peaks, corresponding to ``Bogoliubov's'' quasiparticles in 
a system with dielectric gap $W_{\bf k}$ (in case of long -- range order of
CDW or SDW type). Smooth maxima -- system without long -- range order
(pseudogap behavior), $\xi_{\bf k}=0$ -- at the Fermi surface.
(c) -- same as (b), but for $\xi_{\bf k}>0$, i.e. above the Fermi surface.
Note characteristic asymmetry of maxima in this case.}
\label{sdqual}
\end{figure}

Qualitative form of the spectral density in the Fermi -- liquid picture
is illustrated in Fig. \ref{sdqual}(a).

If some long -- range order (e.g. of SDW(AFM) or CDW type) appears in the 
system, an energy gap $W_{\bf k}$ (of dielectric nature) opens in the spectrum of 
elementary excitations (${\bf k}$ -- dependence stresses the possibility of gap
opening only on the part of the Fermi surface), and single -- particle Green's
function acquires the form of Gorkov's function (where we also add some 
damping $\Gamma_{\bf k}$):
\begin{equation}
G(\omega,{\bf k})=\frac{u^2_{\bf k}}{\omega-E_{\bf k}+i\Gamma_{\bf k}}
+\frac{v^2_{\bf k}}{\omega+E_{\bf k}-i\Gamma_{\bf k}}
\label{GorkGF}
\end{equation}
where the excitation spectrum is now:
\begin{equation}
E_{\bf k}=\sqrt{\xi^2_{\bf k}+W^2_{\bf k}}
\label{EkGrk}
\end{equation}
and we introduced Bogoliubov's coefficients:
\begin{eqnarray}
u^2_{\bf k}=\frac{1}{2}\left(1+\frac{\xi_{\bf k}}{E_{\bf k}}\right) \\
v^2_{\bf k}=\frac{1}{2}\left(1-\frac{\xi_{\bf k}}{E_{\bf k}}\right)
\label{u-v}
\end{eqnarray}
Then the spectral density is:
\begin{equation}
A(\omega{\bf k})=
\frac{u^2_{\bf k}}{\pi}\frac{\Gamma_{\bf k}}
{(\omega-E_{\bf k})^2+\Gamma^2_{\bf k}}+
\frac{v^2_{\bf k}}{\pi}\frac{\Gamma_{\bf k}}
{(\omega+E_{\bf k})^2+\Gamma^2_{\bf k}}+...
\label{GorkSDens}
\end{equation}
where now appear {\em two} peaks, narrow if $\Gamma_{\bf k}$ is small enough,
corresponding to ``Bogoliubov's'' quasiparticles. 

If there is no long -- range order, but only strong scattering by fluctuations
of short -- range order, characterized by some correlation length $\xi$, is
present in our system, it is easy to imagine, that spectral density possesses 
(in the same region of momentum space and energy) some precursor structure,
in the form of characteristic ``double -- hump'' structure,
as it is shown qualitatively in Fig. \ref{sdqual}. The widths of these maxima
are naturally determined by  parameter $v_F/\xi$, i.e. inverse time of flight
of an electron through the region of the size of $\xi$, where ``dielectric''
ordering is effectively conserved. Below we shall see that rigorous analysis
leads just to these results. In this sense, the further theoretical 
discussion will be devoted to justification of this qualitative picture.

\begin{figure}
\epsfxsize=9cm
\epsfysize=12cm
\epsfbox{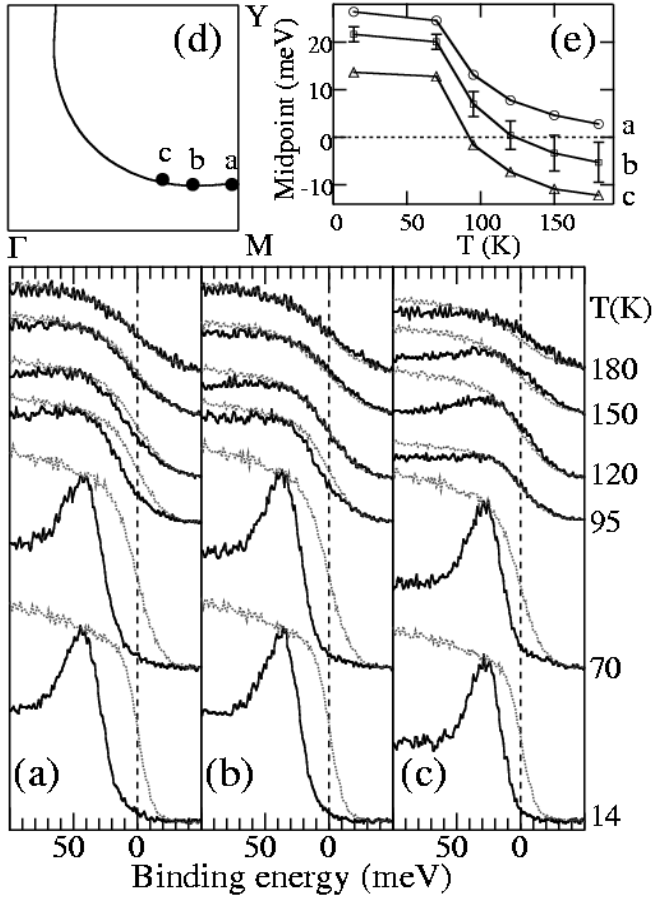}
\caption{ARPES spectra in three different points $a,b,c$ on the Fermi surface
of $Bi-2212$ for underdoped sample with $T_c=85K$. Thin curves --- spectrum of
the reference sample of $Pt$.}
\label{ARPES}
\end{figure}

In Fig. \ref{ARPES} we show ARPES data for $Bi_2Sr_2CaCu_2O_8$ \cite{Nrm}, 
obtained in three different points on the Fermi surface for different
temperatures.
The presence of the gap (pseudogap) manifests itself by the shift (to the left)
of the leading edge of energy distribution of photoelectrons in comparison to
the reference spectrum of a good metal $(Pt)$. It is seen that this gap is
closed at different temperatures for different values of ${\bf k}$, and the
gap width diminishes as we move from $(0,0)-(0,\pi)$ direction in the 
Brillouin zone. Pseudogap is completely absent in the direction of zone
diagonal $(0,0)-(\pi,\pi)$. At low temperatures this is in complete 
accordance with the picture of $d$ -- wave pairing, which is confirmed in
cuprates by many experiments \cite{Legg,Kirt}. Important thing, however, is
that this ``gap'' in ARPES data is observed also at temperatures significantly
higher than the temperature of superconducting transition $T_c$.

\begin{figure}
\epsfxsize=9cm
\epsfysize=12cm
\epsfbox{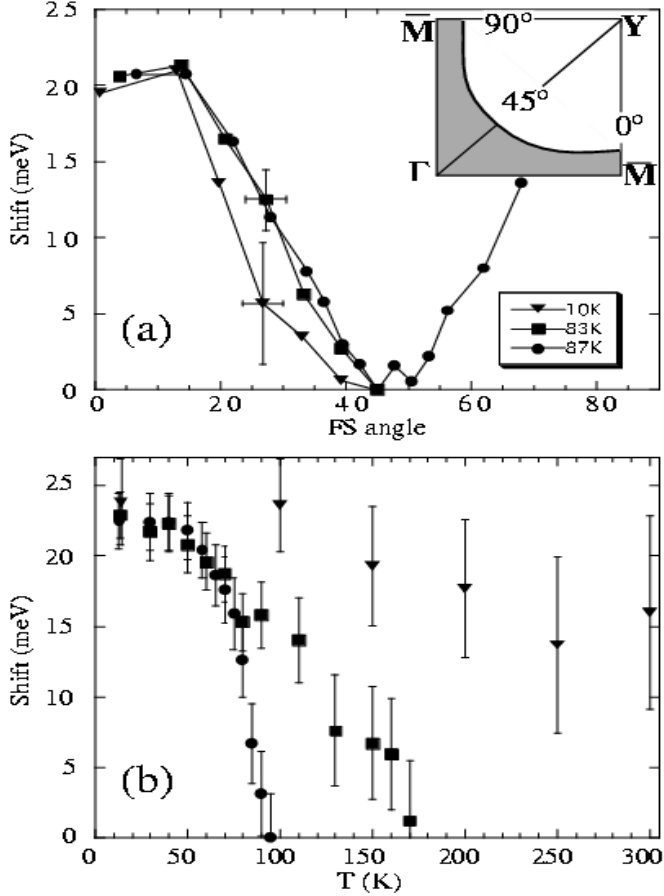}
\caption{Angular and temperature dependences of energy gap in $Bi-2212$,
obtained from ARPES data for samples with $T_c=87K$ (nearly optimally 
doped), $T_c=83K$ and $T_c=10K$ (underdoped): (a) -- the value of ARPES gap,
measured at different points of the Fermi surface (shown at the insert),
with positions determined by polar angle, measured from the direction
$\Gamma\bar M$, $d$ -- wave symmetry of the gap is obvious. 
(b) -- temperature dependence of the maximal gap, measured close to point
$\bar M$.
}
\label{ARPGAP}
\end{figure}

In Fig. \ref{ARPGAP} we show angular dependence of the gap in the Brillouin
zone and temperature dependence of its maximal value, obtained from ARPES
\cite{Din} for several samples of $Bi-2212$ with different compositions.
It is seen that the general $d$ -- wave like symmetry is conserved and
the gap in optimally doped sample vanishes practically at $T=T_c$, while for
underdoped samples we observe typical ``tails'' in the temperature
dependence of the gap for $T>T_c$. Qualitatively we may say that the 
formation of an anisotropic pseudogap at $T>T_c$, which smoothly transforms
into superconducting gap for $T<T_c$, leads to ``destruction'' parts of the 
Fermi surface of underdoped samples, close to the point $(0,\pi)$ (and
symmetrical to it), already for $T<T^*$, and the sizes of these parts grow
as temperature lowers \cite{Nrm}. 

\begin{figure}
\epsfxsize=7cm
\epsfysize=7cm
\epsfbox{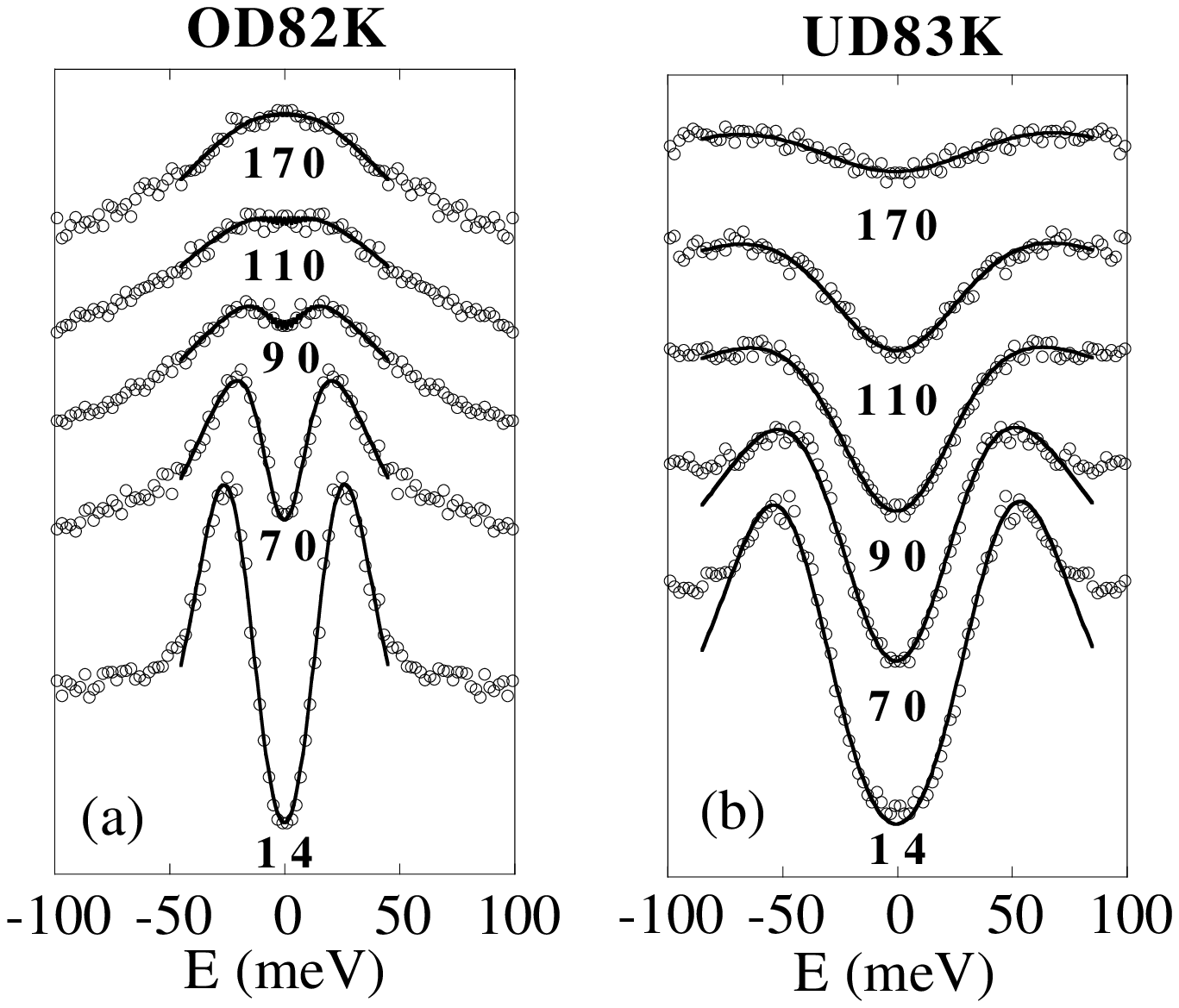}
\caption{Symmetrized ARPES spectra for overdoped sample of $Bi-2212$ with
$T_c=82K$ (a) and underdoped sample with $T_c=83K$ (b) at the point of 
intersection of the Fermi surface with the border of Brillouin zone
$(0,\pi)-(\pi,\pi)$. }
\label{Isym}
\end{figure}

Of central interest is, of course, the evolution of the spectral density
$A({\bf k}_F,\omega)$ at the Fermi surface. Under rather weak assumptions it
can be directly determined from ARPES data \cite{Nrm}. Assuming electron --
hole symmetry (always valid close enough to the Fermi surface, in reality
for $|\omega|$ less than some tenths of $meV$) we have
$A({\bf k}_F,\omega)=A({\bf k}_F,-\omega)$, so that taking into account
$f(-\omega)=1-f(-\omega)$, from (\ref{IARP}) and for ${\bf k}={\bf k}_F$ 
we obtain $I(\omega)+I(-\omega)=A({\bf k}_F,\omega)$. Thus, the spectral
density at the Fermi surface can be directly determined using symmetrized
experimental data $I(\omega)+I(-\omega)$. As an example of such analysis,
in Fig. \ref{Isym} we show the dtat of Ref. \cite{NorRan} for underdoped
sample of $Bi-2212$ with $T_c=83K$ and overdoped with $T_c=82K$ at different
temperatures. It is seen that pseudogap existence clearly manifests itself
in characteristic ``double -- humps'' structure of spectral density, which
appears (in an underdoped system) for temperatures significantly higher 
than $T_c$.
We see that these data completely correspond to the expected qualitative 
form of spectral density in the pseudogap state.

\begin{figure}
\epsfxsize=14cm
\epsfysize=14cm
\epsfbox{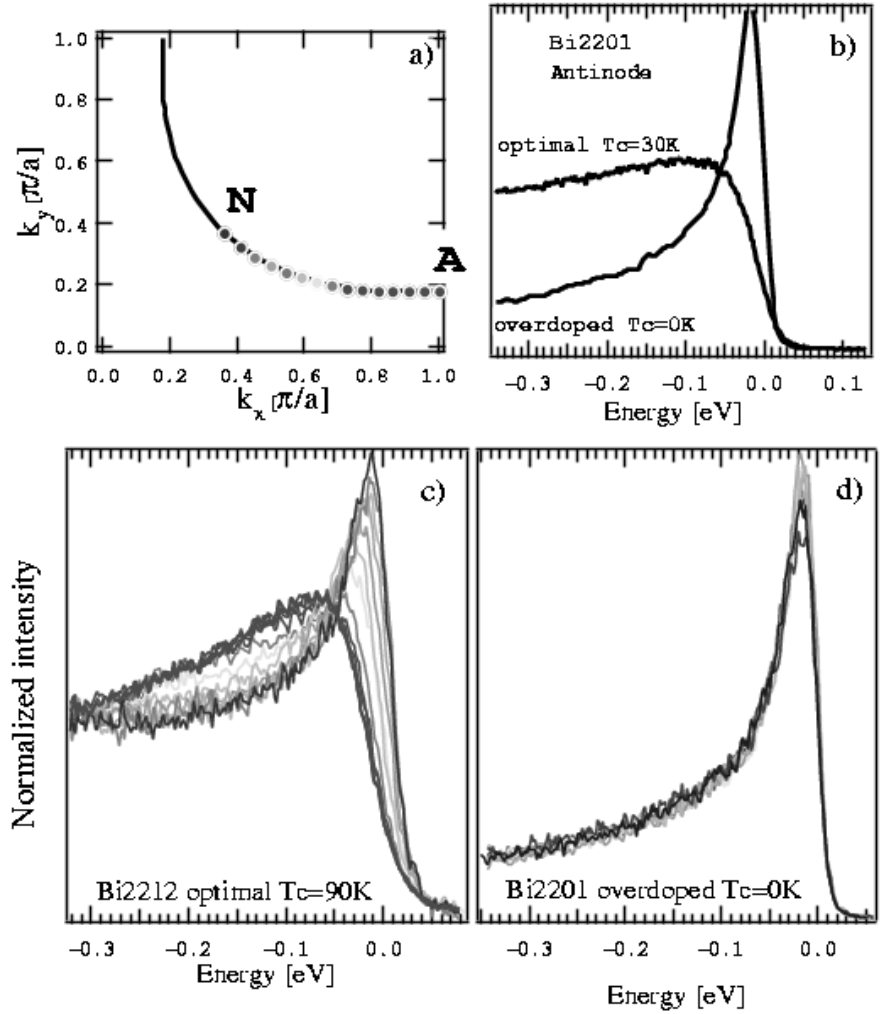}
\caption{ARPES spectra, measured at different point at the Fermi surface, 
moving from the diagonal direction in Brillouin zone to $(\pi,0)$: 
(a) -- points, where data were taken;
(b) -- comparison of data in в ``antinodal'' point $A$ (vicinity of $(\pi,0)$) 
in optimally doped and overdoped $Bi2201$;
(c) -- data for optimally doped $Bi2212$ with $T_c=90K$, obtained at $T=140K$;
(d) -- similar data for overdoped $Bi2201$ with $T_c=0$, obtained at $T=140K$.
} 
\label{nodal} 
\end{figure}

\begin{figure} 
\epsfxsize=8cm 
\epsfbox{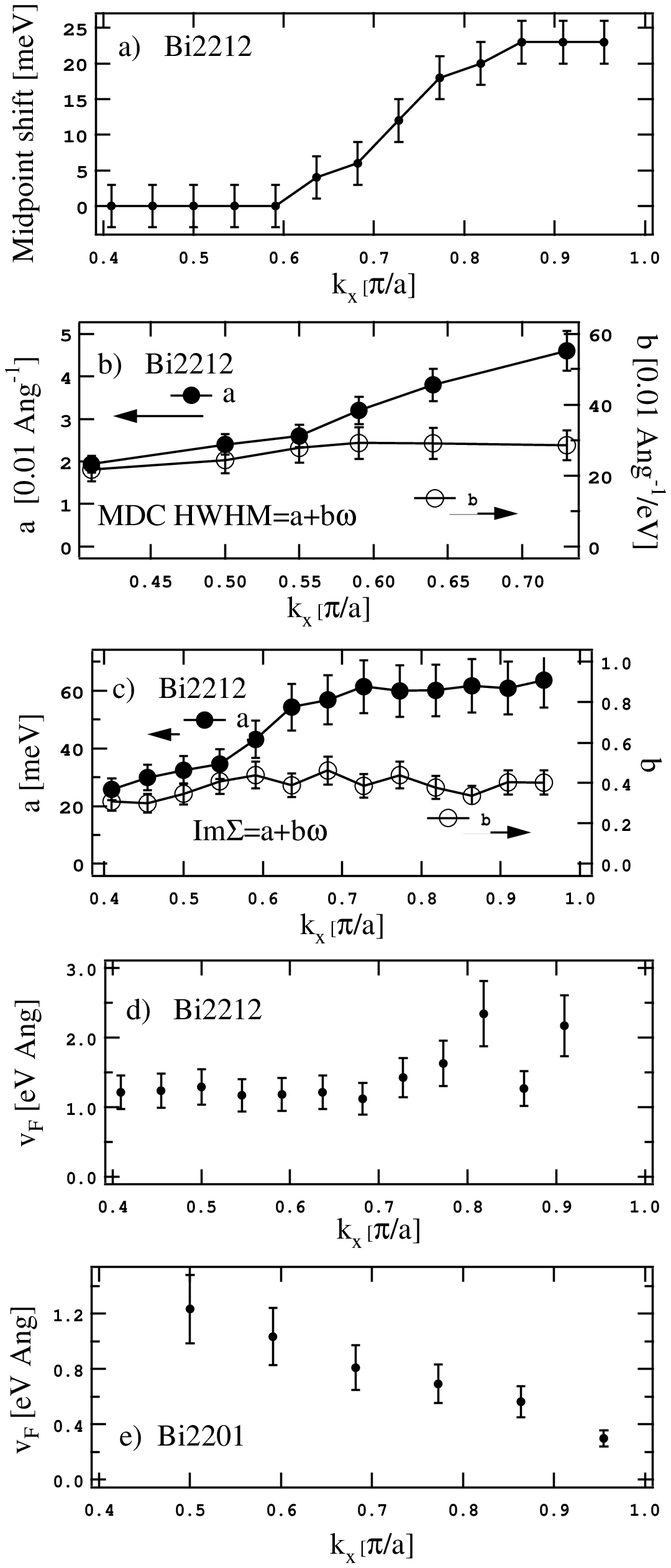}
\caption{The fit of data on ARPES spectrum width to ``marginal'' dependence
of the imaginary part of self -- energy: $Im\Sigma=a+b\omega$. Clearly seen is
anisotropy of static (quasistatic) scattering in optimally doped $Bi2212$, 
while dynamic (inelastic) scattering remains isotropic (c). Behavior of the
velocity on the Fermi surface as shown on (d) and (e) (overdoped $Bi2201$).} 
\label{gam_anis} 
\end{figure}

Let us stress once again, that well defined quasiparticles correspond to 
narrow enough peak in the spectral density $A({\bf k}_F,\omega)$ at
$\omega=0$. Such behavior, until recently, was in practically never observed in 
copper oxides. However, it was discovered some time ago that in superconducting 
phase, at $T\ll T_c$, there exists sharp enough peak of the spectral density,
corresponding to well defined quasiparticles, in the vicinity of an
intersection of the Fermi surface with diagonal of the Brillouin zone
(direction $(0,0)-(\pi,\pi)$), where superconducting gap vanishes  \cite{Kami}. 
At the same time, close to the point  $(0,\pi)$ the Fermi surface remains
``destroyed'' by both superconducting gap and the pseudogap. The studies of
these ``nodal'' quasiparticles id of great importance and lead to some
clarification of the problem of Fermi -- liquid behavior. Most recent data
show, that quasiparticle peak in diagonal direction persists also at 
temperatures much higher than $T_c$. This is clearly seen from the data of
Ref. \cite{Kam}, shown in Fig. \ref{nodal}, where we can see the evolution of
this peak as we move along the Fermi surface. It is seen that quasiparticle
behavior is valid everywhere for overdoped samples and only in the vicinity
of diagonal for underdoped (and optimal).  Quasiparticle peak in diagonal
direction persists even for strongly underdoped samples, which are on the
edge of metal -- insulator transition \cite{Zhou}.
From data shown in Fig. \ref{gam_anis} \cite{Kam}, we can also see significant
anisotropy of static (or more precisely quasistatic within limits of ARPES
resolution) scattering, growing as we move to the vicinity of $(0,\pi)$, 
which can obviously be related to pseudogap formation.  

\begin{figure}
\epsfxsize=12cm
\epsfysize=6cm
\epsfbox{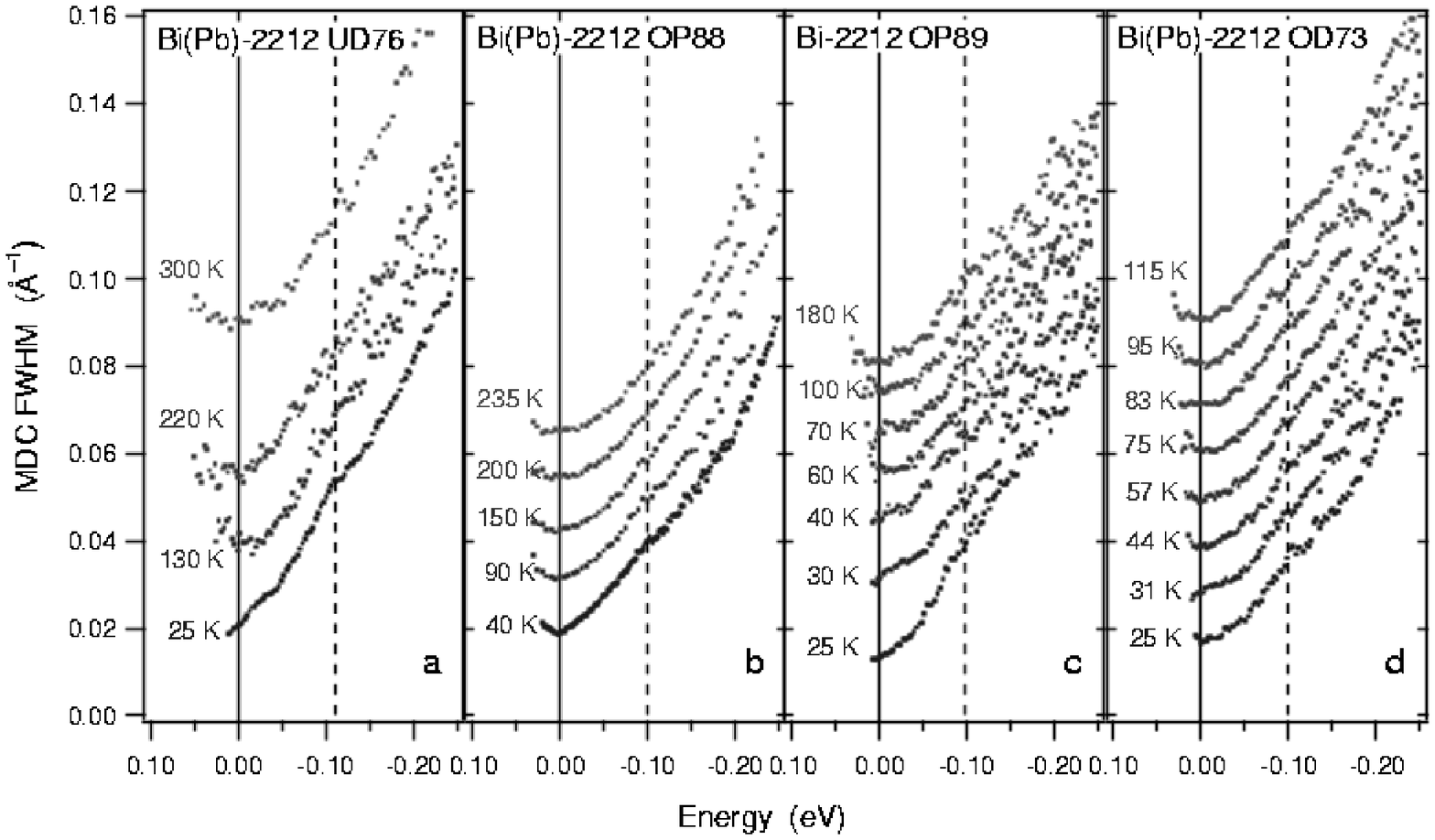}
\caption{Experimental data on the width of ARPES peaks in
$Bi(Pb)2212$ and $Bi2212$ for different temperatures, as doping level changes
from underdoped sample with $T_c=76K$ (a) to overdoped with с $T_c=73K$ (d).} 
\label{damp} 
\end{figure}

Finally, in Fig. \ref{damp} we show spectacular results of Ref. \cite{Kord}, 
obtained from ARPES measurements with significantly improved resolution,
from which, even by the ``naked eye'', we can see that effective damping
of ``nodal'' quasiparticles is in general (except possibly underdoped samples)
more or less quadratic in energy, as we can expect for a standard Fermi --
liquid. Only close to optimal doping and well into underdoped region we can
see some additional contribution to scattering, most probably of magnetic
nature.
These newest results provide quite new information on much discussed problem
of Fermi -- liquid behavior in HTSC -- oxides. Apparently we observe 
typical Fermi -- liquid behavior in overdoped samples, but it is ``destroyed''
as we move to optimal doping and into underdoped region, though only in parts
of the momentum space (in the vicinity of the points like $(0,\pi)$), 
where pseudogap appears, leading to additional strong scattering (damping). 

The growth of pseudogap anomalies as we move to the region around $(0,\pi)$,
where also becomes maximal the amplitude of superconducting $d$ -- wave gap,
is often interpreted as an evidence of $d$ -- wave symmetry of the pseudogap
and of its superconducting nature, smoothly transforming into the real gap 
for $T<T_c$. However, there is lot of evidence that pseudogap actually competes
with superconductivity and is, most probably, of dielectric nature.
Detailed discussion of this evidence can be found e.g. in Refs. \cite{MS,TL}.
Here we only limit ourselves to the most general argument --- pseudogap
anomalies in HTSC grow as we move deep into underdoped region, where
superconductivity just vanish. It is difficult to imagine, how this type of
behavior can be due to precursor Cooper pairing at $T>T_c$.

\begin{figure}
\epsfxsize=6cm
\epsfbox{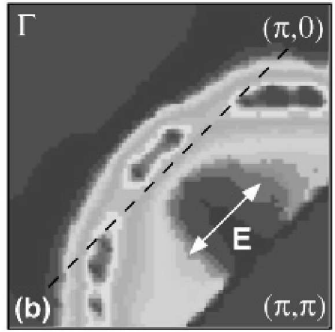}
\caption{Fermi surface of electronic superconductor 
$Nd_{1.85}Ce_{0.15}CuO_4$ in one quadrant of the Brillouin zone, obtained by
direct integration of ARPES spectra. Dashed line denotes the border of
antiferromagnetic Brillouin zone.} 
\label{nd_ce} 
\end{figure}

Of course, there are experimental data, giving direct evidence for 
``dielectric'' nature of the pseudogap. In Fig. \ref{nd_ce} we present
ARPES data on the Fermi surface of electronic superconductor
$Nd_{1.85}Ce_{0.15}CuO_4$ \cite{Arm}, which clearly show, that the
``destruction'' of the Fermi surface due to pseudogap formation takes place
in the region around ``hot spots'', appearing at intersections of the Fermi
surface with borders of the ``future'' antiferromagnetic Brillouin zone, which
would have appeared after the establishment of antiferromagnetic long -- range
order. Real gap in this case is obviously of the usual ``band'' or
``insulating'' nature, nothing to do with Cooper pairing of $d$ -- wave
symmetry. Most ARPES experiments in HTSC -- oxides are performed on systems
with hole -- like conductivity, where the distance (in momentum space) between
sides of the Fermi surface in the vicinity of $(0,\pi)$ is just smaller, than
in $NdCeCuO$. Also smaller in these systems is characteristic energy scale 
(width) of the pseudogap, as was mentioned previously during the discussion
of optical data shown in Fig. \ref{optcond}. Accordingly, the resolution of 
ARPES is apparently insufficient to resolve separate ``hot spots'', which are
too close to each other in the vicinity of $(0,\pi)$. In my opinion, the
results of Ref. \cite{Arm} practically solve the problem in favor of
``dielectric'' scenario of pseudogap formation in cuprates.

In conclusion, we must stress that in different experiments discussed above,
characteristic temperature $T^*$, defining crossover into the pseudogap state
can somehow change, depending on the property which is being studied. However,
in all cases there is some systematic dependence of $T^*$ on the doping level
and this temperature vanishes at some concentration of carriers slightly
higher than optimal.
\begin{figure}
\epsfxsize=8cm
\epsfbox{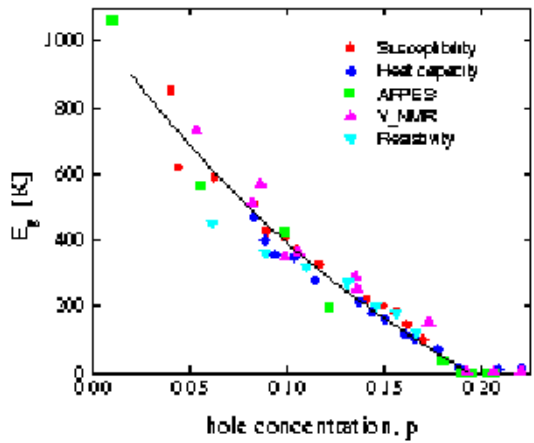}
\caption {Dependence of the energy width of the pseudogap $E_g$ in $YBCO$
on hole concentration determined from different experiments.}
\label{Eg}
\end{figure}
In Fig. \ref{Eg} we show a compendium of data (derived from a number of
different experiments) on the energy width of the pseudogap $E_g$ for
$YBCO$ as a function of hole concentration \cite{TL} 
(in this reference it was assumed, rather arbitrarily, that $E_g=2.5T^*$). 
It is seen that pseudogap  vanishes at the critical concentration 
$p_c\approx 0.19$, slightly higher than optimal concentration of carriers,
which, by the way, is an additional argument in favor of non superconducting
nature of the pseudogap in cuprates. 

More detailed review of miscellaneous experiments on the pseudogap formation
in high temperature superconductors can be found in Refs. \cite{Tim,TL}.


\section{Theoretical considerations -- variants of simplified model.}

As we mentioned above, there are two alternative scenarios to explain pseudogap
anomalies in HTSC -- systems. The first one is based on the model of Cooper 
pair formation already above the temperature of superconducting transition
(precursor pairing), while the second assumes, that the origin of the pseudogap
state is due to scattering by fluctuations of short -- range order of
``insulating'' type (e.g. antiferromagnetic (SDW) or charge density wave (CDW)),
existing in underdoped region of copper oxides. This second scenario seems to
be more attractive, both due to a number of experimental evidences and due to
a simple fact, that all pseudogap anomalies become stronger, as carrier
concentration drops and the system moves farther away from optimal
concentration for superconductivity towards dielectric (antiferromagnetic)
phase.

Consider typical Fermi surface of electrons moving in the $CuO_2$ plane, 
shown in Fig. \ref{f_surf}. If we neglect fine details, the observed 
(e.g. in ARPES) Fermi surface topology (and also the spectrum of elementary
excitations) in $CuO_2$ plane, in the first approximation are well enough
described by the usual tight -- binding model:
\begin{equation}
\varepsilon_{\bf k}=-2t(\cos k_xa+\cos k_ya)-4t'\cos k_xa\cos k_ya
\label{spectr}
\end{equation}
where $t\approx0.25eV$ is the nearest neighbor transfer integral, while
$t'$ is the transfer integral between second -- nearest neighbors, which
can change between  $t'\approx -0.45t$ for $YBa_2Cu_3O_{7-\delta}$ and 
$t'\approx -0.25t$ for $La_{2-x}Sr_xCuO_4$, $a$ is the square lattice constant.

\begin{figure} 
\epsfxsize=14cm
\epsfbox{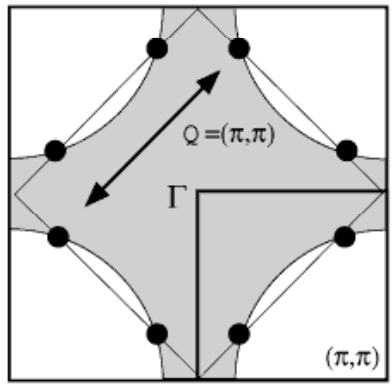}
\caption{Fermi surface in the Brillouin zone and ``hot spots'' model. 
Magnetic zone appears in the presence of antiferromagnetic long --
range order. ``Hot spots'' correspond to intersections of its borders
with Fermi surface and are connected by the scattering vector of the order
of ${\bf Q}=(\frac{\pi}{a},\frac{\pi}{a})$.}
\label{hspots}
\end{figure}

Phase transition to antiferromagnetic state induces lattice period doubling
and leads to the appearance of ``antiferromagnetic'' Brillouin zone in inverse
space as shown in Fig. \ref{hspots}. If the spectrum of carriers is given by
(\ref{spectr}) with $t'=0$ and we consider the half -- filled case, Fermi
surface becomes just a square conciding with the borders of antiferromagnetic
zone and we have a complete ``nesting'' --- flat parts of the Fermi surface
match each other after the translation by vector of antiferromagnetic
ordering ${\bf Q}=(\pm\pi/a,\pm\pi/a)$. In this case and for $T=0$ the 
electronic spectrum is unstable, energy gap appears everywhere on the
Fermi surface and the system becomes insulator, due to the formation of
antiferromagnetic spin density wave (SDW)\footnote{Analogous dielectrization
is realized also in the case of the formation of the similar charge density
wave (CDW).}. This picture corresponds to one of the popular schemes to
explain antiferromagnetism in cuprates, see e.g. Ref. \cite{SWZ} and a 
review paper \cite{Izy}, though, as was noted above, experimental situation is
well described by simple Heisenberg model with localized spins. 
In the case of the Fermi surface shown in Fig.\ref{hspots} the appearance of
antiferromagnetic long - range order, in accordance with general rules of the
band theory, leads to the appearance of discontinuities of 
isoenergetic surfaces (e.g. Fermi surface) at crossing points with borders of 
new (magnetic) Brillouin zone due to gap opening at points connected by 
vector ${\bf Q}$. Note similarity of this picture with experimental data
shown in Fig. \ref{nd_ce}.

In the region of cuprate phase diagram of interest to us antiferromagnetic
long -- range order is absent, however, a number of experiments support the
existence (everywhere below the $T^*$ -- line) of well developed fluctuations of
antiferromagnetic short -- range order which scatter electrons with
characteristic momentum transfer of the order of ${\bf Q}$. In principle, it
is not very important to consider AFM(SDW) fluctuations, similar effects may
be due CDW fluctuations. There is no doubt, that such scattering processes
exist (and dominate!) in cuprates in the whole pseudogap region. To convince
yourself, just look at the data shown in Fig. \ref{nodal}! 

For concreteness consider, however, the model of 
``nearly antiferromagnetic'' Fermi -- liquid \cite{Mont1,Mont2}, where the
effective interaction of electrons with spin fluctuations is described by
dynamic spin susceptibility $\chi_{\bf q}(\omega)$, the form of which was
determined by fitting to NMR experiments \cite{MilMon,Iz}:
\begin{equation}
V_{eff}({\bf q},\omega)=g^2\chi_{\bf q}(\omega)\approx
\frac{g^2\xi^2}{1+\xi^2({\bf q-Q})^2-i\frac{\omega}{\omega_{sf}}}
\label{V}
\end{equation}
where $g$ is coupling constant, $\xi$ -- correlation length of spin 
fluctuations, ${\bf Q}=(\pm\pi/a,\pm\pi/a)$ is vector of antiferromagnetic
ordering in dielectric phase,  $\omega_{sf}$ -- characteristic frequency of
spin fluctuations.

Dynamical spin susceptibility $\chi_{\bf q}(\omega)$ is peaked around
wave vectors $(\pm\pi/a,\pm\pi/a)$, and this leads to appearance of
``two types'' of quasiparticles --- ``hot'' one, with momenta in the vicinity
of  ``hot spots'' on the Fermi surface (Fig.\ref{hspots}) and energies 
satisfying the inequality ($v_F$ -- velocity at the Fermi surface):
\begin{equation}
|\varepsilon_{\bf k}-\varepsilon_{\bf k+Q}|<v_F/\xi,
\label{hsp}
\end{equation} 
and  ``cold'' one with momenta close to the parts of the Fermi surface
surrounding diagonals of Brillouin zone $|p_x|=|p_y|$ and not satisfying
(\ref{hsp}). This terminology is connected with strong scattering of
quasiparticles in the vicinity of ``hot spots'' with momentum transfer of
the order of ${\bf Q}$ due to interaction with spin fluctuations (\ref{V}), 
while for quasiparticles with momenta far from  ``hot spots'' this interaction
is weak enough. In the following we shall call this ``hot spots'' model.
Correlation length of fluctuations of short -- range antiferromagnetic order
$\xi$, described by (\ref{V}), is an important parameter of this theory. 
Note that in real HTSC -- systems this length is not very large, usually
$2a<\xi<8a$ \cite{BarzP,P97}. 

Characteristic frequency of spin fluctuations $\omega_{sf}$, depending on the
compound and doping level, is usually in the limits of $10-100K$ 
\cite{BarzP,P97}, so that in most part of the pseudogap region on the phase
diagram we have $2\pi T \gg \omega_{sf}$ and actually can neglect spin
dynamics, limiting ourselves to quasistatic approximation:
\begin{equation}
V_{eff}({\bf q})=W^2\frac{\xi^2}{1+\xi^2({\bf q-Q})^2}
\label{Vef}
\end{equation}
where $W$ is an effective parameter with dimensions of energy, which in the
model of AFM fluctuations can be written as \cite{Sch}:
\begin{equation}
W^2=g^2\frac{<{\bf S}_i^2>}{3}=g^2<(n_{i\uparrow}-n_{i\downarrow})^2>
\label{dd}
\end{equation}
where $g$ is interaction constant of electrons and spin fluctuations,
$<{\bf S}_i^2>$ is the average square of spin on a lattice site, 
$n_{i\uparrow}$,\ $n_{i\downarrow}$ -- operators of a number of electrons on
a given site with appropriate spin direction. In this approximation, dynamic
field of spin fluctuations is just replaced by the static Gaussian random
field of ``quenched''\footnote{In this case all ``loop'' insertions into
interaction lines of perturbation theory and corresponding to quantum 
corrections of higher orders to spin (or charge) fluctuations, and inevitably 
present in full dynamical problem, just vanish.} spins, antiferromagnetically 
correlated on lengths of the order of $\xi$.

It is clear that in the framework of our semiphenomenological approach, both
correlation length $\xi$ and parameter $W$ are to be considered as some
functions of carrier concentration (and temperature) to be determined from
the experiment. In particular, $W$ determines the effective width of the
pseudogap. Full microscopic theory of the pseudogap state is not our aim here,
and in the following we shall deal only with simple modelling of appropriate
transformation of electronic spectrum and its influence on different physical
properties, e.g. on superconductivity.

Considerable simplification of calculations can be achieved if we substitute
(\ref{Vef}) by model interaction of the following form \cite{KS} (similar
simplification was first used in Ref. \cite{Kamp}):  
\begin{equation} 
V_{eff}({\bf q})=W^2\frac{\xi^{-1}}{\xi^{-2}+(q_x-Q_x)^2} 
\frac{\xi^{-1}}{\xi^{-2}+(q_y-Q_y)^2}
\label{Veff}
\end{equation}
In fact  (\ref{Veff}) is qualitatively quite similar to (\ref{Vef}) and
almost do not differ from it quantitatively in most interesting region of
$|{\bf q-Q}|<\xi^{-1}$. This introduces effective one -- dimensionality 
into our problem.

\begin{figure}
\epsfxsize=14cm
\epsfysize=9cm
\epsfbox{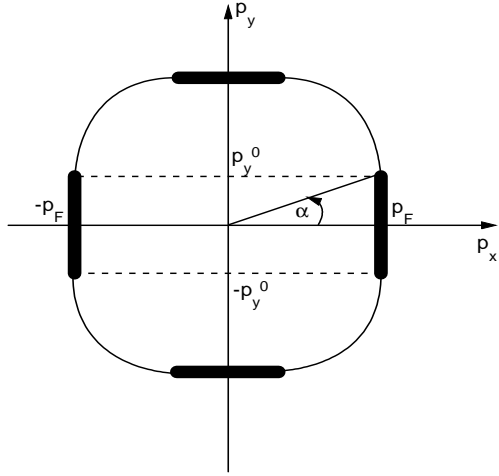}
\caption{Fermi surface with ``hot patches'', which are shown by thick
lines of the width $\sim\xi^{-1}$. The angle $\alpha$ determines the size of
a ``hot patch'', $\alpha=\pi/4$  corresponds to the square Fermi surface.}
\label{hpatches}
\end{figure}

Scattering by antiferromagnetic fluctuations in HTSC -- oxides is not always
most intensive at commensurate vector ${\bf Q}=(\pi/a,\pi/a)$, in general case
${\bf Q}$ may correspond to incommensurate scattering (see e.g. insert 
at Fig. \ref{f_surf}). Taking into account the observed topology of the Fermi 
surface with flat parts, as shown in Fig. \ref{f_surf}, we can introduce 
another model of scattering by fluctuations of short -- range order, which
we call the ``hot patches'' model \cite{PS}. In this model we assume the
Fermi surface of two -- dimensional electronic system as shown in 
Fig. \ref{hpatches}. The size of ``hot patches'' is determined by the
angular parameter $\alpha$. It is well known that flat parts of the Fermi
surface usually lead to instabilities towards formation of charge (CDW)
or spin (SDW) density wave and formation of the appropriate long -- range
order and (dielectric) energy gap at these flat parts. We are interested in
fluctuation region, when long -- range order is not yet established.
Fluctuations of the short -- range order are again assumed to be static
and Gaussian, and effective interaction is determined by (\ref{Veff}), 
with scattering vectors $Q_x=\pm 2p_F$,\ $Q_y=0$ or $Q_y=\pm 2p_F$,\ $Q_x=0$.
It is also assumed that fluctuations interact only with electrons from these
flat (``hot'') parts of the Fermi surface, shown in Fig. \ref{hpatches}, 
so that this scattering is in fact one -- dimensional. In the case of
$\alpha=\pi/4$ we have just a square Fermi surface and purely one --
dimensional problem. For $\alpha<\pi/4$ there are also ``cold'' parts of the
Fermi surface, where scattering is either absent or small.
The choice of scattering vector  ${\bf Q}=(\pm 2k_F,0)$ or 
${\bf Q}=(0,\pm 2p_F)$ corresponds, in general, to the case of incommensurate
fluctuations, as the Fermi momentum $p_F$ has no direct relation to the period
of inverse lattice. Commensurate case can also be analyzed within this model
\cite{PS}.

Thus, the main idea of models under discussion reduces to the assumption
of strong scattering by fluctuations of short -- range order, which, 
according to (\ref{V}), (\ref{Veff}), is effective in a limited part of 
momentum space with characteristic size of the order of $\xi^{-1}$ around
``hot'' spots or patches, which leads to pseudogap transformation of the
spectrum in these regions. It will be seen in the following, that within our 
assumptions these models can be solved  ``nearly exactly'', and much of the
remaining discussion will be devoted to a description of this solution. Mostly
we shall pay our attention to the discussion of the ``hot spots'' model as
more ``realistic'' and not described in detail in previous reviews.
As to ``hot patches'' model -- detailed discussion and further references
can be found in Ref. \cite{MS}.

\section{Elementary (``toy'') model of the pseudogap.}

Before we go to the analysis of ``realistic hot spots model'' it may be 
useful to consider an elementary one -- dimensional model of the pseudogap,
which allows an exact solution in analytic form \cite{S74}. Consider an
electron moving in one dimension in a random field of Gaussian fluctuations
with correlation function (in momentum representation and identified with
an interaction line in appropriate diagram technique) of the following form:
\begin{equation}
V_{eff}(Q)=2W^2\left\{\frac{\kappa}{(Q-2p_F)^2+\kappa^2}+
\frac{\kappa}{(Q+2p_F)^2+\kappa^2}\right\}
\label{VeffD}
\end{equation}
where $\kappa=\xi^{-1}(T)$. The choice of scattering vector $Q\sim\pm 2p_F$,  
corresponds to the case of incommensurate fluctuations. An exact solution can
be obtained in the asymptotic limit of $\xi\to\infty$ ($\kappa\to 0$), i.e. for
very large correlation length of fluctuations of short range order\footnote{
Let stress, that this limit here does not mean the establishment of any
long -- range order. Electron moves in the Gaussian random field with
special pair correlator, not in periodic system.}. Now we can sum {\em all}
Feynman diagrams of perturbation theory for an ``interaction'' of the form of
(\ref{VeffD}), which in this limit reduces to:
\begin{equation}
V_{eff}(Q)=2\pi W^2\{\delta(Q-2p_F)+\delta(Q+2p_F)\}
\label{SQ}
\end{equation}
Consider the simplest contribution to self -- energy of an electron, 
described by diagram shown in Fig. \ref{veff}, which we write in Matsubara 
representation:
\begin{figure} 
\epsfxsize=5cm 
\epsfbox{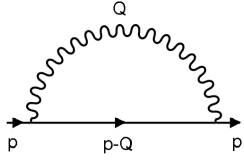}
\caption{Simplest diagram for self -- energy part of an electron.
Wavy line denotes interaction $V_{eff}(Q)$.} 
\label{veff} 
\end{figure} 
\begin{eqnarray}
\Sigma(\varepsilon_n p)=\int\frac{dQ}{2\pi}V_{eff}(Q)
\frac{1}{i\varepsilon_n-\xi_{p-Q}}\approx 
2W^2\int_{-\infty}^{\infty}\frac{dx}{2\pi}\frac{\kappa}
{x^2+\kappa^2}\frac{1}{i\varepsilon_n+\xi_p-v_Fx}=\nonumber\\
=2W^2\int_{-\infty}^{\infty}\frac{dx}{2\pi}\frac{\kappa}{(x-i\kappa)
(x+i\kappa)}\frac{1}{i\varepsilon_n+\xi_p-v_Fx}=\nonumber\\
=\frac{W^2}{i\varepsilon_n+\xi_p+iv_F\kappa}
\label{slfLRA}
\end{eqnarray}
where, for definiteness, we assumed $p\sim +p_F$,\ $\varepsilon_n>0$ and
defined a new integration variable $x$ as $Q=2p_F+x$. We also used here
``nesting'' property of the spectrum $\xi_{p-2p_F}=-\xi_p$, which is always
valid in one dimension for the standard $\xi_p=v_F(|p|-p_F)$.

\begin{figure} 
\epsfxsize=6cm 
\epsfbox{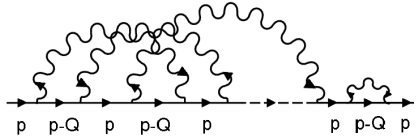}
\caption{Diagram of an arbitrary order for the single -- electron Green's
function.}
\label{veff_n} 
\end{figure}

The limit of $\xi(T)\to\infty$ ($\kappa\to 0$) should be understood as: 
\begin{equation}
v_F\kappa=v_F\xi^{-1}\ll Max\{2\pi T,\xi_p\}
\label{limxi}
\end{equation}
or
\begin{equation}
v_F\kappa=v_F\xi^{-1}\ll 2\pi T,\qquad \xi(T)\gg |p-p_F|^{-1}
\label{limxiT}
\end{equation}
Then (\ref{slfLRA}) reduces to:
\begin{equation}
\Sigma(\varepsilon_n p)\approx
\frac{W^2}{i\varepsilon_n+\xi_p}
\label{slfSK}
\end{equation}
Now, for an ``interaction'' of the form of (\ref{SQ}), there is no problem to
write down the contribution of an arbitrary diagram of the type shown in
Fig. \ref{veff_n}. In such diagram, in the $n$ -- th order in $V_{eff}(Q)$ 
we have  $2n$ vertices, connected, in all possible ways, by interaction lines. 
These lines alternatively\footnote{This alternation is important to guarantee
that an electron remains close to the Fermi surface  (points $\pm p_F$), or
large denominators will appear in terms of perturbation theory. This is not
important in the case of commensurate fluctuations, like period doubling,
when we are dealing with tight -- binding spectrum and ``bringing'' or 
``taking away'' of any number of momenta $Q=(\pi/a,\pi/a)$ does not take
electron far from the Fermi surface. In this case combinatorics of diagrams
is different \cite{Wonn}.} either ``take away'' or ``bring'' the
momenta $Q=2p_F$. As a result, the appropriate analytic expression for our
diagram contains alternating Green's functions 
$\frac{1}{i\varepsilon_n-\xi_p}$ (entering $n$ times) and 
$\frac{1}{i\varepsilon_n+\xi_p}$ (also entering $n$ times) plus an extra 
(initial) $\frac{1}{i\varepsilon_n-\xi_p}$\footnote{Of course, similar analysis
applies to the problem with an arbitrary scattering vextor $Q$, when we have
alternating $\frac{1}{i\varepsilon_n-\xi_p}$ and
$\frac{1}{i\varepsilon_n-\xi_{p-Q}}$. We have taken $Q=2p_F$ only to make
formula more compact and to put the pseudogap precisely at the Fermi level.}. 
Also we have to take an account of the factor $W^{2n}$. Finally, we can see,
that contributions of all diagrams in a given order just coincide and their
sum can be determined from pure combinatorics and is determined just by their
number, which is equal to $n!$. It is obvious --- there are  $2n$ points 
(vertices) with ``ingoing'' or ``outgoing'' interaction lines. Out of these, 
$n$ points are connected with ``outgoing'' lines, which can ``enter'' the
remaining ``free'' $n$ vertices in any of $n!$ ways. Use now the 
identity\footnote{In mathematics this is called Borel summation.}:
\begin{equation}
\sum_{n=0}^{\infty}n!z^n=\sum_{n=0}^{\infty}\int_{0}^{\infty}d\zeta e^{-\zeta}
(\zeta z)^n=\int_{0}^{\infty}d\zeta e^{-\zeta}\frac{1}{1-\zeta z}
\label{Borel}
\end{equation}
Then we can easily sum the {\em whole} series for the Green's function and
obtain the following exact solution:
\begin{eqnarray} 
G(\varepsilon_l p)=\sum_{n=0}^{\infty}\frac{W^{2n}n!}{(i\varepsilon_l- 
\xi_p)^n(i\varepsilon_l+\xi_p)^n(i\varepsilon_l-\xi_p)}\equiv
\sum_{n=0}^{\infty}n!z^n(\varepsilon_l,\xi_p)G_0(\varepsilon_l\xi_p)=\nonumber\\
=\int_{0}^{\infty}d\zeta e^{-\zeta}\frac{i\varepsilon_l+\xi_p}
{(i\varepsilon_l)^2-\xi^2_p-\zeta W^2}\equiv <G_{\zeta W^2}(
\varepsilon_l \xi_p)>_{\zeta},\qquad \varepsilon_l=(2l+1)\pi T
\label{GrSK}
\end{eqnarray}
where we have used the notation:
\begin{equation}
z(\varepsilon_l,\xi_p)=W^2G_0(\varepsilon_l,\xi_p)G_0(\varepsilon_l,-\xi_p)
\label{zGG}
\end{equation}
Now, what has appeared is just the ``normal'' Green's function of an insulator
(of Peierls type):
\begin{equation}
G_{W^2}(\varepsilon_l p)=\frac{i\varepsilon_l+\xi_p}
{(i\varepsilon_l)^2-\xi^2_p-W^2}
\label{GDnor}
\end{equation}
under the ``averaging'' procedure of the form:
\begin{equation}
<...>_{\zeta}=\int_{0}^{\infty}d\zeta e^{-\zeta}...
\label{zetaav}
\end{equation}
It is easy to convince yourself (formal proof, as well as many other details
on this model can be found in Ref. \cite{Diagr}) that (\ref{GrSK}) is just the
Green's function of an electron moving in an external field of the form
$2V\cos(2p_Fx+\phi)$, with amplitude ``fluctuating'' according to the so called
Rayleigh distribution\footnote{This distribution is well known in statistical
radiophysics, see e.g.: S.M.Rytov. Introduction to Statistical Radiophysics.\  
Part I.\ ``Nauka'',\ Moscow, 1976.}:  
\begin{equation}
{\cal P}(V)=\frac{2V}{W^2}e^{-\frac{V^2}{W^2}}
\label{Rayleigh}
\end{equation}
while the phase $\phi$ is distributed homogeneously on the interval from $0$ 
to $2\pi$.

\begin{figure} 
\epsfxsize=6cm 
\epsfysize=6cm 
\epsfbox{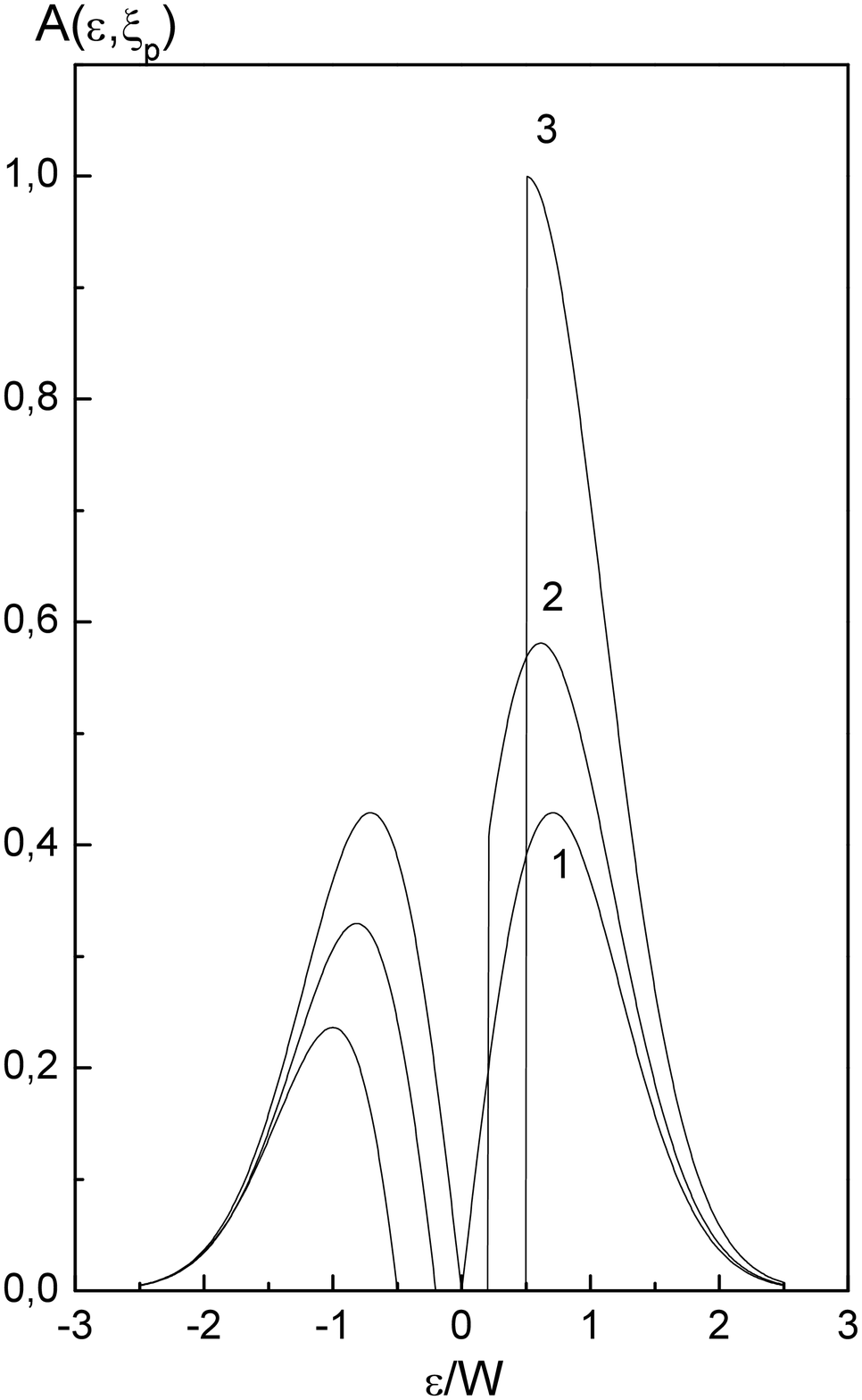}
\caption{Spectral density in the model of pseudogap state:\ 
(1)---$\xi_p=0$;\ (2)---$\xi_p=0.1 W$;\ (3)---$\xi_p=0.5 W$. }
\label{spdnsk} 
\end{figure}

Performing analytical continuation $i\varepsilon_l\to\varepsilon\pm i\delta$
from (\ref{GrSK}) we obtain (for $\varepsilon>0$):
\begin{eqnarray}
ImG^{R,A}(\varepsilon\xi_p)=\mp\pi(\varepsilon+\xi_p)\int_{0}^{\infty}d\zeta
e^{-\zeta}\delta(\varepsilon^2-\xi_p^2-\zeta W^2)=\nonumber\\
=\mp\frac{\pi}{W^2}(\varepsilon+\xi_p)\theta(\varepsilon^2-\xi_p^2)
e^{-\frac{\varepsilon^2-\xi^2_p}{W^2}}
\label{ImGSK}
\end{eqnarray}
so that the spectral density
\begin{equation}
A(\varepsilon\xi_p)=-\frac{1}{\pi}ImG^R(\varepsilon\xi_p)
\label{spdnsSK}
\end{equation}
has ``non Fermi -- liquid like '' form, shown in Fig. \ref{spdnsk}.
\begin{figure} 
\epsfxsize=6cm 
\epsfysize=6cm 
\epsfbox{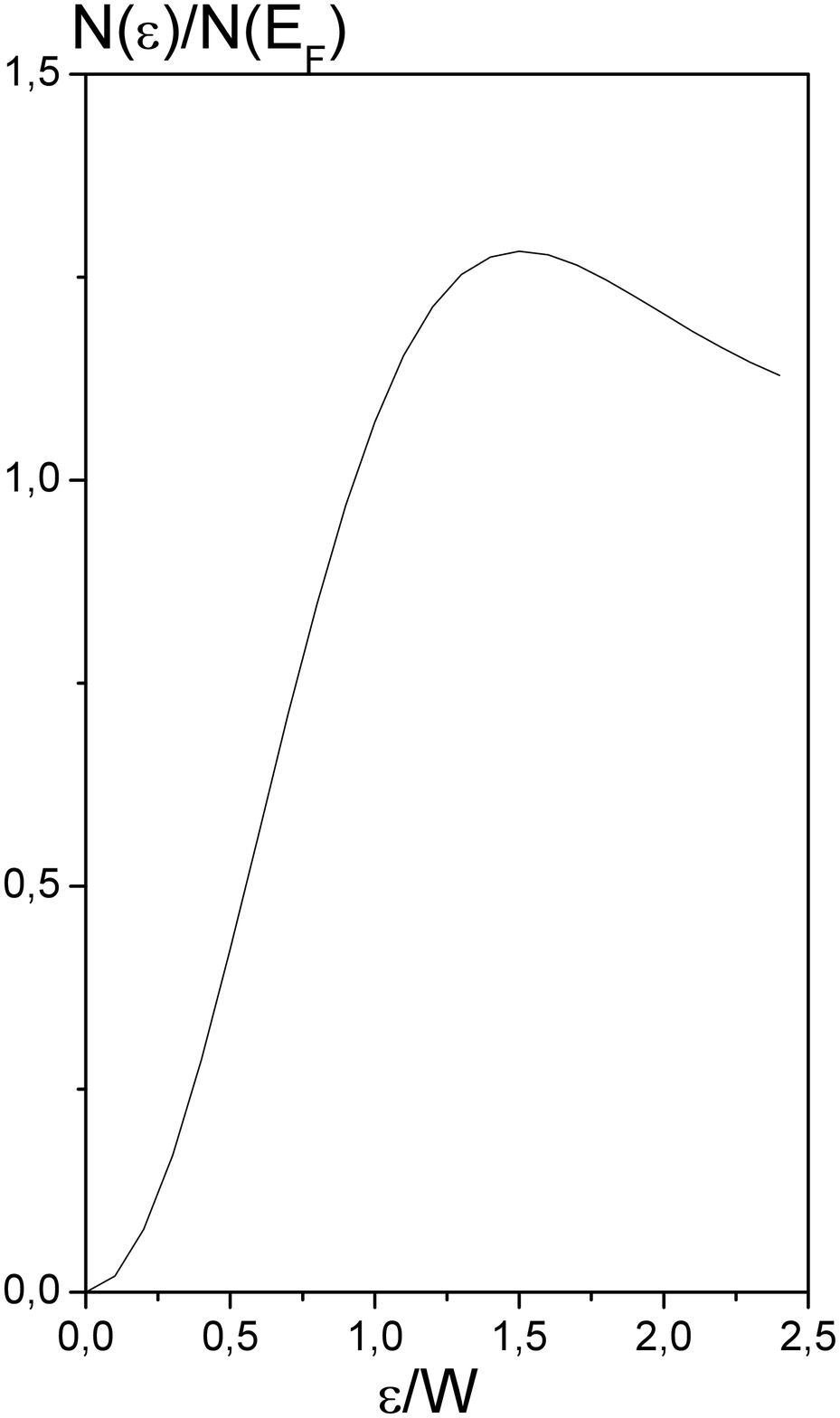}
\caption{Density of states with pseudogap.}
\label{dos_sk} 
\end{figure} 
Let us stress, that our Green's function does not have any poles on the
real axis of $\varepsilon$, which may correspond to quasiparticle energies as
required in Fermi -- liquid theory.

Electronic density of states has the following form:
\begin{eqnarray}
\frac{N(\varepsilon)}{N_0(E_F)}=\left|\frac{\varepsilon}{W}\right|
\int_{0}^{\frac{\varepsilon^2}{W^2}}d\zeta\frac{e^{-\zeta}}
{\sqrt{\frac{\varepsilon^2}{W^2}-\zeta}}=2\left|\frac{\varepsilon}
{W}\right|\exp\left(-\frac{\varepsilon^2}{W^2}\right)
Erfi\left(\frac{\varepsilon}{W}\right)=\nonumber\\
=\left\{\begin{array}{l}
1\qquad\mbox{при}\qquad |\varepsilon|\to\infty\\
\frac{2\varepsilon^2}{W^2}\qquad\mbox{при}\qquad |\varepsilon|\to 0
\end{array}
\right.
\label{PSGap}
\end{eqnarray}
where $N_0(E_F)$ is the density of states of free electrons at the Fermi level,
and $Erfi(x)=\int_{0}^{x}dxe^{x^2}$ is error function of imaginary argument.
Characteristic form of this density of states is shown in Fig. \ref{dos_sk} 
and demonstrate the presence of a ``soft'' pseudogap around the Fermi level.
Obviously, this is just the density of states of one -- dimensional insulator
with the energy gap $2V$, averaged over fluctuations of this gap with
probability distribution (\ref{Rayleigh}).

\begin{figure} 
\epsfxsize=6cm 
\epsfbox{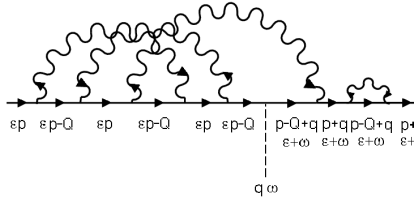}
\caption{Diagram of an arbitrary order for the vertex part of an interaction
with an external electromagnetic field.}
\label{vertx_n} 
\end{figure}

Remarkable property of this model is the possibility of obtaining an
exact solution (sum {\em all} diagrams) also for the response function to an 
external electromagnetic field \cite{S74,Diagr}. The arbitrary diagram for
the vertex part, describing the response to an external field, can be obtained
from an arbitrary diagram for the Green's function (of the type shown in
Fig. \ref{veff_n}) by an ``insertion'' of an external field line into any 
of electronic lines, as shown in Fig. \ref{vertx_n}.  
Making such ``insertions'' into all diagrams of the series (\ref{GrSK}) it is
possible (after some long, but direct calculations) to sum the whole series
for the vertwx part and obtain closed expression for response functions, e.g.
for polarization operator. Details can be found in Refs. \cite{S74,Diagr}). 
However, the structure of an answer is clear without any calculations ---
you must just calculate the response of an insulator with fixed gap
$2V$ and afterwards average the result over gap fluctuations with distribution
function (\ref{Rayleigh}). In particular, for polarization operator we obtain 
the following elegant expression $(\omega_m=2\pi mT)$:
\begin{eqnarray}
\Pi({\bf q}\omega_m)=\int_{0}^{\infty}d\zeta e^{-\zeta}2T\sum_n
\int_{-\infty}^{\infty}\frac{dp}{2\pi}\left\{
G_{\zeta W^2}(\varepsilon_n{\bf p})
G_{\zeta W^2}(\varepsilon_n+\omega_m{\bf p+q})\right.+ \nonumber\\    
\left.+F_{\zeta W^2}(\varepsilon_n{\bf p})
F^+_{\zeta W^2}(\varepsilon_n+\omega_m{\bf p+q})\right\}=
<\Pi_{\zeta W^2}({\bf q}\omega_m)>_{\zeta}
\label{PiSK}
\end{eqnarray}
where automatically appears the product of two ``anomalous'' Green's functions:
\begin{equation}
F(\varepsilon_n p)=\frac{W^*}{(i\varepsilon_n)^2-\xi_p^2-|W|^2},\quad
F^+(\varepsilon_n+\omega_m p)=
\frac{W}{(i\varepsilon_n+i\omega_m)^2-\xi_p^2-|W|^2}
\label{FGMP}
\end{equation}
describing Umklapp processes in a system with long -- range order \cite{Diagr}. 
Due to the absence of any long -- range order in our model, the value of 
(\ref{FGMP}) vanishes after averaging over phase, while the average of pair
of these functions in two -- particle response (\ref{PiSK}) is non zero.
Finally, under the averaging procedure over the gap fluctuations we have
simply the polarization operator of an insulator (of Peierls type).

\begin{figure} 
\epsfxsize=6cm 
\epsfysize=6cm 
\epsfbox{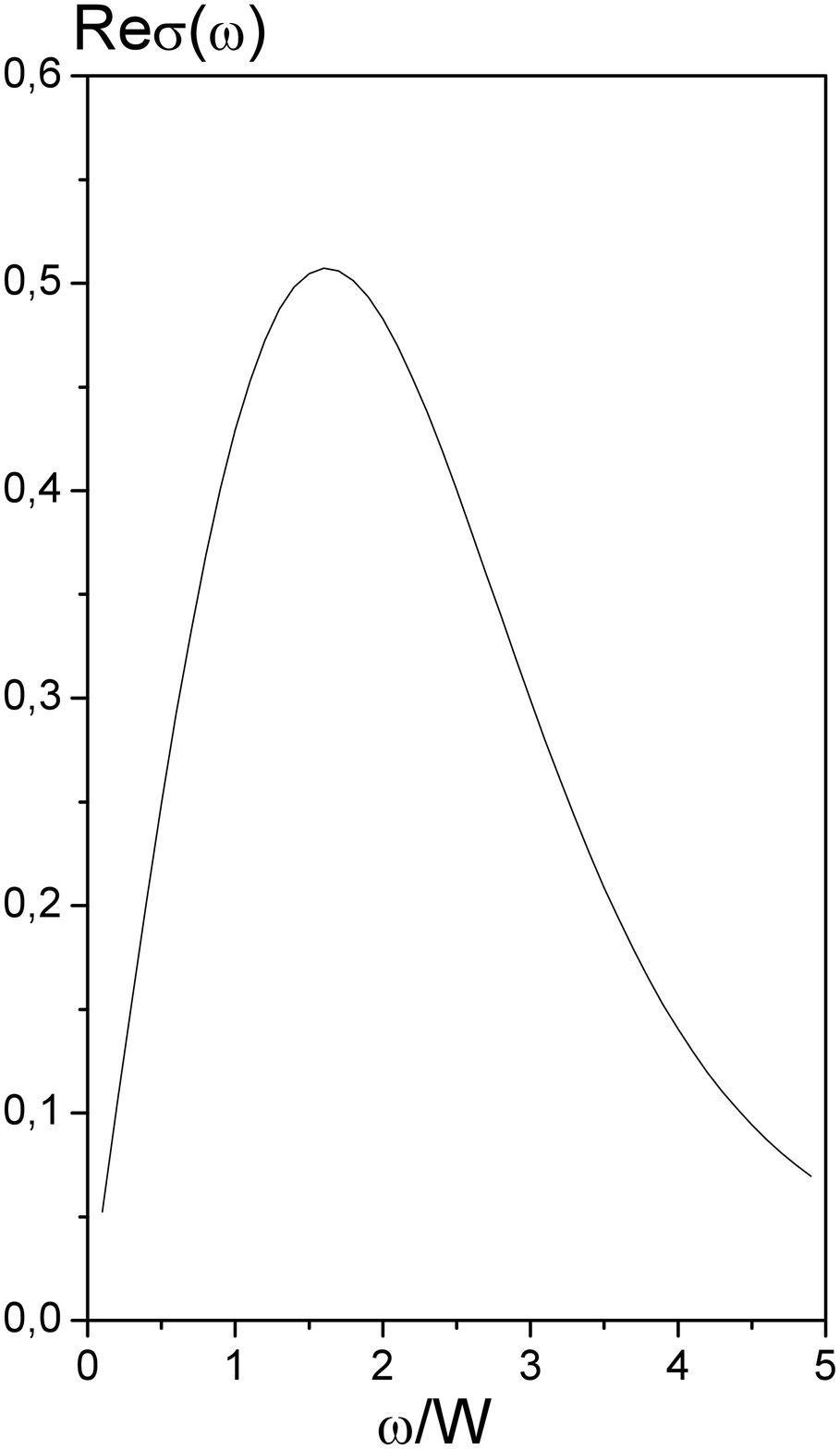}
\caption{Frequency dependence of the real part of conductivity in the
pseudogap state. Conductivity is given in units of 
$\frac{\omega_p^2}{4\pi W}$.} 
\label{opt_sk} 
\end{figure}

The real part of conductivity for such one -- dimensional insulator with
fixed gap $2W$ has the following form \cite{Diagr}:  
\begin{equation} 
Re\sigma_{W^2}(\omega)=\left\{
\begin{array}{l}
\frac{ne^2}{m\omega}
\frac{\pi}{\sqrt{\frac{\omega^2}{4W^2}-1}}\frac{W}{\omega}\quad
\mbox{при}\quad |\omega|>2W\\
0\qquad\mbox{при}\qquad  |\omega|<2W
\end{array}
\right.
\label{RecndD}
\end{equation}
It is seen, that absorption of electromagnetic energy is going through
quasiparticle excitation via the energy gap and is non zero for $\omega>2W$. 
In the pseudogap state this expression must be averaged over fluctuations
of $W$, described by distribution function (\ref{zetaav}) or (\ref{Rayleigh}).
So finally, from (\ref{RecndD}) we get:
\begin{equation}
Re\sigma(\omega)=\frac{\omega_p^2}{4}\frac{W}{\omega^2}
\int_{0}^{\frac{\omega^2}{4W^2}}d\zeta e^{-\zeta}\frac{\zeta}
{\sqrt{\frac{\omega^2}{4W^2}-\zeta}}
\label{condPSG}
\end{equation}
Appropriate frequency dependence id shown in Fig. \ref{opt_sk}.
We can see characteristic smooth maximum of absorption through the pseudogap.

This elementary model of the pseudogap state is very useful for an analysis of
a number of problems. It is easily generalized to two -- dimensional case for
the ``hot patches'' model \cite{PS}. This allows the analysis of the problem
of formation of superconducting state on the ``background'' of this
(dielectric) pseudogap \cite{PS,KS1,KS3,KS2,AC}. In particular, due to the
possibility of obtaining an exact solution in closed analytic form, it is
possible to study rather fine problems of the absence of self -- averaging
property of superconducting order parameter in the random field of pseudogap
fluctuations \cite{KS1}, showing the possible mechanism of formation of
local inhomogeneities (``superconducting drops'') at temperatures higher than
the mean -- field critical temperature of superconducting transition. This may
help to explain e.g. experimentally observed manifestations of 
superconductivity at these high temperatures (like anomalous Nernst effect),
which are usually interpreted in the spirit of superconducting scenario
of pseudogap formation. Possible direct connection with the picture of
inhomogeneous superconductivity, observed in STM experiments \cite{Pan,Davis},
is also obvious.

However, advantages of the model determine also its deficiencies. 
In particular, absolutely unrealistic is asymptotics of infinite correlation
length of pseudogap fluctuations. In real systems, as was noted above, this
correlation length is usually not larger than few interatomic spacings.
Besides that, the growth of correlation length inevitably leads to the 
breaking of our assumption of the Gaussian nature of pseudogap fluctuations.
Analysis of effects of finiteness of correlation length is actually a  
complicated problem. For one -- dimensional model such generalization of the
model under discussion was proposed in Ref. \cite{S79}. It was shown, that
as correlation length $\xi$ becomes smaller, it leads to smooth ``filling''
of the pseudogap, due to the growth of the scattering parameter $v_F/\xi$, i.e.
of the inverse time of flight of an electron through the region of the
size of $\sim\xi$, where effectively we have ``dielectric'' ordering. The
method used in Ref. \cite{S79} forms the basis of appropriate generalization
to two dimensions \cite{KS,Sch}, which will be discussed below during our
analysis of ``hot spots'' model. Let us also mention the simplified version
of one -- dimensional model with finite correlation length, similar in spirit
to the model of Ref. \cite{S74,S79}, proposed in Ref. \cite{BK} and used in
Ref. \cite{KS2} to analyze the problems of self -- averaging properties of
superconducting order parameter in the ``hot patches'' model.

\section{``Hot spots'' model.}

\subsection{``Nearly exact'' solution for one -- particle 
Green's function.}

Let us now describe our ``nearly exact'' solution for ``hot spots'' model.
Consider first -- order (in $V_{eff}$ (\ref{Veff})) contribution to electron
self -- energy, corresponding to simplest diagram, shown in Fig. \ref{veff}:
\begin{equation}
\Sigma(\varepsilon_n{\bf p})=\sum_{\bf q}V_{eff}({\bf q})
\frac{1}{i\varepsilon_{n}-\xi_{\bf p+q}}
\label{sig}
\end{equation}
For large enough correlation lengths $\xi$, the main contribution to the sum
over ${\bf q}$ comes from the vicinity of ${\bf Q}=(\pi/a,\pi/a)$. 
Then we can write:
\begin{equation}
\xi_{\bf p+q}=\xi_{\bf p+Q+k}\approx \xi_{\bf p+Q}+{\bf v_{\bf p+Q}k}
\label{linsp}
\end{equation}
where ${\bf v}_{\bf p+Q}=\frac{\partial\xi_{\bf p+Q}}{\partial{\bf p}}$ is the
appropriate velocity of a quasiparticle on the Fermi surface.
Then (\ref{sig}) is easily calculated and we get:
\begin{equation}
\Sigma(\varepsilon_n{\bf p})=\frac{W^2}{i\varepsilon_n-\xi_{\bf p+Q}
+i(|v^x_{\bf p+Q}|+|v^y_{\bf p+Q}|)\kappa sign\varepsilon_n}
\label{sigm}
\end{equation}
where $\kappa=\xi^{-1}$. Let us stress that both here and below 
``linearization'' of the quasiparticle spectrum (\ref{linsp}) under the 
integral (\ref{sig}) is performed only over small, due to large enough $\xi$, 
correction (of the order of $v_F\xi^{-1}$) to the spectrum close to 
the Fermi surface, while the forms of the spectrum itself
$\xi_{\bf p}$ and $\xi_{\bf p+Q}$ are given by the general expression
(\ref{spectr}) with $\xi_{\bf p}=\varepsilon_{\bf p}-\mu$.

\footnotesize

\begin{figure} 
\epsfxsize=6cm 
\epsfbox{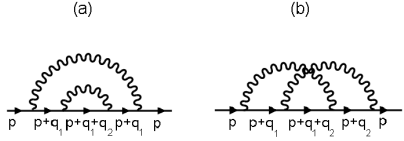}
\caption{Diagrams of second order in effective interaction with pseudogap
fluctuations.}
\label{sec_ord} 
\end{figure}

Consider now correction of the second order, shown in Fig. \ref{sec_ord}. 
Using (\ref{Veff}) we obtain:
\begin{eqnarray}
\Sigma(a)=\Delta^4\int\frac{d{\bf k_1}}{\pi^2}\int\frac{d\bf k_2}{\pi^2}
\frac{\kappa}{\kappa^2+k_{1x}^2}\frac{\kappa}{\kappa^2+k_{1y}^2}
\frac{\kappa}{\kappa^2+k_{2x}^2}\frac{\kappa}{\kappa^2+k_{2y}^2} \nonumber \\
\frac{1}{i\varepsilon_n-\xi_{\bf p+Q}-v^x_{\bf p+Q}k_{1x}-v^y_{\bf p+Q}k_{1y}}
\frac{1}{i\varepsilon_n-\xi_{\bf p}-v^x_{\bf p}(k_{1x}+k_{2x})
-v^y_{\bf p}(k_{1y}+k_{2y})}
\nonumber \\
\frac{1}{i\varepsilon_n-\xi_{\bf p+Q}-v^x_{\bf p+Q}k_{1x}-v^y_{\bf p+Q}k_{1y}}
\label{sig1}
\end{eqnarray}

\begin{eqnarray}
\Sigma(b)=\Delta^4\int\frac{d{\bf k_1}}{\pi^2}\int\frac{d\bf k_2}{\pi^2}
\frac{\kappa}{\kappa^2+k_{1x}^2}\frac{\kappa}{\kappa^2+k_{1y}^2}
\frac{\kappa}{\kappa^2+k_{2x}^2}\frac{\kappa}{\kappa^2+k_{2y}^2} \nonumber \\
\frac{1}{i\varepsilon_n-\xi_{\bf p+Q}-v^x_{\bf p+Q}k_{1x}-v^y_{\bf p+Q}k_{1y}}
\frac{1}{i\varepsilon_n-\xi_{\bf p}-v^x_{\bf p}(k_{1x}+k_{2x})
-v^y_{\bf p}(k_{1y}+k_{2y})}
\nonumber \\
\frac{1}{i\varepsilon_n-\xi_{\bf p+Q}-v^x_{\bf p+Q}k_{2x}-v^y_{\bf p+Q}k_{2y}}
\label{sig2}
\end{eqnarray}
where we have used the explicit form of the spectrum (\ref{spectr}), from which
it follows, in particular, that
$\xi_{\bf p+2Q}=\xi_{\bf p}$, ${\bf v}_{\bf p+2Q}={\bf v}_{\bf p}$
for ${\bf Q}=(\pi/a,\pi/a)$. If the signs of $v_{\bf p}^x$ and $v_{\bf p+Q}^x$, 
as well as of $v_{\bf p}^y$ and $v_{\bf p+Q}^y$ coincide, the integrals in
(\ref{sig1}) and (\ref{sig2}) are completely determined by contributions from
the poles of Lorentzians, which describe interaction with fluctuations of
short -- range order, so that after elementary contour integration we get:
\begin{equation}
\Sigma(a)=\Sigma(b)=\frac{1}{[i\varepsilon_n-\xi_{\bf p+Q}+i(|v_{\bf p+Q}^x|
+|v_{\bf p+Q}^y|)\kappa]^2}\frac{1}{i\varepsilon_n-\xi_{\bf p}+
i2(|v_{\bf p}^x|+|v_{\bf p}^y|)\kappa}
\label{sig3}
\end{equation}
Here and in the following, for definiteness, we assume $\varepsilon_n>0$. 

It is not difficult to convince yourself, that in case of coinciding signs
of velocity projections at ``hot spots'', similar calculation is valid for
an arbitrary diagram of higher order.

\normalsize

Thus, in case of coinciding signs of velocity projections at the Fermi surface
$v^x_{\bf p}$ and $v^x_{\bf p+Q}$, and those of $v^y_{\bf p}$ and 
$v^y_{\bf p+Q}$, Feynman integrals in any diagram of arbitrary order are
determined only by contributions form the poles of Lorentzians in 
(\ref{Veff}) and are easily calculated. Similar situation holds also in the  
case of velocities at ``hot spots'', connected by vector ${\bf Q}$, are 
perpendicular to each other. In this case the contribution of an arbitrary
diagram of $N$ -- th order in (\ref{Veff}) for electron self -- energy has
the following form:
\begin{equation}
\Sigma^{(N)}(\varepsilon_n{\bf p})=W^{2N}\prod_{j=1}^{2N-1}
\frac{1}{i\varepsilon_n-\xi_{j}({\bf p})+in_jv_j\kappa}
\label{Ansatz}
\end{equation}
where $\xi_j({\bf p})=\xi_{\bf p+Q}$ and 
$v_j=|v_{\bf p+Q}^{x}|+|v_{\bf p+Q}^{y}|$ 
for odd $j$ and $\xi_j({\bf p})=\xi_{\bf p}$ and $v_{j}=
|v_{\bf p}^x|+|v_{\bf p}^{y}|$ for even $j$. 
Here $n_j$ is the number of interaction lines, surrounding $j$ -- th
Green's function (counting from the first one in diagram) and we again take 
$\varepsilon_n>0$.
  
In Ref. \cite{KS} we have studied in detail when these conditions on velocity
projections are satisfied in the points of the Fermi surface connected by 
vector ${\bf Q}$ (``hot spots'') and have presented explicit examples of
appropriate geometries of the fermi surface, which can be realized for
specific relations between parameters $t$ and $t'$ in (\ref{spectr}). 
In these cases expression (\ref{Ansatz}) is exact, with only limitation being 
related to our use of ``linearization'' (\ref{linsp}). In all other cases
(for other relations between $t$ and $t'$) we use (\ref{Ansatz}) as rather
successful {\em Ansatz} for the contribution of an arbitrary order, obtained
by simple continuation of spectrum parameters $t$ and $t'$ to the region of
interest to us. Even in most inappropriate one -- dimensional case \cite{S79}, 
corresponding to square Fermi surface, appearing for (\ref{spectr}) if $t'=0$ 
and $\mu=0$, the use of this {\em  Ansatz} produces results (e.g. for the 
density of state) which are very close quantitatively \cite{Sad} to the 
results of an exact numerical simulation of this problem \cite{Brt}. 

Generally speaking, note that in standard diagrammatic approaches we usually
perform summation some {\em subseries} of diagrams, which are considered as
dominant over some smallness parameter. Here we are dealing with much more
rare situation --- we can sum the {\em whole} diagram series, though 
contribution of each diagram is calculated probably approximately. It is in
this sense that we use the term ``nearly exact'' solution.

Using {\em Ansatz} (\ref{Ansatz}) we can see, that  {\em the contribution
of an arbitrary diagram with intersecting interaction lines is actually equal 
to the contribution of some diagram of the same order without intersections of
these lines} \cite{S79}. Thus, in fact we can limit ourselves to consideration
of only  diagrams without intersecting interaction lines, taking the 
contribution of diagrams with intersections into account with the help of
additional combinatorial factors, which are attributed to ``initial'' vertices
or just interaction lines \cite{S79}. As a result we obtain the following
recursion relation (continuous fraction representation \cite{S79}), which
gives an effective algorithm for numerical computations \cite{KS}:
\begin{eqnarray}
G_{k}(\varepsilon_n\xi_{\bf p})=\frac{1}{i\varepsilon_n-
\xi_k({\bf p})+ikv_k\kappa
-\Sigma_{k+1}(\varepsilon_n\xi_{\bf p})}\equiv\nonumber\\
\equiv\left\{G^{-1}_{0k}(\varepsilon_n\xi_{\bf p})
-\Sigma_{k+1}(\varepsilon_n\xi_{\bf p})\right\}^{-1}
\label{G}
\end{eqnarray}
\begin{equation}
\Sigma_{k}(\varepsilon_n\xi_{\bf p})=W^2\frac{v(k)}
{i\varepsilon_n-\xi_k({\bf p})+ikv_k\kappa-
\Sigma_{k+1}(\varepsilon_n\xi_{\bf p})}
\label{rec}
\end{equation}
\begin{figure} 
\epsfxsize=6cm 
\epsfbox{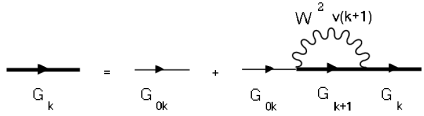}
\caption{Graphical representation of recursion relation for the single --
particle Green's function.}
\label{recurr} 
\end{figure} 
Graphically this recursion relation for the Green's function is shown
in Fig. \ref{recurr}. The ``physical'' Green's function is obtained as:\ 
$G(\varepsilon_n\xi_{\bf p})=G_{k=0}(\varepsilon_n\xi_{\bf p})$.  
In (\ref{G}) we also defined: 
\begin{equation} G_{0k}(\varepsilon_n\xi_{\bf p})=
\frac{1}{i\varepsilon_n-\xi_k({\bf p})+ ikv_k\kappa} 
\label{G0k} 
\end{equation}
Combinatorial factor:
\begin{equation}
v(k)=k
\label{vcomm}
\end{equation}
for the case of commensurate fluctuations with
${\bf Q}=(\pi/a,\pi/a)$ \cite{S79}, if we don not take into spin structure od
interaction (CDW -- type fluctuations). 
For incommensurate CDW fluctuations \cite{S79}:
\begin{equation} 
v(k)=\left\{\begin{array}{cc}
\frac{k+1}{2} & \mbox{for odd $k$} \\
\frac{k}{2} & \mbox{for even $k$}
\end{array} \right.
\label{vinc}
\end{equation}

If we take into account (Heisenberg) spin structure of interaction with 
pseudogap fluctuations in  ``nearly antiferromagnetic Fermi -- liquid'' (spin --
fermion model \cite{Sch}), combinatorics of diagrams becomes more complicated.
Thus, spin -- conserving scattering processes obey commensurate combinatorics,
while spin -- flip scattering is described by diagrams of incommensurate
case (``charged'' random field in terms of Ref. \cite{Sch}). In this model
recursion relation for the Green's function again is given by (\ref{rec}), 
but combinatorial factor $v(k)$ takes now the following form \cite{Sch}:  
\begin{equation} 
v(k)=\left\{\begin{array}{cc}
\frac{k+2}{3} & \mbox{for odd $k$} \\
\frac{k}{3} & \mbox{for even $k$}
\end{array} \right.
\label{vspin}
\end{equation}

The obtained solution for the single -- particle Green's function is
asymptotically exact in the limit of $\xi\to\infty$, when solution can be
found also in analytical form  \cite{S74,Sch}. It is also exact in trivial
limit of  $\xi\to 0$, when for fixed values of  $W$ interaction (\ref{Veff}) 
just vanishes. For all intermediate values of $\xi$ our solution gives, as
already noted, very good interpolation, being practically exact for certain
geometries of the Fermi surface, appearing for certain relations between
parameters of the spectrum (\ref{spectr}) \cite{KS}. Note also that our
formalism can be easily used also to describe pseudogap within
superconducting scenario \cite{KS,Ya}. Our preference for ``dielectric'' 
scenario is based mainly on physical considerations.

Using (\ref{G}) we can easily perform numerical calculations of  
single -- electron spectral density:
\begin{equation}
A(E{\bf p})=-\frac{1}{\pi}ImG^R(E{\bf p})
\label{spdns}
\end{equation}
which can also be determined from ARPES experiments \cite{MS}.
In (\ref{spdns}) $G^R(E{\bf p})$ represents retarded Green's function
obtained by the usual analytic continuation of (\ref{G}) from Matsubara
frequencies to the real axis of $E$. Analogously we cam compute single --
particle density of states as:
\begin{equation}
N(E)=\sum_{\bf p}A(E{\bf p})=-\frac{1}{\pi}\sum_{\bf p}ImG^R(E{\bf p})
\label{NE}
\end{equation}
Details of these calculations and discussion of the results for our
two -- dimensional model can be found in Refs. \cite{Sch,KS}. Here we shall
demonstrate only few most important results.

\begin{figure}
\epsfxsize=7cm
\epsfysize=7cm
\epsfbox{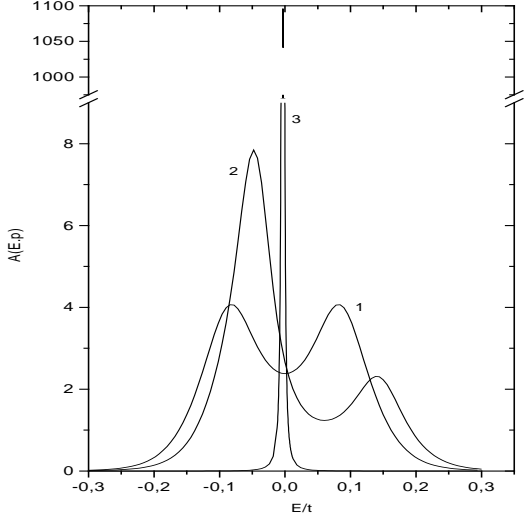}
\caption{Spectral density in the ``hot spots'' model, the case of
incommensurate fluctuations, $\kappa a=0.01$:\ 
(1) -- at the ``hot spot'' 
$p_xa/\pi=0.142, p_ya/\pi=0.587$,\ 
(2) -- close to ``hot spot'' at 
$p_xa/\pi=0.145, p_ya/\pi=0.843$,\ 
(3) -- far from ``hot spot'' at
$p_xa/\pi=p_y/\pi=0.375$.}
\label{sdnhs}
\end{figure}

\begin{figure}
\epsfxsize=7cm
\epsfysize=7cm
\epsfbox{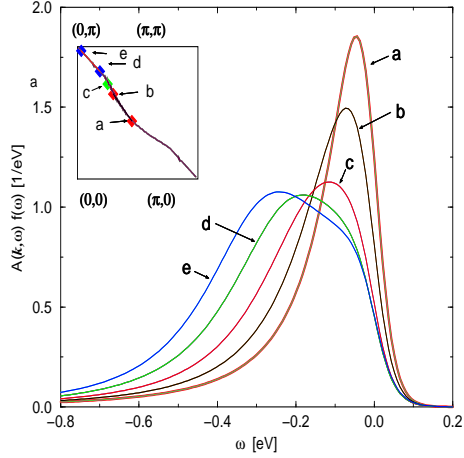}
\caption{The product of spectral density and Fermi distribution function at
different points on the Fermi surface, shown at the insert.
Spin -- fermion model, correlation length $\xi=3a$.}
\label{fspdns}
\end{figure}

As a typical example in Fig. \ref{sdnhs} we show the results \cite{KS} for
spectral density of electrons for the incommensurate (CDW) case.
We can see that spectral density close to the ``hot spot'' has an expected
non Fermi -- liquid like form and there are no well defined quasiparticle
peaks. Far from the ``hot spot'' spectral density is characterized by a narrow
peak, corresponding to well defined quasiparticles (Fermi -- liquid).
In Fig. \ref{fspdns} from Ref. \cite{Sch} we show the product of the Fermi
distribution and spectral density at different points of the
``renormalized'' Fermi surface, defined by the equation
$\varepsilon_{\bf p}-Re\Sigma(E=0{\bf p})=0$, where the ``bare'' spectrum
$\varepsilon_{\bf p}$ is given by (\ref{spectr}) with
$t=-0.25eV, t'=-0.35t$ and for hole concentration $n_h=0.16$, with coupling
constant in (\ref{V}) $g=0.8eV$ and correlation length $\xi=3a$ (commensurate
case, spin -- fermion model). We can clearly see complete qualitative 
agreement with ARPES data discussed above with quite different behavior
close and far from the ``hot spot''.
\begin{figure}
\epsfxsize=7cm
\epsfysize=7cm
\epsfbox{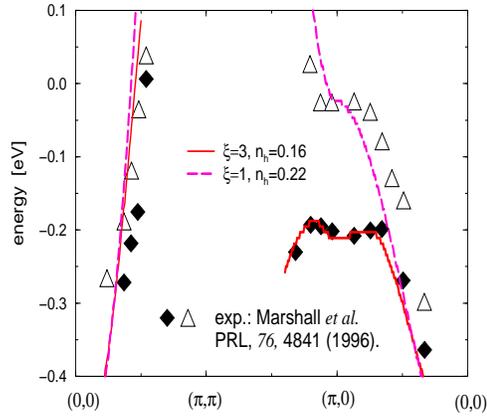}
\caption{Positions of the maximum of the spectral density for different
values of correlation length $\xi$ and hole concentrations, calculated for
spin -- fermion model and compared with ARPES data for
$Bi_2Sr_2Ca_{1-x}Dy_{x}Cu_2O_{8+\delta}$ with $x=1$ (triangles) and 
$x=0.175$ (diamonds).}
\label{locmax}
\end{figure}
Finally, in Fig. \ref{locmax} taken from Ref. \cite{Sch} we show calculated
(for spin fermion -- model) positions of the maximum of
$A(\omega{\bf k})$ for two different concentrations of holes compared with
appropriate experimental ARPES data of Ref. \cite{Marsh} for
$Bi_2Sr_2Ca_{1-x}Dy_{x}Cu_2O_{8+\delta}$. The point is, that positions of the
maxima of spectral density in the plane of $(\omega,{\bf k})$, determined from 
ARPES, in an ideal system of the Fermi -- liquid type define dispersion 
(spectrum) of quasiparticles (cf. Fig. \ref{sdqual}(a)). 
For overdoped system the 
values of $n_h=0.22$ and $\xi=a$ were assumed during these calculations. 
The results obtained demonstrate rather well defined dispersion curves
both along the diagonal of Brillouin zone, and in the direction 
$(0,0)-(\pi,0)$. For underdoped system it was assumed that $n_h=0.16$ and 
$\xi=3a$. In this case, in diagonal direction we again see spectral curve
crossing the Fermi level, while close to ``hot spots'' (in the vicinity
of $(\pi,0)$) there is seen only a smeared maximum of spectral density remaining
approximately $200\ meV$ below the Fermi level (pseudogap). In general, 
agreement between theory and experiment is rather satisfactory.

\begin{figure}
\epsfxsize=6cm
\epsfysize=5cm
\epsfbox{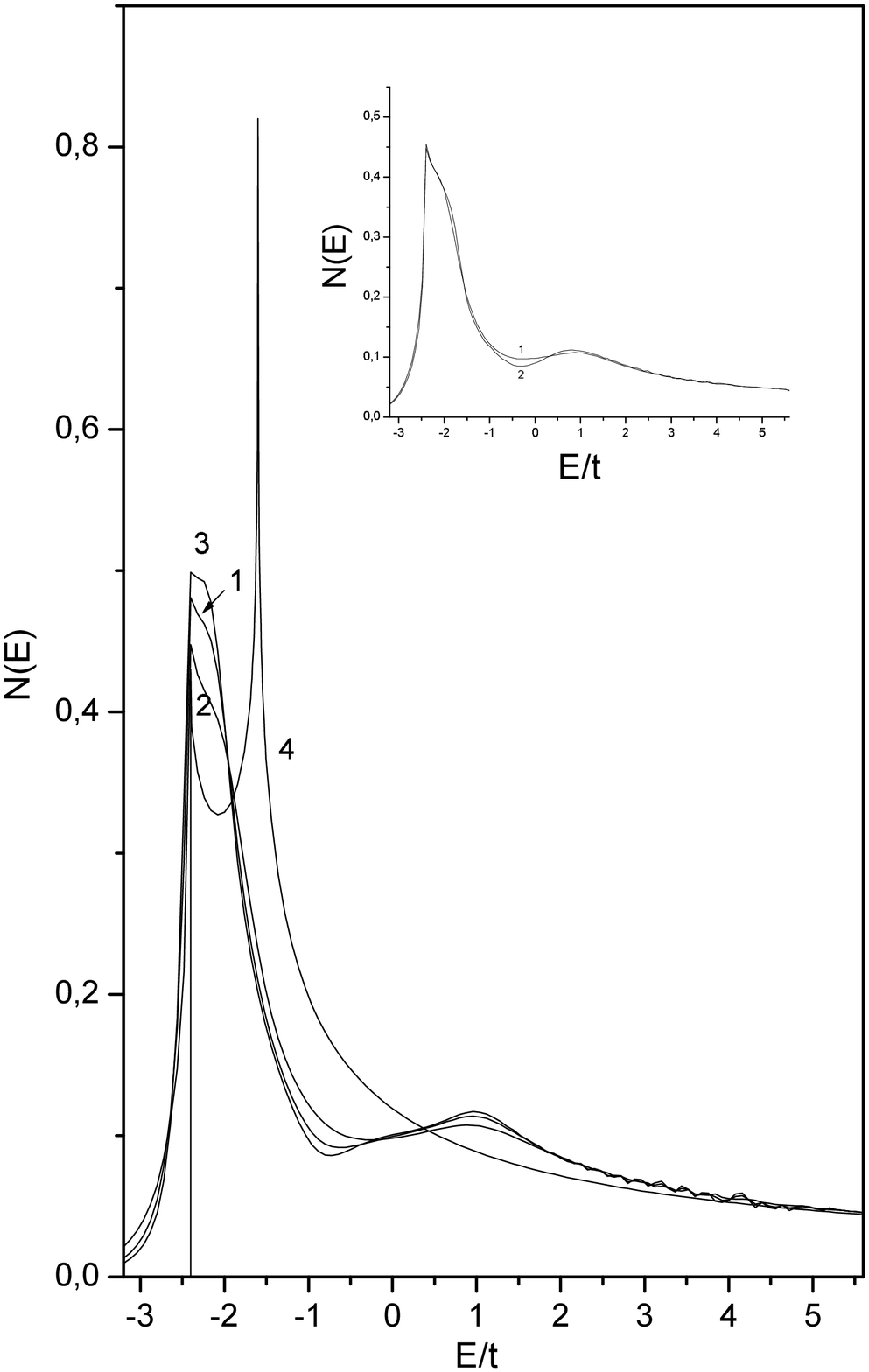}
\caption{Single -- electron density of states (in units of $1/ta^2$) for 
for different combinatorics of diagrams and $t'/t=-0.4$, $\mu /t=-1.3$, 
which is typical for HTSC -- cuprates:\ 
(1) --- incommensurate case,\ (2) --- commensurate case,\ 
(3) --- combinatorics of spin -- fermion model,\
(4) --- in the absence of pdeudogap fluctuations.\
We assume $W/t=1$ and correlation length $\kappa a=0.1$.
At the insert:\ Density of states for commensurate combinatorics and:\ 
(1)---$\kappa a=0.1$; (2)---$\kappa a=0.01$
}
\label{dos_hs}
\end{figure}

Ley us now consider the single -- electron density of states, defined by the
integral of the spectral density $A(E{\bf p})$ over the whole Brillouin zone.
Detailed calculations of the density of states in ``hot spots'' model were
performed in Ref. \cite{KS}. As an example, in Fig. \ref{dos_hs} we show
appropriate data for the Fermi surface topology typical for HTSC -- systems.
We can see, that for typical value  $t'/t=-0.4$ there is a shallow minimum
in the density of states (pseudogap), which is only slightly dependent on the
value of correlation length $\xi$. At the same time, e.g. for $t'/t=-0.6$ 
(which is untypical for HTSC -- cuprates) there are ``hot spots'' on the Fermi
surface, but pseudogap in the density of states is practically unobservable
\cite{KS}. We can see only smearing of Van Hove singularity, which is 
present for an ideal case in the absence of pseudogap scattering. In this
sense most clear manifestations of the pseudogap behavior are not in the
density of states, but in spectral density, which is in general accordance
with experiments.


\subsection{Recurrence relations for the vertex part and 
optical conductivity.}

\begin{figure}
\epsfxsize=5cm
\epsfbox{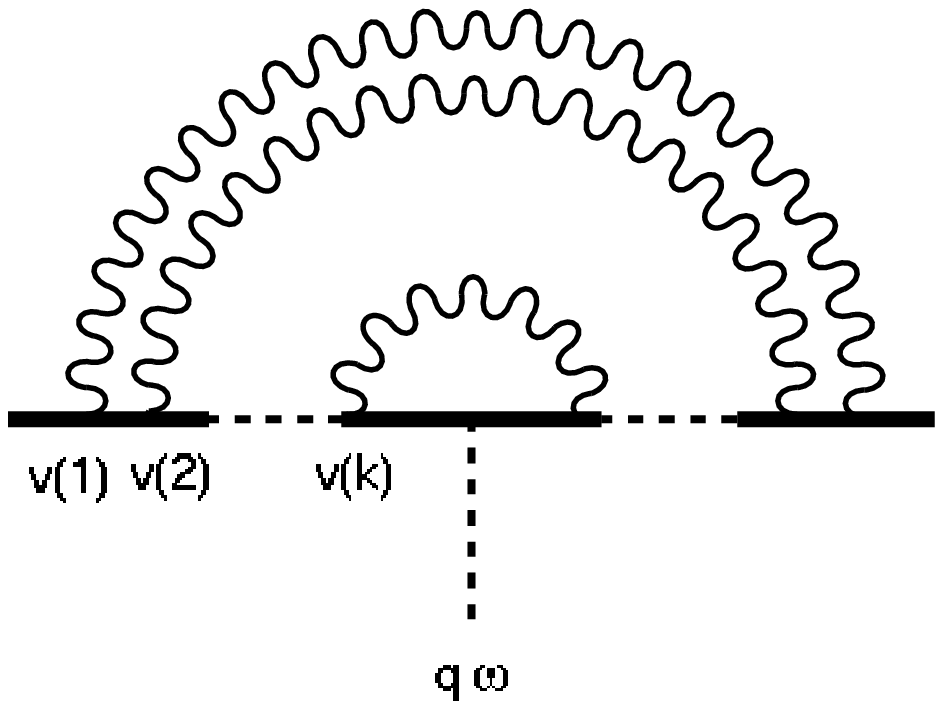}
\caption{General form of higher order correction to the vertex part.}
\label{vert}
\end{figure}

To calculate optical conductivity we need to know the vertex part,
describing electromagnetic response of our system. This vertex part can be
found using the method, proposed for similar one -- dimensional model in
Refs. \cite{S91,ST91}. Below we are following Ref. \cite{SS}. Above we have 
seen, that any diagram for irreducible vertex part can be obtained by the
insertion of an external field line in appropriate diagram for electron
self -- energy \cite{S74}. As in our model it is sufficient to take account
only of self -- energy diagrams without intersecting interaction lines with
additional combinatorial factors $v(k)$ in ``initial'' vertices, to calculate
vertex corrections it is possible to limit ourselves only to diagrams of the
type shown in Fig. \ref{vert}. Diagrams with intersecting interaction lines
are also accounted for automatically.
\begin{figure}
\epsfxsize=10cm
\epsfbox{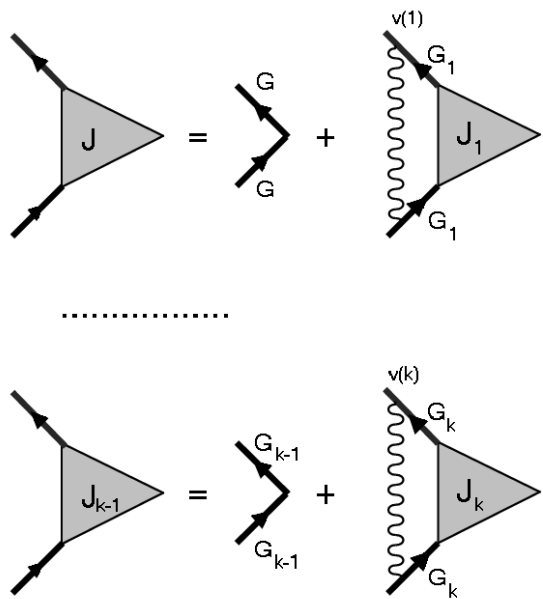}
\caption{Recursion relations for the vertex part.}
\label{recvertx}
\end{figure} 
\begin{figure}
\epsfxsize=8cm
\epsfbox{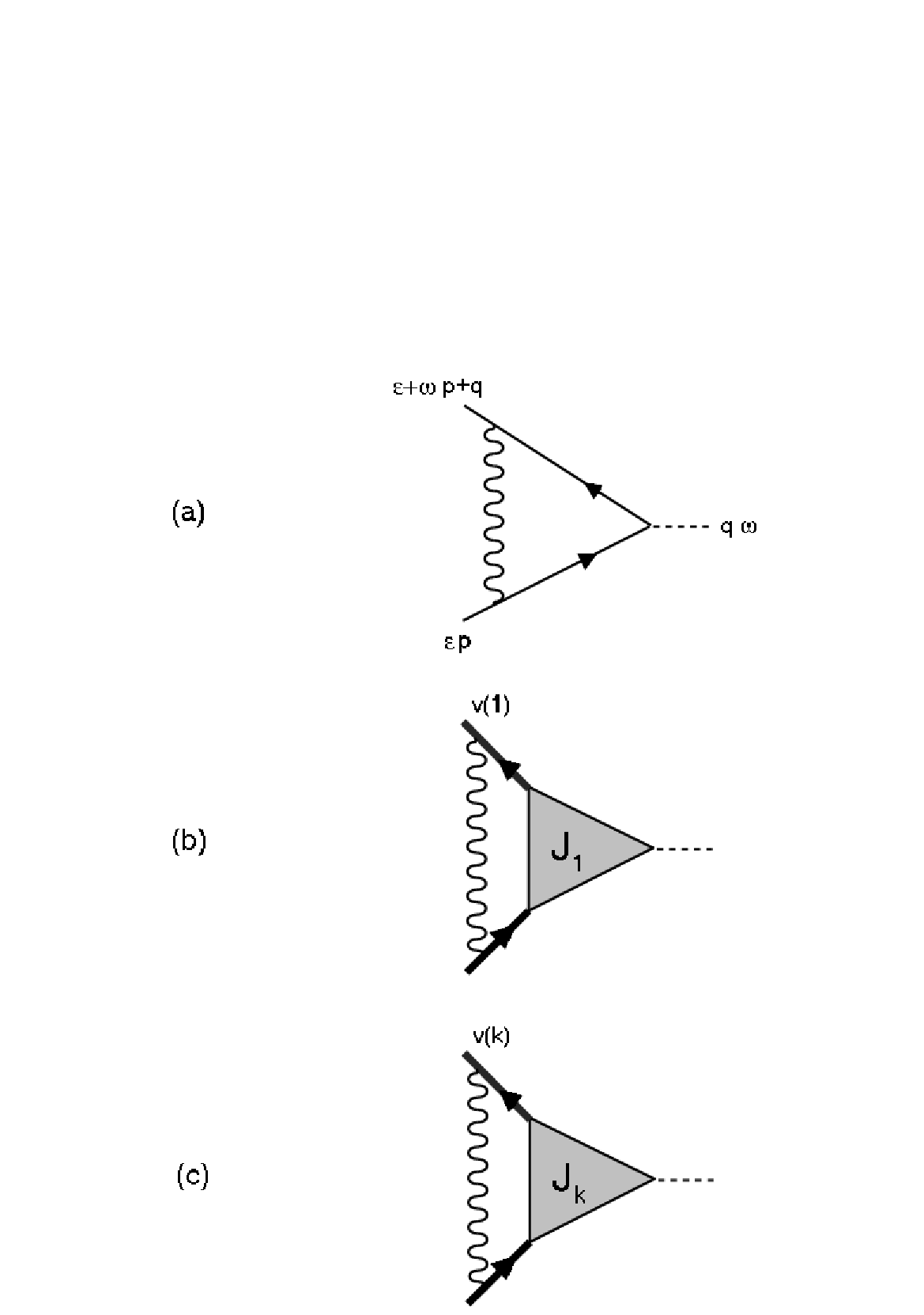}
\caption{Simplest corrections to the vertex parts.}
\label{vertcorr}
\end{figure} 
Now we immediately obtain the system of recursion equations for the vertex
parts, shown graphically in Fig. \ref{recvertx}. To obtain appropriate
analytic expressions, consider the simplest vertex correction, shown in
Fig. \ref{vertcorr} (a). Performing calculations for $T=0$ in $RA$ -- channel, 
we can easily obtain its contribution as:

\footnotesize
\begin{eqnarray}
&&{\cal J}_1^{(1)RA}(\varepsilon{\bf p};\varepsilon+\omega{\bf p+q})=
\sum_{\bf K}V_{eff}({\bf K})G_{00}^A(\varepsilon\xi_{\bf p-K})
G_{00}^R(\varepsilon+\omega\xi_{\bf p-K+q})=\nonumber\\
&&=W^2\left\{G_{00}^A(\varepsilon,\xi_1({\bf p})+iv_1\kappa)-
G_{00}^R(\varepsilon+\omega,\xi_1({\bf p+q})-iv_1\kappa)\right\}
\frac{1}{\omega+\xi_1({\bf p})-\xi_1({\bf p+q})}=\nonumber\\
&&=W^2G_{00}^A(\varepsilon,\xi_1({\bf p})+iv_1\kappa)
G_{00}^R(\varepsilon+\omega,\xi_1({\bf p+q})-iv_1\kappa)
\left\{1+\frac{2iv_1\kappa}{\omega+\xi_1({\bf p})-\xi_1({\bf p+q})}
\right\}\equiv
\nonumber\\
&&\equiv W^2G_{01}^A(\varepsilon,\xi_{\bf p})
G_{01}^R(\varepsilon+\omega,\xi_{\bf p+q})
\left\{1+\frac{2iv_1\kappa}{\omega+\xi_1({\bf p})-\xi_1({\bf p+q})}\right\}
\label{J10}
\end{eqnarray}
\normalsize
where, during calculations of integrals, we have used the following identity,
valid for free -- particle Green's functions:
\begin{equation}
G_{00}^A(\varepsilon\xi_{\bf p})G_{00}^R(\varepsilon+\omega\xi_{\bf p+q})=
\left\{G_{00}^A(\varepsilon\xi_{\bf p})-G_{00}^R(\varepsilon+
\omega\xi_{\bf p+q})\right\}\frac{1}{\omega-\xi_{\bf p+q}+\xi_{\bf p}}
\label{ident0}
\end{equation}
``Dressing'' internal electronic lines we obtain diagram, shown in Fig. 
\ref{vertcorr} (b), and using the identity:
\begin{eqnarray}
G^A(\varepsilon\xi_{\bf p})G^R(\varepsilon+\omega\xi_{\bf p+q})=
\left\{G^A(\varepsilon\xi_{\bf p})-G^R(\varepsilon+\omega
\xi_{\bf p+q})\right\}\times\nonumber\\
\times\frac{1}{\omega-\xi_{\bf p+q}+\xi_{\bf p}
-\Sigma^R_1(\varepsilon+\omega\xi_{\bf p+q})+\Sigma^A_1(\varepsilon\xi_{\bf p})}
\label{ident} 
\end{eqnarray} 
valid for full Green's functions, we can write down the contribution of this
diagram as:
\footnotesize
\begin{eqnarray}
&&{\cal J}_1^{RA}(\varepsilon{\bf p};\varepsilon+\omega{\bf p+q})=
W^2v(1)G_{1}^A(\varepsilon,\xi_{\bf p})
G_{1}^R(\varepsilon+\omega,\xi_{\bf p+q})
\Biggl\{1+\nonumber\\
&&\left.+\frac{2iv_1\kappa}{\omega-\xi_1({\bf p+q})+\xi_1({\bf p})
-\Sigma_2^R(\varepsilon+\omega\xi_{\bf p+q})
+\Sigma^A_2(\varepsilon\xi_{\bf p})} 
\right\}J_1^{RA}(\varepsilon{\bf p};\varepsilon+\omega{\bf p+q})\nonumber\\ 
\label{J1} 
\end{eqnarray}
\normalsize
Here we assumed, that interaction line in the vertex correction diagram of
Fig. \ref{vertcorr} (b) ``transforms'' self -- energies
$\Sigma_1^{R,A}$ of internal electronic lines into $\Sigma_2^{R,A}$, in
accordance with our main approximation for self -- energies used above
(cf. Fig. \ref{recurr})\footnote{This assumption is justified by the fact,
that it guarantees validity of certain exact relation, following from the Ward 
identity (see below) \cite{S91}.}.

Now it is not difficult to write down similar expression for the diagram of
general form, shown in Fig. \ref{vertcorr} (c):
\footnotesize
\begin{eqnarray}
&&{\cal J}_k^{RA}(\varepsilon{\bf p};\varepsilon+\omega{\bf p+q})=
W^2v(k)G_{k}^A(\varepsilon,\xi_{\bf p})
G_{k}^R(\varepsilon+\omega,\xi_{\bf p+q})
\Biggl\{1+\nonumber\\
&&\left.+\frac{2iv_k\kappa k}{\omega-\xi_k({\bf p+q})+\xi_k({\bf p})
-\Sigma_{k+1}^R(\varepsilon+\omega\xi_{\bf p+q})
+\Sigma^A_{k+1}(\varepsilon\xi_{\bf p})} 
\right\}J_k^{RA}(\varepsilon{\bf p};\varepsilon+\omega{\bf p+q}) 
\nonumber\\
\label{Jk} 
\end{eqnarray}
\normalsize
Accordingly, we can write dowm the following fundamental recursion relation
for the vertex part, shown in Fig. \ref{recurr}:
\footnotesize
\begin{eqnarray}
&&J_{k-1}^{RA}(\varepsilon{\bf p};\varepsilon+\omega{\bf p+q})=
1+W^2v(k)G_{k}^A(\varepsilon,\xi_{\bf p})
G_{k}^R(\varepsilon+\omega,\xi_{\bf p+q})
\Biggl\{1+\nonumber\\
&&\left.+\frac{2iv_k\kappa k}{\omega-\xi_k({\bf p+q})+\xi_k({\bf p})
-\Sigma_{k+1}^R(\varepsilon+\omega\xi_{\bf p+q})
+\Sigma^A_{k+1}(\varepsilon\xi_{\bf p})} 
\right\}J_k^{RA}(\varepsilon{\bf p};\varepsilon+\omega{\bf p+q}) 
\nonumber\\
\label{Jrec} 
\end{eqnarray}
\normalsize
``Physical'' vertex $J^{RA}(\varepsilon{\bf p};\varepsilon+\omega{\bf p+q})$
is defined as $J^{RA}_{k=0}(\varepsilon{\bf p};\varepsilon+
\omega{\bf p+q})$. Recursion procedure (\ref{Jrec}) takes into account all
diagrams of perturbation theory for the vertex part. 
For $\kappa\to 0\quad (\xi\to\infty)$ (\ref{Jrec}) redoces to the series.
studied in Refs. \cite{S74} (see also Ref. \cite{Sch}), which can be summed
exactly in analytic form. Standard ``ladder'' approximation is obtained from
our scheme putting all combinatorial factors $v(k)$ in (\ref{Jrec}) to unity
\cite{ST91}.

\begin{figure}
\epsfxsize=6cm
\epsfbox{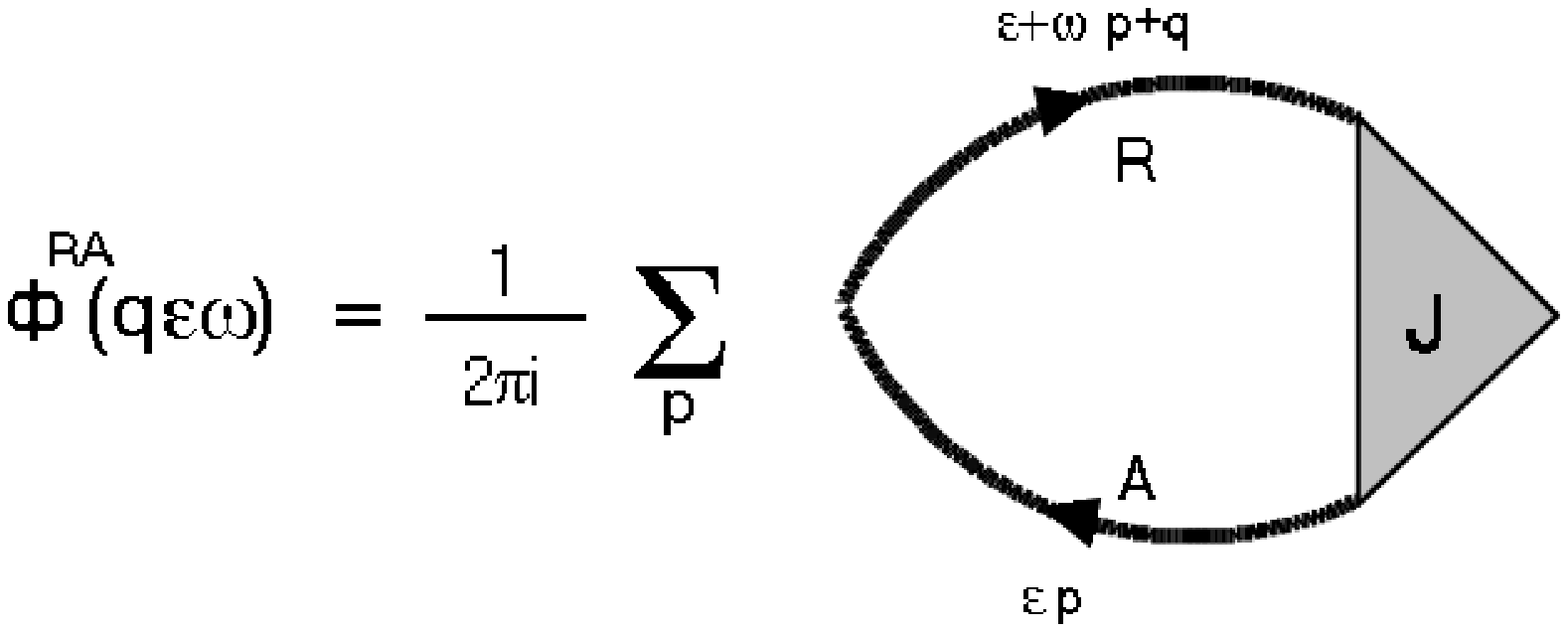}
\caption{Diagrammatic representation for two -- particle function
$\Phi^{RA}(q\omega)$.}
\label{loop}
\end{figure}

Conductivity is expressed via retarded density -- density response function
$\chi^R(q\omega)$ \cite{VW}:
\begin{equation}
\sigma(\omega)=e^2\lim_{q\to 0}\left(-\frac{i\omega}{q^2}\right)
\chi^R(q\omega)
\label{conduc}
\end{equation}
where $e$ is electronic charge,
\begin{equation}
\chi^R(q\omega)=\omega\left\{\Phi^{RA}(0q\omega)-\Phi^{RA}(00\omega)\right\}
\label{chiR}
\end{equation}
where two -- particle Green's function $\Phi^{RA}(\varepsilon q\omega)$ is
defined by loop diagram shown in Fig. \ref{loop}.

Direct numerical computations confirm, that the recurrence procedure
(\ref{Jrec}) satisfies an exact relation, which directly follows (for 
$\omega\to 0$) from the Ward identity \cite{VW}:
\begin{equation}
\Phi^{RA}(00\omega)=-\frac{N(E_F)}{\omega}
\label{Ward}
\end{equation}
where $N(E_F)$ is the density of states at the Fermi level $E_F=\mu$. 
Actually, this is our main motivation for the {\em Ansatz}, used during the
derivation of (\ref{J1}),\ (\ref{Jk}) and (\ref{Jrec}).

Finally, conductivity is written in the following symmetrized form,
convenient for numerical computations:
\footnotesize
\begin{eqnarray}
&&\sigma(\omega)=\frac{e^2\omega^2}{\pi}\lim_{q\to 0}\frac{1}
{q^2}\sum_{\bf p}\left\{G^R\left(\frac{\omega}{2},{\bf p}
+\frac{{\bf q}}{2}\right)J^{RA}\left(\frac{\omega}{2},{\bf p}
+\frac{{\bf q}}{2};-\frac{\omega}{2},{\bf p}-\frac{{\bf q}}{2}\right)
G^A\left(-\frac{\omega}{2},{\bf p}-\frac{{\bf q}}{2}\right)-\right.\nonumber\\
&&\left.-G^R\left(\frac{\omega}{2},{\bf p}\right)
J^{RA}\left(\frac{\omega}{2},{\bf p};-\frac{\omega}{2},{\bf p}\right)
G^A\left(-\frac{\omega}{2},{\bf p}\right)\right\}\nonumber\\ 
\label{optconduct} 
\end{eqnarray}
\normalsize
where we also accounted for an extra factor of 2, due to summation over spin.

Direct numerical calculations \cite{SS} were performed, for different values of
parameters of the ``bare'' spectrum (\ref{spectr}), using
(\ref{optconduct}),\ (\ref{Jrec}),\ (\ref{G}), with recursion procedure
starting at some high value of $k$, where all $\Sigma_k$ and $J_k$ were 
assumed to be zero. Integration in (\ref{optconduct}) was made over the 
Brillouin zone. Integration momenta are naturally made dimensionless with the
help of lattice constant $a$, all energies below are given in units of transfer
integral $t$. Conductivity is measured in units of universal conductivity of
two -- dimensional system: $\sigma_0=\frac{e^2}{\hbar}=2.5\ 10^{-4}$
Ohm$^{-1}$, and density of states --- in units of $1/ta^2$. For definiteness,
below we always take $W=t$.

\begin{figure}
\epsfxsize=9cm
\epsfysize=12cm
\epsfbox{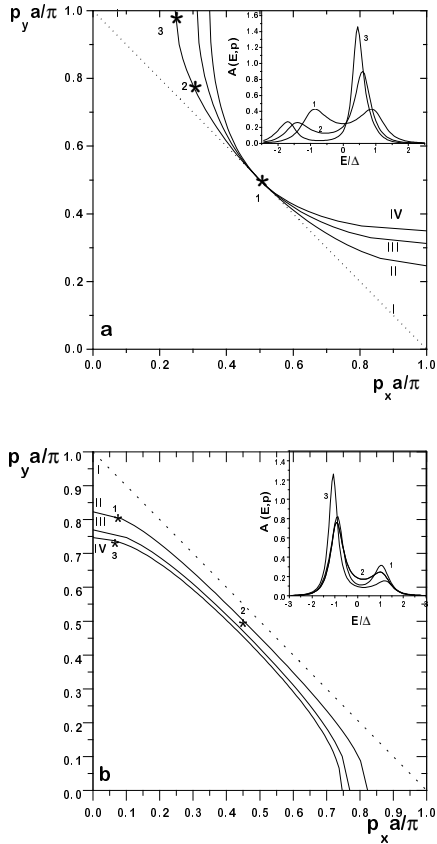}
\caption{Fermi surfaces for different values of $t'$ and chemical
potential $\mu$.
(a) corresponds to $\mu=0$ and the following values of $t'/t$:\
0 -- I;\ -0.2 -- II;\ -0.4 -- III;\ -0.6 -- IV.
(b) corresponds to  $t'=0$ and the following values of $\mu/t$:
0 -- I;\ -0.3 -- II;\ -0.5 -- III;\ -0.6 -- IV.
At the inserts we show energy dependences of spectral densities for spin --
fermion model for $\kappa a=0.1$ at points of the momentum space denotes
by stars.}
\label{FSsqr}
\end{figure}

\begin{figure}
\epsfxsize=7cm
\epsfysize=7cm
\epsfbox{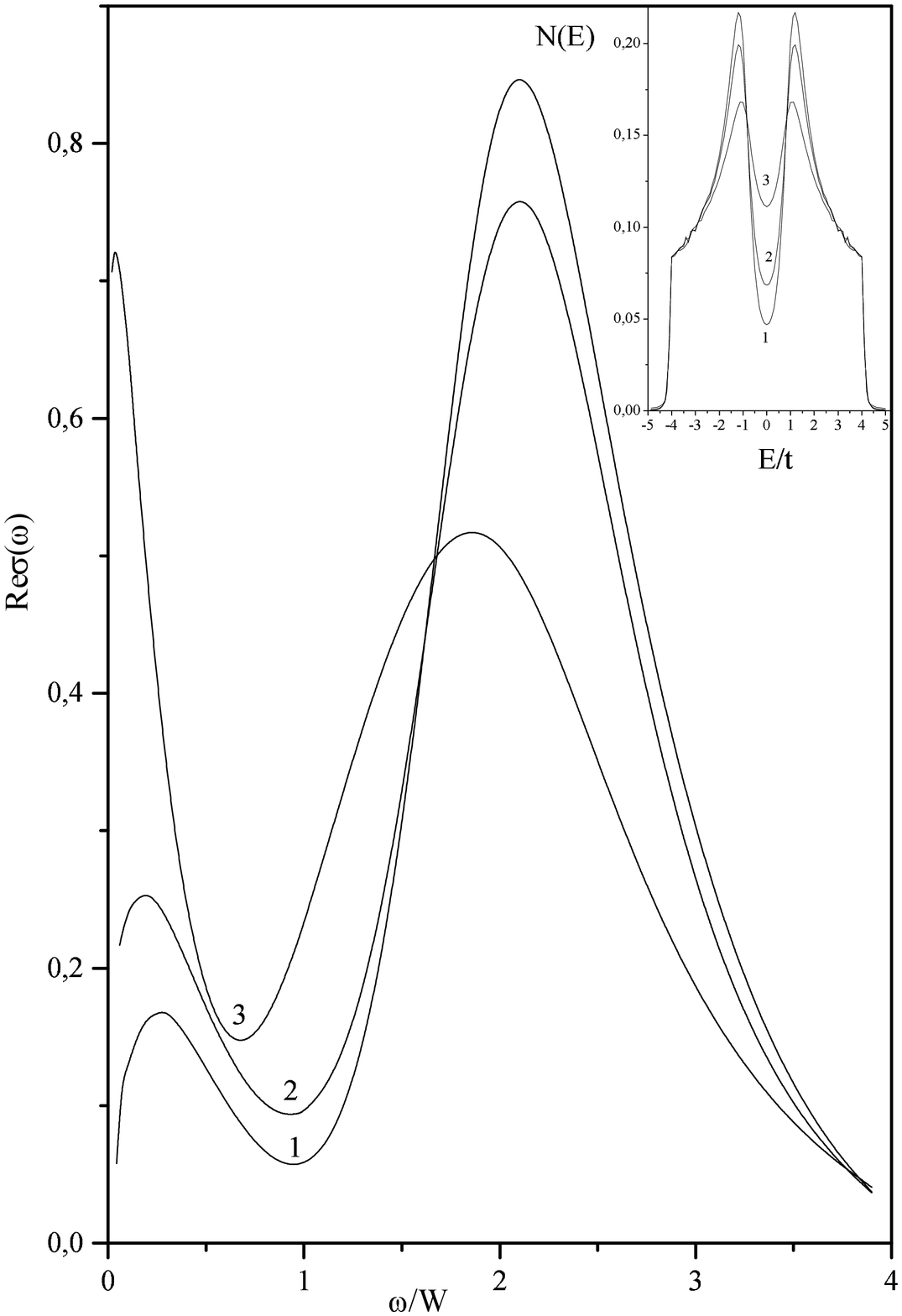}
\caption{Real part of optical conductivity in spin -- fermion model
for the case of square Fermi surface ($\mu=0$, $t'=0$) for different values
of inverse correlation length ${\kappa a}$:\
0.1 -- 1;\ 0.2 -- 2;\ 0.5 -- 3.
At the insert we show appropriate densities of states.}
\label{condsqr}
\end{figure}

First let us consider Fermi surfaces close to the case of half -- filled
band $\mu=0$ and $t'=0$, shown (in the first quadrant of Brillouin zone) 
in Fig. \ref{FSsqr} (a).
We know that for $\mu=0$ and $t'=0$ Fermi surface is simply square 
(complete "nesting"), so that we practically are dealing with one --
dimensional case, analyzed long ago in Refs. \cite{S74,S91,ST91}.
Results of our calculations for real part of optical conductivity
in our two -- dimensional model, for the case of spin -- fermion diagram
combinatorics and different values of correlation length of AFM short -- range
order (parameter $\kappa=\xi^{-1}$, where  $\xi$ is measured in units of
lattice constant $a$) are shown in Fig. \ref{condsqr}. 
Qualitative behavior of conductivity is quite similar to those found for
one -- dimensional model (for the case of incommensurate fluctuations of 
CDW -- type) in Refs. \cite{S91,ST91}. It is characterized by the presence
of large pseudogap absorption maximum 
(appropriate densities of states with pseudogap close to the Fermi level are
shown at the insert in Fig. \ref{condsqr}) at $\omega\sim 2W$ and also by a
maximum at small frequencies, connected with carrier localization in static 
(in our approximation) random field of AFM fluctuations. Localization nature
of this maximum is confirmed by its transformation into characteristic
``Drude -- like'' peak (with maximum at $\omega=0$) if we perform calculations
in ``ladder'' approximation, when combinatorial factors $v(k)=1$, which
corresponds to ``switching off'' the contribution from diagrams with
intersecting interaction lines, leading to two -- dimensional Anderson
localization \cite{VW,GLK}. Qualitative form of conductivity in this case is
also quite similar to those found in Ref. \cite{ST91}. Narrowing of 
localization peak with diminishing correlation length of fluctuations can be
explained, as was noted in Ref. \cite{ST91}, by suppression of effective 
interaction (\ref{Veff}) for small $\xi$ (with fixed value of $W$), leading
to general suppression of scattering, also on the ``cold'' part of the Fermi
surface. Note that general behavior of the density of states and optical
conductivity obtained here is in complete qualitative agreement with the
results, obtained for the similar two -- dimensional model of Peierls
transition by quantum Monte -- Carlo calculations in Ref. \cite{ScM}.

\begin{figure}
\epsfxsize=7cm
\epsfysize=7cm
\epsfbox{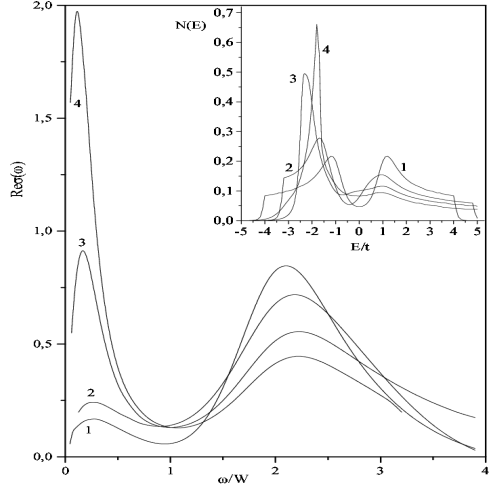}
\caption{Real part of optical conductivity in spin -- fermion model for 
$\mu=0$ and ${\kappa a}=0.1$  and for different Fermi surfaces, obtained from
the square after ``switching on'' the transfer integral $t'/t$:\
0 -- 1;\ -0.2 -- 2;\ -0.4 -- 3;\ -0.6 -- 4.
At the insert --- appropriate densities of states.}
\label{condFSt}
\end{figure}

Now, if we keep $\mu=0$ and ``switch on'' the transfer integral $t'$ between
second nearest neighbors in (\ref{spectr}), we shall obtain Fermi surfaces
different from square one, as shown in Fig. \ref{FSsqr} (a). At the insert on
this figure we also show the energy dependence of spectral density 
(\ref{spdns}) at several typical points on these Fermi surfaces. It is seen 
that spectral density demonstrates characteristic ``non Fermi -- liquid'' 
behavior, of the type studied in Refs. \cite{Sch,KS}, practically everywhere
on the Fermi surface, until this surface is not very different from the square,
despite the fact that the ``hot spot'' in this case lies precisely at the
intersection of the Fermi surface with diagonal of the Brillouin zone.
Appropriate dependences of the real part of optical conductivity are shown in
Fig. \ref{condFSt}.
At the insert in this figure we show appropriate densities of states. It is 
seen that as we move away from the complete ``nesting'', pseudogap absorption
maximum becomes more shallow, while localization peak (in accordance with
the general sum rule for conductivity) grows. Note, however, that pseudogap
absorption remains noticeable even when pseudogap in the density of states is
practically invisible (curves 4 in Fig.  \ref{condFSt}). 

\begin{figure}
\epsfxsize=7cm
\epsfysize=7cm
\epsfbox{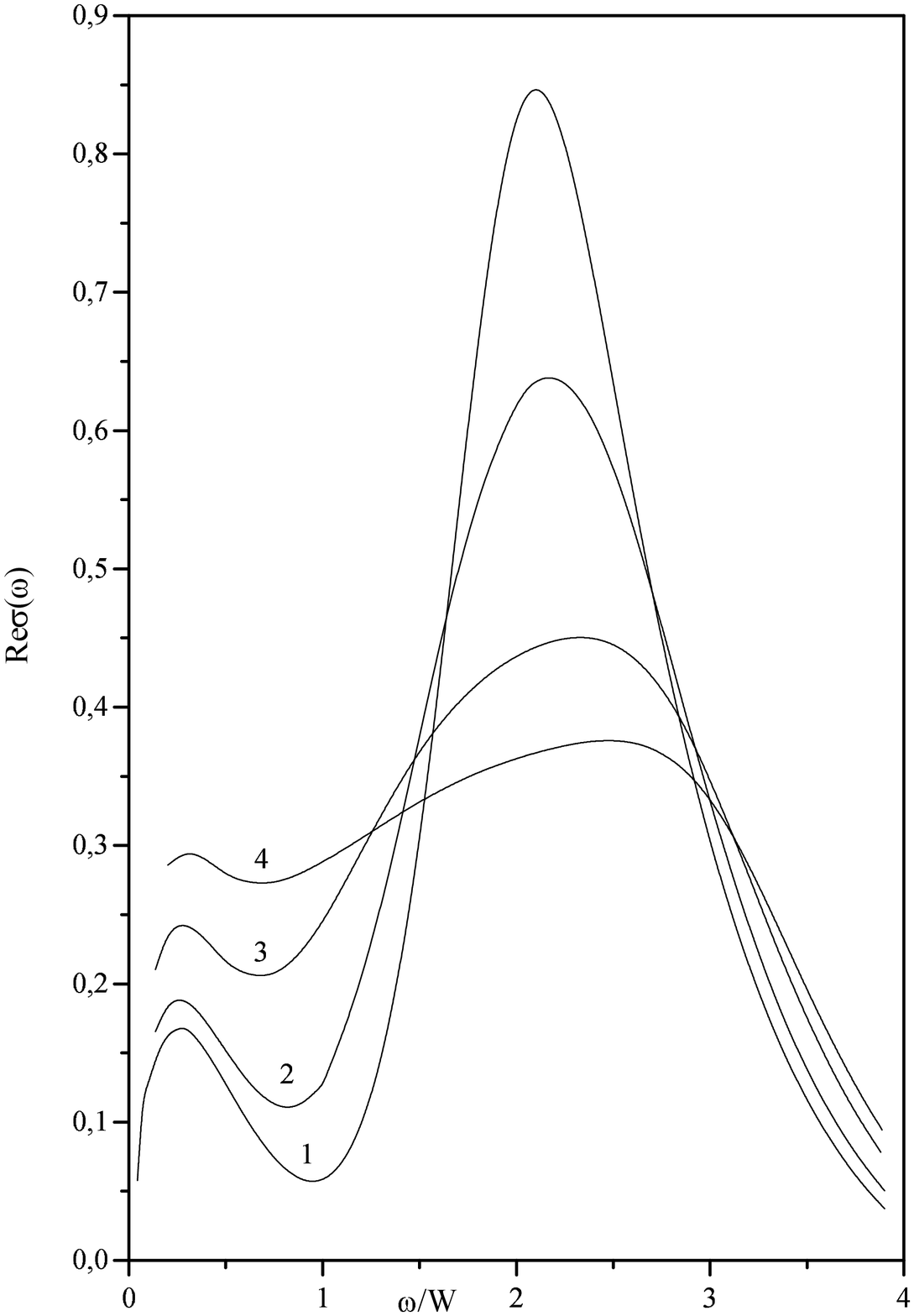}
\caption{Real part of optical conductivity in spin -- fermion model for 
$t'=0$ and ${\kappa a}=0.1$  and for different Fermi surfaces, obtained from
square as we move from the case of half -- filled band.
Chemical potential $\mu/t$:\
0 -- 1;\ -0.3 -- 2;\ -0.5 -- 3;\ -0.6 -- 4.}
\label{condFSm}
\end{figure}

Let us return now to the case of $t'=0$ and change the value of $\mu$, so
that Fermi surfaces are close to the square, as shown in Fig. \ref{FSsqr} (b). 
Strictly speaking, ``hot spots'' on these surfaces are absent, but spectral
density shown at the insert on Fig. \ref{FSsqr} (b), is still typically 
pseudogap like. Appropriate dependences of the real part of optical 
conductivity are shown in Fig. \ref{condFSm}. 

\begin{figure}
\epsfxsize=7cm
\epsfysize=7cm
\epsfbox{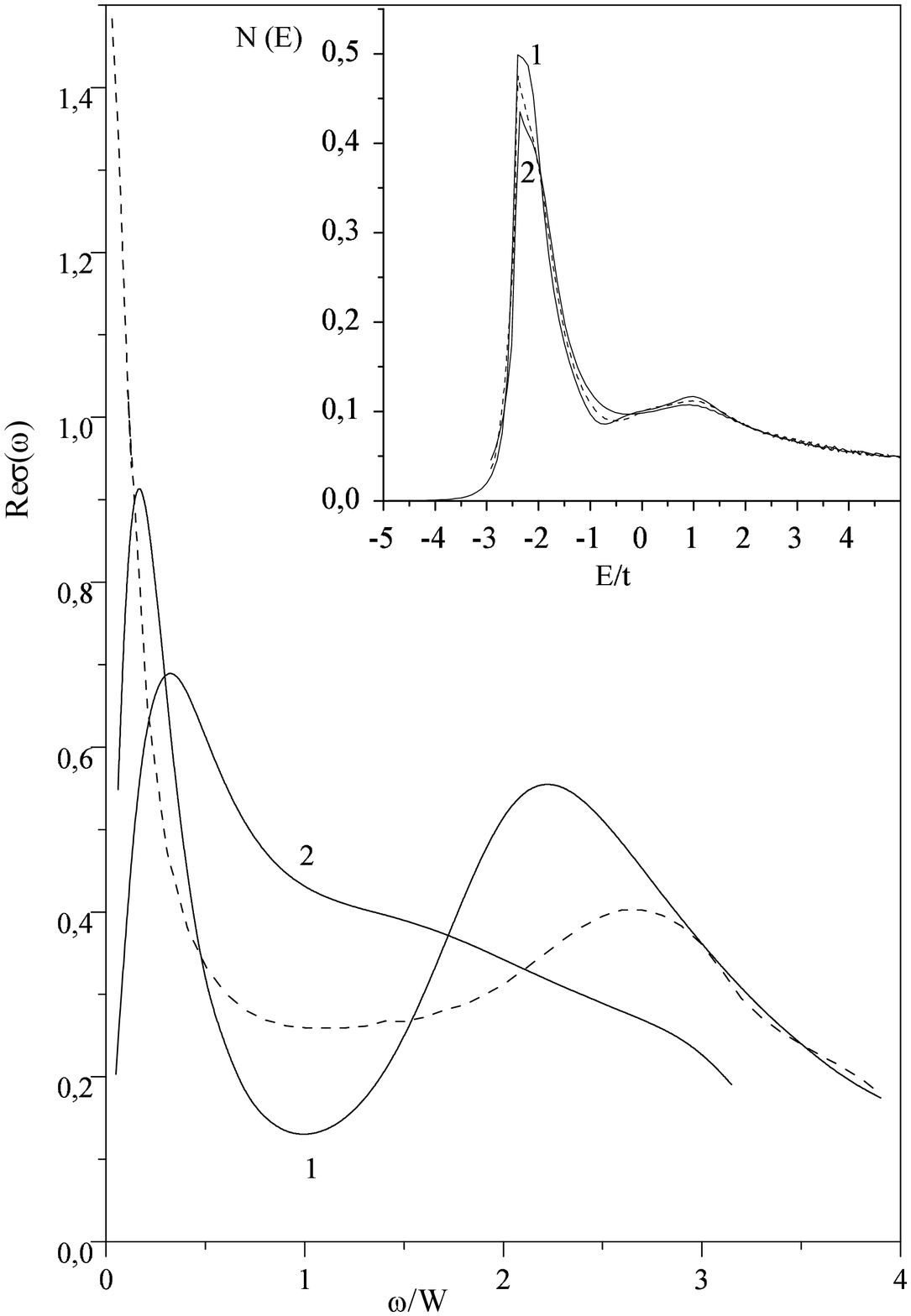}
\caption{Real part of optical conductivity for $t'/t=-0.4$ and $\mu=0$
for ${\kappa a}=0.1$ and for different combinatorics of diagrams:\
1 -- spin -- fermion combinatorics;\
2 -- commensurate case.\ Dashed curve --- ``ladder'' approximation. 
At the insert --- appropriate densities of states.}
\label{condFSga}
\end{figure}

Consider now geometry of the Fermi surface with ``hor spots'' typical for
most of HTSC -- oxides, like that shown in Fig. \ref{hspots}. First, in
Fig. \ref{condFSga} we show the real part of optical conductivity, calculated 
(for different combinatorics of diagrams) for characteristic value of 
$t'=-0.4t$ and for chemical potential $\mu=0$, when ``hot spots'' are on the 
diagonals of the Brillouin zone. It is seen that pseudogap behavior of
conductivity is conserved even in the case of practically absent pseudogap
in the density of states (shown at the insert in Fig. \ref{condFSga}). 
Dashed curve in Fig. \ref{condFSga} shows the result of ``ladder'' 
approximation, demonstrating disappearance of two -- dimensional localization.
In general, as correlation length of the short -- range order diminishes, we
observe ``smearing'' of the pseudogap maximum in conductivity.

\begin{figure}
\epsfxsize=7cm
\epsfysize=7cm
\epsfbox{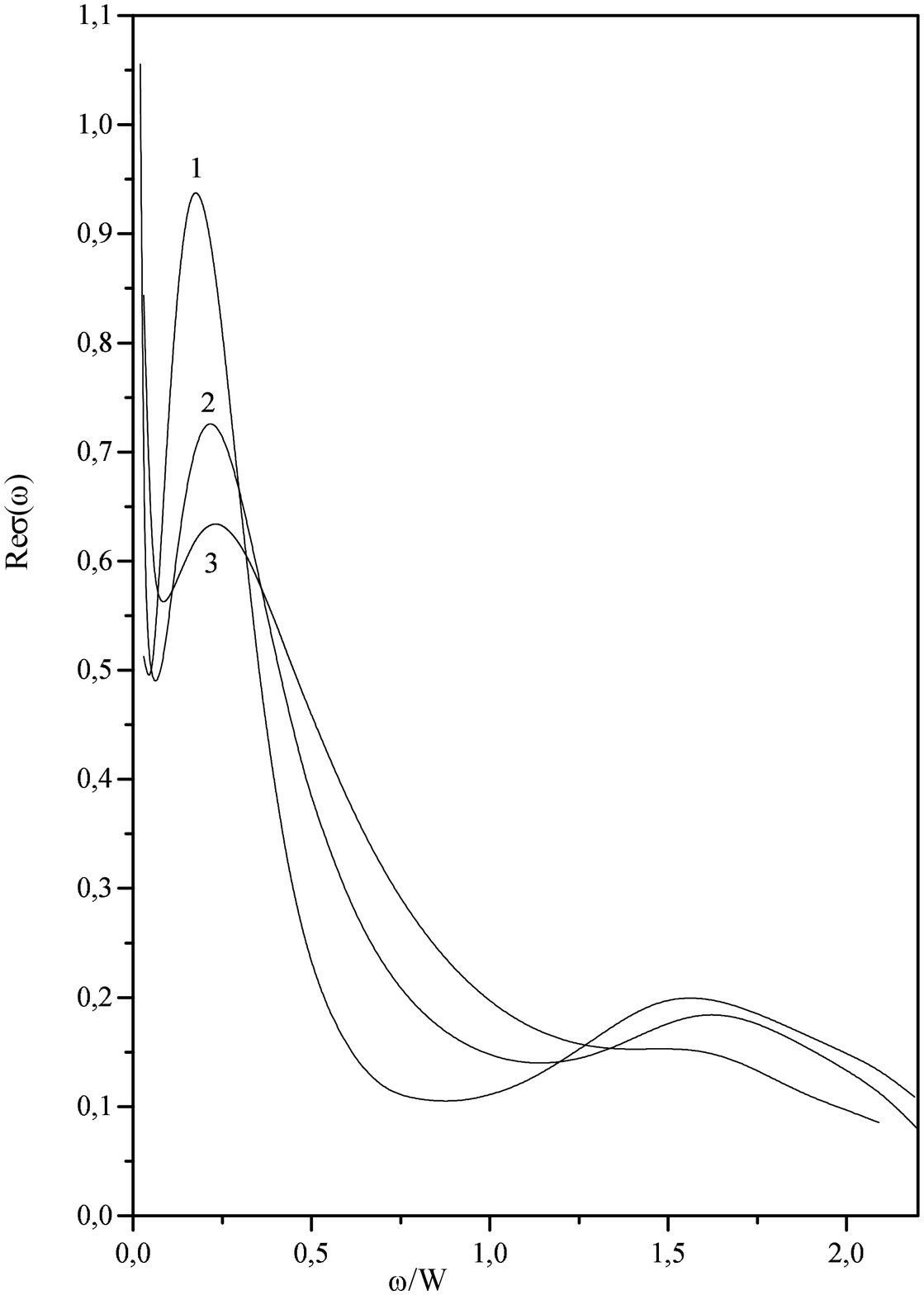}
\caption{Real part of optical conductivity in spin -- fermion model for 
$t'/t=-0,4$ and $\mu/t=-1.3$ and for different values of correlation length
${\kappa a}$:\ 0.05 -- 1;\ 0.1 -- 2;\ 0.2 -- 3.\  
Damping factor ${\gamma/t}=0.005$.}
\label{condHS}
\end{figure}

For most copper oxide superconductors characteristic geometry of the Fermi
surface can be described by $t'=-0.4t$ and $\mu=-1.3t$ \cite{Sch}.
In this case, results of our calculations of optical conductivity for 
different values of inverse correlation length $\kappa$ are shown in 
Fig. \ref{condHS} (for the case of spin -- fermion combinatorics).
here we introduced additional weak scattering due to inelastic processes via
standard replacement $\omega\to\omega+i\gamma$ \cite{GZ}, which leads to the
appearance of a narrow ``Drude -- like'' peak for $\omega < \gamma$ 
(destruction of two -- dimensional localization due to phase decoherence). 
It is easy to see that with the growth of inelastic scattering rate
$\gamma$, localization peak is completely ``smeared'' and transformed to the 
``usual'' Drude -- like peak at small frequencies. Pseudogap absorption maximum
becomes more pronounced with the growth of correlation length
$\xi$ (diminishing $\kappa$). Similar results for the ``hot patches'' model
were obtained in Ref. \cite{S99}

Direct correspondence of these results with experimental data shown in
Fig. \ref{optcond} \cite{Onose} is obvious. Simple estimates show that
characteristic values of conductivity in theory and experiment are also of
the  same order of magnitude. In principle, it is quite possible to make
quantitative fit to the experiment, varying parameters of our model.

\subsection{Interaction vertex for superconducting fluctuations.}

Now we are going to discuss superconductivity formation on the ``background'' 
of pseudogap fluctuations. To take into account pseudogap fluctuations during
the analysis of Cooper instability and derivation of Ginzburg -- Landau
expansion we need knowledge of the vertex part, describing electron interaction
with an arbitrary fluctuation of superconducting order parameter (gap) of a 
given symmetry \cite{KSS}:
\begin{equation} 
\Delta({\bf p},{\bf q})=\Delta_{\bf q}e({\bf p}) 
\label{Deltq}
\end{equation}
where the symmetry factor determining the type (symmetry) of pairing is: 
\begin{equation}
e({\bf p})=
\left\{
\begin{array}{ll}
1 & (\mbox{ $s$ -- wave pairing})\\ 
\cos p_xa-\cos p_ya & (\mbox{ $d_{x^2-y^2}$ -- pairing})
\end{array}.
\right.
\label{ephi}
\end{equation}
and we always assume singlet pairing.

\begin{figure}
\epsfxsize=10cm
\epsfbox{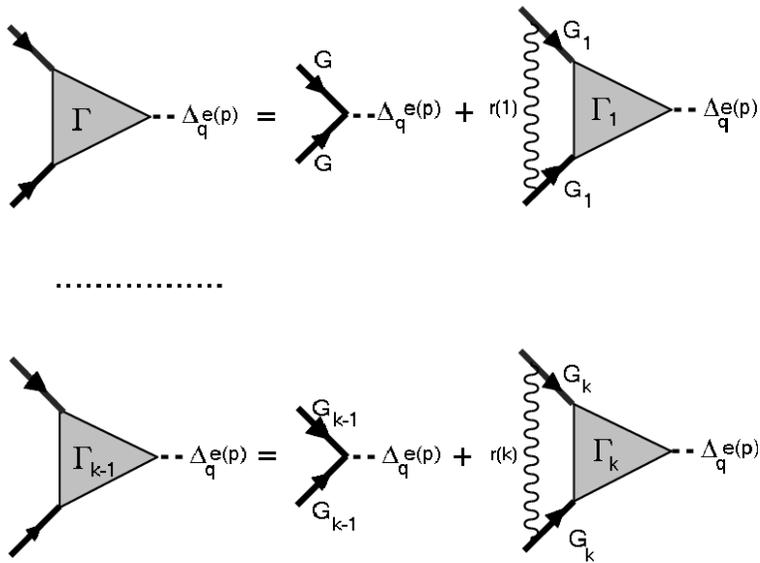}
\caption{Recursion equations for ``triangular'' vertex in Cooper channel.}
\label{vert_c}
\end{figure}

Almost immediately we can write down recursion relations \cite{KSS,KKS} 
for ``triangular'' vertices in Cooper channel, similar to those introduced
above during the analysis of response to electromagnetic field \cite{SS}. 
The vertex part of interest to us can be defined as:
\begin{equation}
\Gamma(\varepsilon_n,-\varepsilon_n,{\bf p},{\bf -p+q})\equiv
\Gamma_{\bf p}(\varepsilon_n,-\varepsilon_n,{\bf q})e({\bf p})
\label{Gamephi}
\end{equation}
Then $\Gamma_{\bf p}(\varepsilon_n,-\varepsilon_n,{\bf q})$ is determined by
the recursion procedure of the following form:
\begin{eqnarray}
\Gamma_{{\bf p}k-1}(\varepsilon_n,-\varepsilon_n,{\bf q})=1 \pm 
W^2r(k)G_k(\varepsilon_n,{\bf p+q}) G_k(-\varepsilon_n,{\bf p})
\Biggl\{1+\nonumber\\
+\frac{2ik\kappa v_k}
{G^{-1}_{k}(\varepsilon_n,{\bf p+q})-G^{-1}_{k}(-\varepsilon_n,{\bf p})
-2ik\kappa v_k}\Biggr\}\Gamma_{{\bf p}k}(\varepsilon_n,-\varepsilon_n,{\bf q}) 
\label{Gamma} 
\end{eqnarray}
which is represented by diagrams shown in Fig. \ref{vert_c}. 
``Physical'' vertex corresponds to 
$\Gamma_{{\bf p}k=0}(\varepsilon_n,-\varepsilon_n,{\bf q})$. Additional
combinatorial factor $r(k)=v(k)$ for the simplest case of charge
(or Ising like spin) pseudogap fluctuations analyzed in Ref. \cite{KSS}. 
For the most interesting case of Heisenberg spin (SDW) fluctuations, which are
mainly considered below, this factor is given by \cite{Sch,KKS}:
\begin{equation}
r(k)=\left\{\begin{array}{ll}
k & \mbox{for even $k$} \\
\frac{k+2}{9} & \mbox{for odd $k$}
\end{array} \right.
\label{rk}
\end{equation}
The choice of the sign before  $W^2$ in the r. h. s. of (\ref{Gamma}) depends
on the symmetry of superconducting order parameter and the type of pseudogap
fluctuations \cite{KSS,KKS}. Summary of all variants is given in Table I.
\begin{center}
{\sl Table I.\ The choice of sign in recursion procedure for the vertex part.}
\begin{tabular}{|c|c|c|c|}
\hline
Pairing & CDW & SDW (Ising) 
&  SDW (Heisenberg) \\ 
\hline 
$s$ & $+$ & $-$ & $+$     \\ 
\hline 
$d$ & $-$ & $+$ & $-$  \\ 
\hline
\end{tabular}
\end{center}
From this table we can see, that in most interesting case of $d$ -- wave
pairing and Heisenberg pseudogap fluctuations this sign is ``$-$'', so that we
have recursion procedure with alternating signs. At the same time, for the case
of $s$ -- wave pairing and the same type of fluctuations we have to take this
sign ``$+$'' and signs in recursion procedure are always the same. 
In Ref. \cite{KSS} it was shown that this difference in types of recursion
procedure leads to two different variants of qualitatively different
behavior of all the main characteristics of superconductors. 

\subsection{Impurity scattering.}

\begin{figure}
\epsfxsize=10cm
\epsfbox{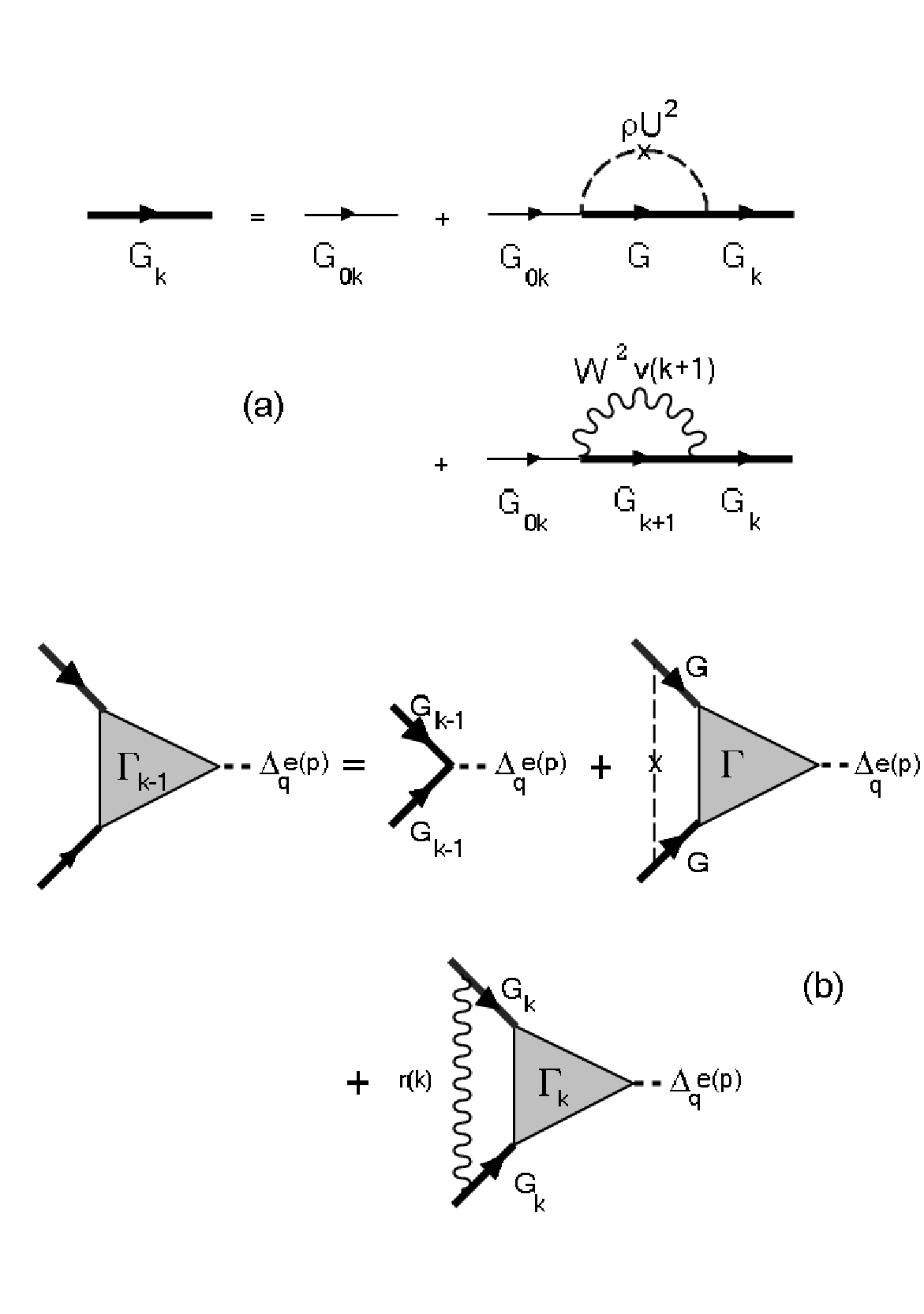}
\caption{Recursion equations for the Green's function (a) and ``triangular'' 
vertex (b) with the account of scattering by impurities.}
\label{Dysonimp}
\end{figure}

Scattering by normal (nonmagnetic) impurities is easily taken into account
in self -- consistent Born approximation, writing down ``Dyson's equation''
shown diagrammatically in Fig. \ref{Dysonimp} (a), where in addition to
Fig. \ref{vert_c}, we have added impurity scattering contribution to electron
self -- energy.

As a result, recursion equation for the Green's function is written as:
\begin{equation}
G_k(\varepsilon_n{\bf p})=\frac{1}{G^{-1}_{0k}(\varepsilon_n{\bf p})-
\rho U^2\sum_{\bf p}G(\varepsilon_n{\bf p})-W^2v(k+1)G_{k+1}(\varepsilon_n
{\bf p})}
\label{Dysrecim}
\end{equation}
where $\rho$ is the concentration of impurities with point -- like potential 
$U$, and impurity scattering self -- energy contains full Green's function
$G(\varepsilon_n{\bf p})=G_{k=0}(\varepsilon_n{\bf p})$, which is to be
determined self -- consistently by our recursion procedure. A contribution to
impurity scattering self -- energy form the real part of Green's function
reduces, as usual, to irrelevant renormalization of the chemical potential,
so that (\ref{Dysrecim}) reduces to:
\begin{equation}
G_k(\varepsilon_n{\bf p})=\frac{1}{i(\varepsilon_n -
\rho U^2\sum_{\bf p}ImG(\varepsilon_n{\bf p})+kv_k\kappa)-\xi_k({\bf p})
-W^2v(k+1)G_{k+1}(\varepsilon_n{\bf p})}
\label{Dysrecimp}
\end{equation}
Thus, in comparison with impurity free case, we have just a substitution
(renormalization):
\begin{eqnarray}
\varepsilon_n\to\varepsilon_n-\rho U^2\sum_{\bf p}ImG(\varepsilon_n{\bf p})
\equiv\varepsilon_n\eta_{\epsilon}
\label{reneps}
\\
\eta_{\epsilon}=1-\frac{\rho U^2 }{\varepsilon_n} \sum_{\bf p}ImG(\varepsilon_n{\bf p})
\label{etaeps}
\end{eqnarray}
If we do not perform fully self -- consistent calculations of impurity
self -- energy, in the simplest approximation we simply have:
\begin{eqnarray}
\varepsilon_n\to\varepsilon_n-\rho U^2\sum_{\bf p}ImG_{00}
(\varepsilon_n{\bf p}) \equiv\varepsilon_n\eta_{\epsilon} 
=\varepsilon_n+\gamma_0 sign\varepsilon_n
\label{renepsi} 
\\ 
\eta_{\epsilon}=1+\frac{\gamma_0}{|\varepsilon_n|}
\label{etaepsi}
\end{eqnarray}
where $\gamma_0=\pi\rho U^2N_0(0)$ is the standard Born impurity scattering rate
($N_0(0)$ is the density of states of free electrons at the Fermi level).

For ``triangular'' vertices of interest to us, recursion equations with the
account of impurity scattering is shown diagrammatically in Fig. \ref{Dysonimp} 
(b). For the vertex, describing electron interacting with superconducting
order parameter fluctuation (\ref{Deltq}) with $d$ -- wave symmetry 
(\ref{ephi}), this equation is considerably simplified, as the contribution
of the second diagram in the r. h. s. of Fig. \ref{Dysonimp} (b) in fact
vanishes due to $\sum_{\bf p}e({\bf p})=0$ (cf. discussion of similar
situation in Ref. \cite{PS}). Then the recursion equation for the vertex
takes the form (\ref{Gamma}), where $G_k(\pm\varepsilon_n{\bf p})$ are given
by (\ref{Dysrecim}),\ (\ref{Dysrecimp}), i.e. are just Green's functions
``dressed'' by impurity scattering, determined gy diagrams of 
Fig. \ref{Dysonimp} (a). For the vertex describing interaction with order
parameter fluctuations with $s$ -- wave symmetry, we have the following
equation:
\begin{eqnarray} 
&&\Gamma_{{\bf p}k-1}(\varepsilon_n,-\varepsilon_n,{\bf q})=1 + 
\rho U^2\sum_{\bf p}G(\varepsilon_n,{\bf p+q})G(-\varepsilon_n,{\bf p})
\Gamma_{\bf p}(\varepsilon,-\varepsilon_n,{\bf q})\pm \nonumber\\
&&\pm W^2r(k)G_k(\varepsilon_n,{\bf p+q}) G_k(-\varepsilon_n,{\bf p})
\Biggl\{1+\label{Gammaimp}\\
&&+\frac{2ik\kappa v_k}
{G^{-1}_{k}(\varepsilon_n,{\bf p+q})-G^{-1}_{k}(-\varepsilon_n,{\bf p})
-2ik\kappa v_k}\Biggr\}\Gamma_{{\bf p}k}(\varepsilon_n,-\varepsilon_n,{\bf q}) 
\nonumber
\end{eqnarray}
where $G_k(\pm\varepsilon_n{\bf p})$ is again given by
(\ref{Dysrecim}),\ (\ref{Dysrecimp}), and the sign before $W^2$ is determined
according to the rules formulated above. The difference with the case of the
vertex, describing interaction with $d$ -- wave fluctuations, is in the
appearance of the second term in the r. h. s. of (\ref{Gammaimp}), so that 
we have to substitute:
\begin{equation}
1\to\eta_{\Gamma}=1+\rho U^2\sum_{\bf p}G(\varepsilon_n,{\bf p+q})
G(-\varepsilon_n,{\bf p})\Gamma_{\bf p}(\varepsilon,-\varepsilon_n,{\bf q})
\label{etagam}
\end{equation}
Now the self -- consistent procedure look as follows. We start from
``zeroth'' approximation $G=G_{00}$,\ $\Gamma_{\bf p}=1$,\ then in Eqs.
(\ref{Dysrecimp}),\ (\ref{Gammaimp}) we just have 
$\eta_{\varepsilon}=\eta_{\Gamma}=1-\rho U^2/\varepsilon_n 
\sum_{\bf p}ImG_{00}(\varepsilon_n{\bf p})$. Then perform recursions
(starting from some big enough value of $k$) and determine new values of
$G=G_{k=0}$ and $\Gamma_{\bf p}=\Gamma_{k=0}$. Again calculate
$\eta_{\varepsilon}$ and $\eta_{\Gamma}$ using (\ref{etaeps}) and
(\ref{etagam}), put these into (\ref{Dysrecimp}),\ (\ref{Gammaimp}) etc.,\ 
until convergence.

While analyzing vertices with  $d$ -- wave symmetry, we have simply to put 
$\eta_{\Gamma}=1$ at all stages of calculation. 
In fact, in these case there is no serious need to perform fully self --
consistent calculation, because it leads only to insignificant corrections to
the results of non self -- consistent calculation, using only simplest
substitution (\ref{renepsi}) \cite{KK}.

\begin{figure}
\epsfxsize=4.5cm
\epsfysize=4.5cm
\epsfbox{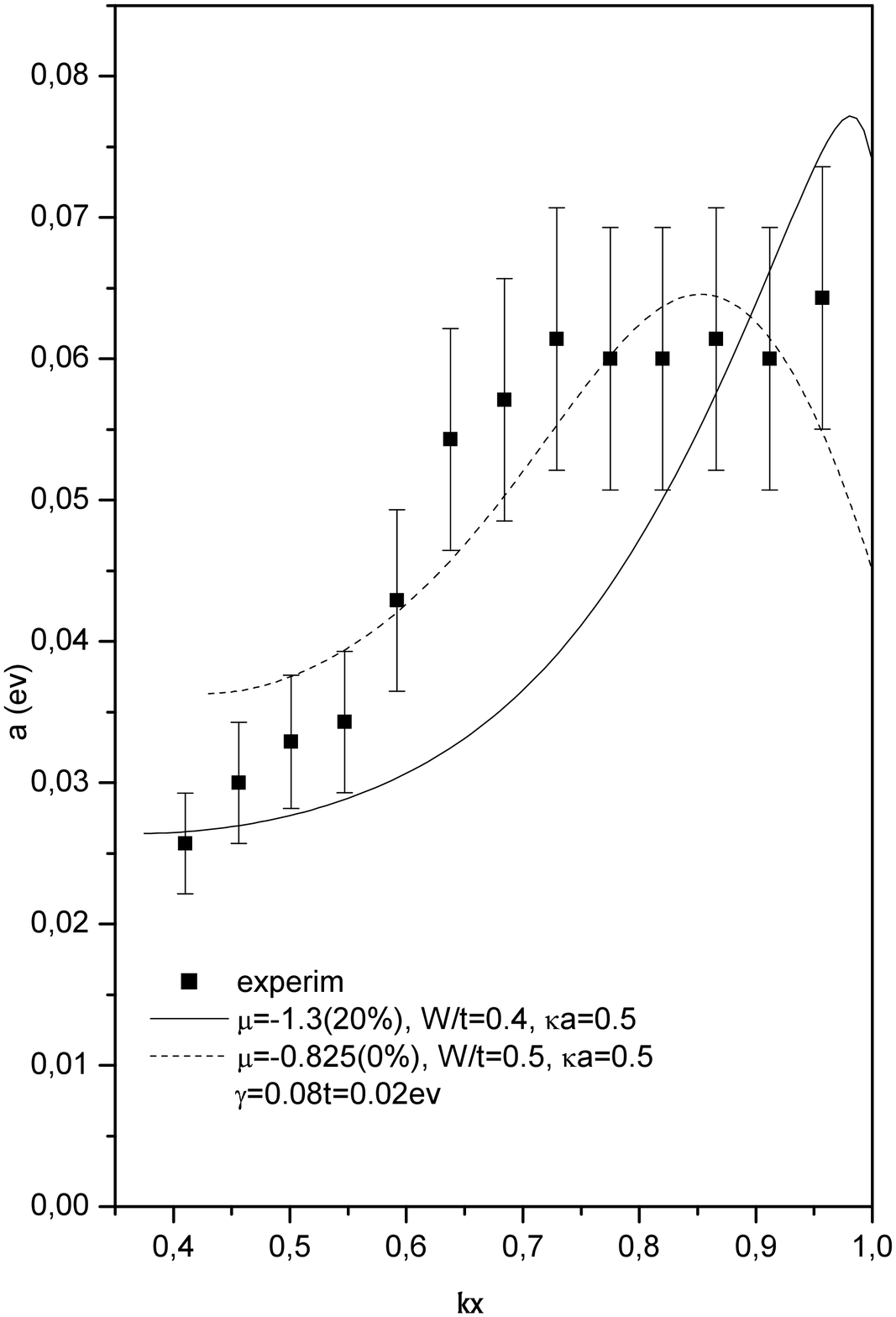}
\caption{Comparison of ARPES data for the static contribution to the
imaginary part of electron self -- energy $a=Im\Sigma(E=0,{\bf p})$ with 
calculated results for the ``hot spots'' model with impurity
scattering, for typical set of model parameters, shown in the figure.
Shown are the results for commensurate combinatorics and two values of
doping: $0\%$ and $20\%$. Momentum $k_x$ is given in units of $\frac{\pi}{a}$, 
and its change between  0 and 1 corresponds to measurement point moving
along the Fermi surface from the diagonal of the Brillouin zone towards the
vicinity of the point $(0,\pi)$.
}
\label{a_exper}
\end{figure}

As an illustration, in Fig. \ref{a_exper} we show comparison of ARPES data of
Ref. \cite{Kam} for momentum dependence of $a=Im\Sigma(E=0,{\bf p})$, 
taken from Fig. \ref{gam_anis} (c), with the results of non self -- consistent
calculation using (\ref{Dysrecimp}), (\ref{renepsi}), (\ref{etaepsi}). 
Assumed values of parameters, typical for HTSC, are shown on the figure,
for the parameters of the spectrum (\ref{spectr}) we have taken
$t=0.25 eV$,\  $t'=-0.4t$, while chemical potential was calculated for two
limiting doping levels. It can be seen that we obtain correct order of
magnitude estimate of anisotropy in momentum space, but the general form
of this dependence is only qualitatively similar to that observed in the 
experiment, which lies in between two calculated curves. More or less
similar results are obtained also in the case of spin -- fermion combinatorics.
In principle, rather approximate agreement of calculated results with
experiment is not very surprising. Our model is certainly oversimplified,
and experimental data are also not very precise. Besides, we practically
know nothing about the values of parameters of the model, appropriate for
the system, studied in these experiments.

\subsection{Superconducting transition temperature and Ginzburg -- Landau
expansion.}

\begin{figure}
\epsfxsize=6cm
\epsfbox{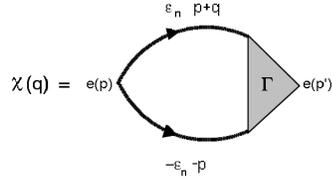}
\caption{Diagrammatic representation for generalized susceptibility 
$\chi ({\bf q})$ in Cooper channel.}
\label{loopc}
\end{figure}

Critical temperature of superconducting transition is determined by the
equation for Cooper instability of the normal phase:
\begin{equation}
1-V\chi(0;T)=0
\label{coopinst}
\end{equation}
where generalized Cooper susceptibility is defined by diagram shown in
Fig. \ref{loopc} and is equal to:
\begin{equation}
\chi({\bf q};T)=-T\sum_{\varepsilon_n}\sum_{\bf p}G(\varepsilon_n{\bf p+q})
G(-\varepsilon_n,-{\bf p})e^2({\bf p})
\Gamma_{\bf p}(\varepsilon_n,-\varepsilon_n,{\bf q})
\label{chiq}
\end{equation}
Pairing coupling constant $V$ is assumed to be nonzero in some layer of the
width of $2\omega_c$ around the Fermi level and determines the ``bare''
transition temperature $T_{c0}$ in the absence of the pseudogap fluctuations
via standard BCS equation\footnote{We do not discuss the microscopic nature
of this interaction -- it can be due to an exchange by AFM spin fluctuations, 
phonons, or combination of both.}:  
\begin{equation} 
1=\frac{2VT}{\pi^2}\sum_{n=0}^{\bar m}\int_{0}^{\pi/a}dp_x\int_{0}^{\pi/a}dp_y 
\frac{e^2({\bf p})}{\xi^2_{\bf p}+\varepsilon^2_n} 
\label{TcBCS} 
\end{equation} 
where $\bar m=[\frac{\omega_c}{2\pi T_{c0}}]$ is dimensionless cut -- off
parameter of the sum over Matsubara frequencies. All calculations \cite{KSS,KKS} 
were performed for the typical spectrum of quasiparticles (\ref{spectr}), for
different relations between $t$,\ $t'$ and $\mu$. Choosing, rather arbitrarily,
$\omega_c=0.4t$ and $T_{c0}=0.01t$, we can easily find the appropriate value
of pairing interaction $V$ in (\ref{TcBCS}), leading to this given value of
$T_{c0}$ for different types of pairing. In particular, for
$t'/t=-0.4$, $\mu /t=-1.3$ and for $s$ -- wave pairing we get
$\frac{V}{ta^2}=1$, while for $d_{x^2-y^2}$ -- pairing we obtain 
$\frac{V}{ta^2}=0.55$.

To determine $T_c$ we need only the knowledge of Cooper susceptibility for
$q=0$ which considerably simplifies all calculations \cite{KSS}. In general
case, e.g. to derive coefficients of Ginzburg -- Landau expansion we need to
know $\chi(q;T)$ for arbitrary (small) $q$.

\begin{figure}
\epsfxsize=10cm
\epsfbox{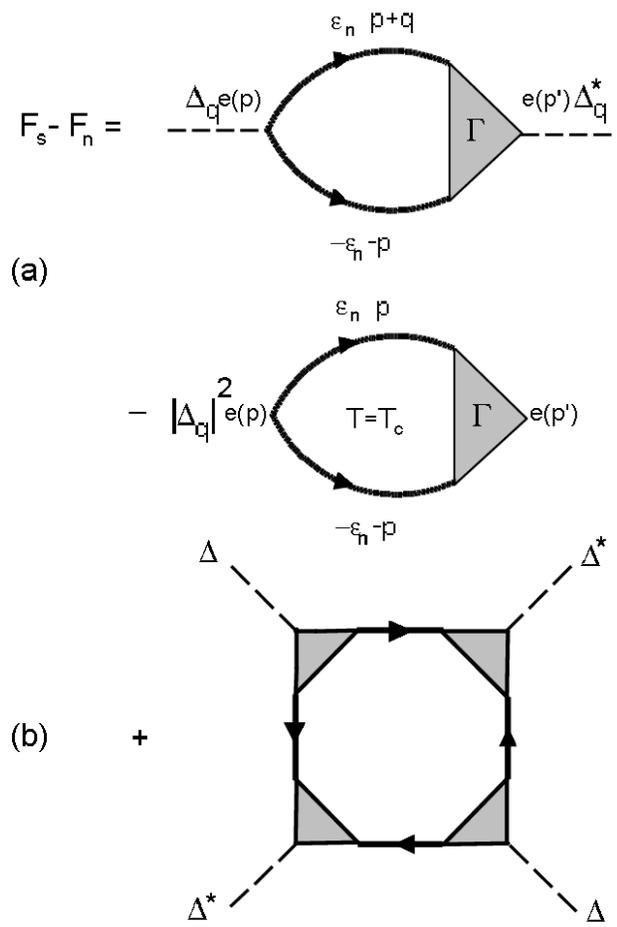}
\caption{Diagrammatic representation of Ginzburg -- Landau expansion.}
\label{GL_exp}
\end{figure}

Ginzburg -- Landau expansion for the difference of free energies of
superconducting and normal states is written in the following standard form:
\begin{equation}
F_{s}-F_{n}=A|\Delta_{\bf q}|^2
+q^2 C|\Delta_{\bf q}|^2+\frac{B}{2}|\Delta_{\bf q}|^4
\label{GiLa}
\end{equation}
and is determined by the loop expansion of free energy of an electron
in the field of fluctuations of suoerconducting order parameter (\ref{Deltq}), 
shown in Fig. \ref{GL_exp}.

It is convenient to normalize Ginzburg -- Landau coefficients $A,B,C$ 
by their values in the absence of pseudogap fluctuations, writing
the following representation \cite{KSS}: 
\begin{equation}
A=A_0K_A;\qquad   C=C_0K_C;\qquad    B=B_0K_B,
\label{ACD}
\end{equation}
\begin{eqnarray} 
&&A_0=N_0(0)\frac{T-T_{c}}{T_{c}}<e^2({\bf p})>;\quad 
C_0=N_0(0)\frac{7\zeta(3)}{32\pi^{2}T_c^2}<|{\bf v}({\bf p})|^2e^2({\bf p)}>; 
\nonumber\\ 
&&B_0=N_0(0)\frac{7\zeta(3)}{8\pi^{2}T_c^2}<e^4({\bf p})>,
\label{ACDf}
\end{eqnarray}
where the angular brackets denote the usual averaging over the Fermi surface:
$<\ldots>=\frac{1}{N_0(0)}\sum_p\delta (\xi_{\bf p})\ldots$,\ where
$N_0(0)$ -- density of states on the Fermi level for free electrons.

Then we obtain the following general expressions \cite{KSS}:
\begin{equation} 
K_A=\frac{\chi(0;T)-\chi(0;T_c)}{A_0}
\label{Ka}
\end{equation}
\begin{equation}
K_C=\lim_{q\to 0}\frac{\chi({\bf q};T_c)-\chi(0;T_c)}{q^2C_0}
\label{Kc}
\end{equation}
\begin{equation}
K_B=\frac{T_c}{B_0}\sum_{\varepsilon_n}\sum_{\bf p}e^4({\bf p})
(G(\varepsilon_n{\bf p})G(-\varepsilon_n,-{\bf p}))^2
(\Gamma_{\bf p}(\varepsilon_n,-\varepsilon_n,0))^4
\label{Kb}
\end{equation}
which were used for direct numerical calculations.
In the presence of impurities all Green's functions and vertices, entering
these expressions, should be calculated using Eqs. (\ref{Dysrecimp}) and 
(\ref{Gammaimp}), derived above.

The knowledge of the coefficients of Ginzburg -- Landau expansion allows to
determination of all major characteristics of a superconductor close to the
transition temperature $T_c$. Coherence length is defined as:
\begin{equation}
\frac{\xi^2(T)}{\xi_{BCS}^2(T)}=\frac{K_C}{K_A},
\label{xiii}
\end{equation}
where $\xi_{BCS}(T)$ is the value of this length in the absence of the 
pseudogap. Penetration depth is:
\begin{equation}
\frac{\lambda(T)}{\lambda_{BCS}(T)}=
\left(\frac{K_{B}}{K_{A}K_{C}}\right)^{1/2},
\label{lm}
\end{equation}
where we again normalized to the value of $\lambda_{BCS}(T)$ in the absence of
pseudogap fluctuations. 
Analogously, normalized slope of the upper critical field close to $T_{c}$ is:  
\begin{equation} 
\frac{\left|\frac{dH_{c2}}{dT}\right|_{T_c}}
{\left|\frac{dH_{c2}}{dT}\right|_{T_{c0}}}= 
\frac{T_{c}}{T_{c0}}\frac{K_A}{K_C}. 
\label{dHc2}
\end{equation}
Specific heat discontinuity at the transition:
\begin{equation} 
\Delta C=\frac{(C_s-C_n)_{T_c}}{(C_s-C_n)_{T_{c0}}}
=\frac{T_c}{T_{c0}}\frac{K_A^2}{K_B}.
\label{Cs}
\end{equation}

Results of these calculations for the cases of charge (CDW) and Ising like
spin (SDW) fluctuations of short range order can be found in Ref. \cite{KSS}. 
Here we shall concentrate mainly on the analysis of most important and 
interesting case of Heisenberg spin (SDW) fluctuations and also on the 
discussion of the role of impurity scattering (disordering) \cite{KKS,KKS1}. 
Due to particular importance of the case of $d$ -- wave pairing in the physics
of copper oxide high -- temperature superconductors, more attention will be
given to this case.

In the following we give the results of calculations made for typical values
of spectrum parameters $t'/t=-0.4$, $\mu /t=-1.3$, while for correlation length
we assume $\kappa a=0.2$. To spare space we do not show the results for
dimensionless GL -- coefficients $K_A,\ K_B,\ K_C$, but restrict ourselves to
demonstration of more important dependences of all major physical 
characteristics.

When we are dealing with dependences on the effective width of the pseudogap,
all characteristics are normalized to their values at $T=T_{c0}$, while in
case of dependences on the impurity scattering rate $\gamma_0$ we normalize
to the values at $T=T_{c0}(W)$, i.e. at the value of the ``bare'' transition
temperature at a given value of $W$, but in the absence of impurity
scattering ($\gamma_0=0$).

\subsubsection{$d$ -- wave pairing.}

\begin{figure}
\epsfxsize=7cm
\epsfysize=8cm
\epsfbox{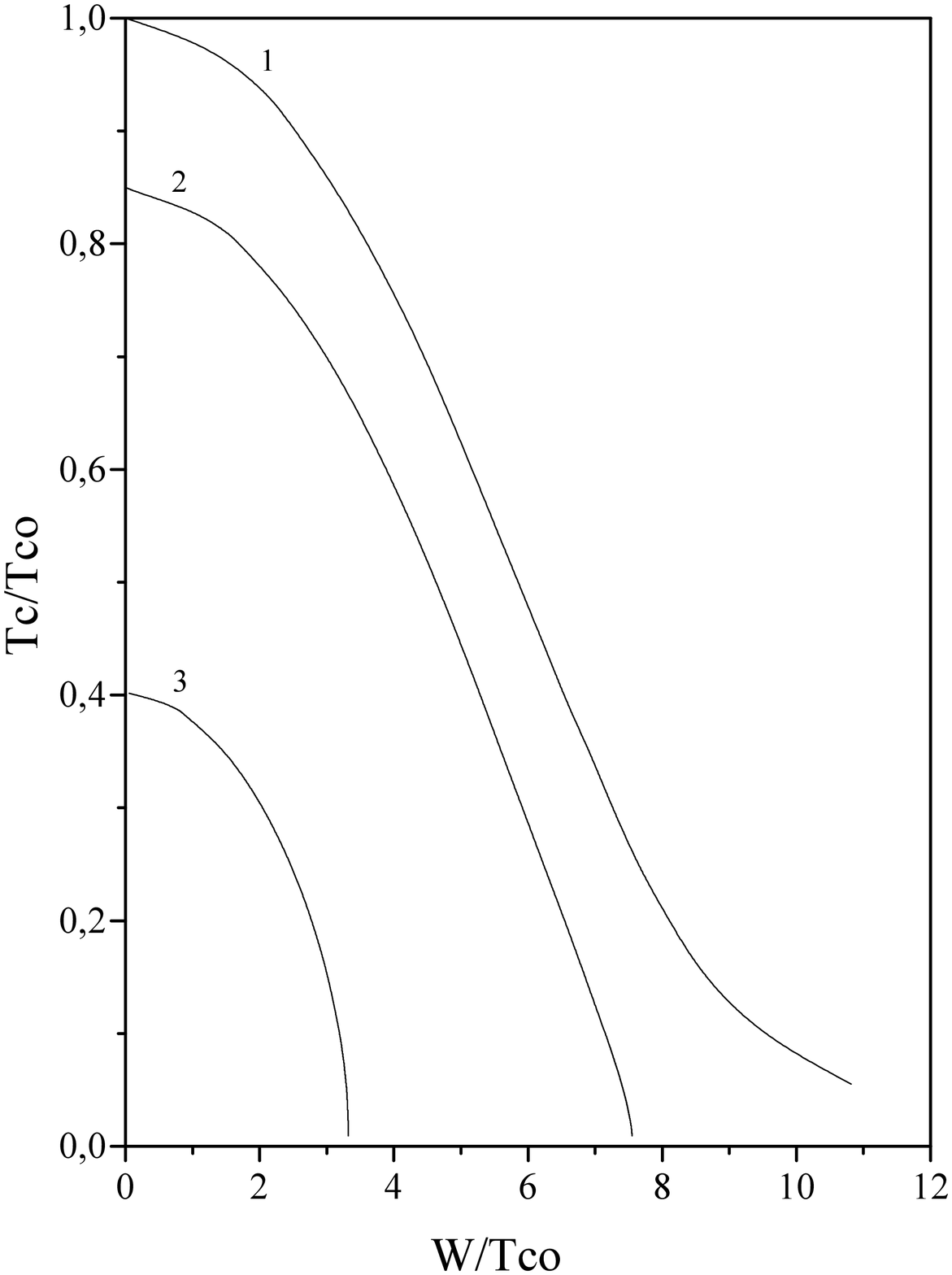}
\caption{Dependence of $T_c$ on the effective width of the pseudogap $W$ in the
case of $d$ -- wave pairing, for several values of impurity scattering rate
$\gamma_0/T_{c0}$: 0 -- 1; 0.18 -- 2; 0.64 -- 3.
Inverse correlation length $\kappa a$=0.2
} 
\label{sc1} 
\end{figure}

In Fig. \ref{sc1} we show the dependence of superconducting transition
temperature $T_c$ on the effective pseudogap width $W$ for several values
of impurity scattering rate. It is seen that pseudogap fluctuations lead to
significant suppression of superconductivity, and in the presence of finite
disorder we always obtain some ``critical'' value of $W$, where the value of
$T_c$ vanishes. This suppression of $T_c$ is naturally due to a partial
``dielectrization'' of electronic spectrum in the vicinity of ``hot spots''.
\cite{Sch,KS}.

\begin{figure}
\epsfxsize=7cm
\epsfysize=6cm
\epsfbox{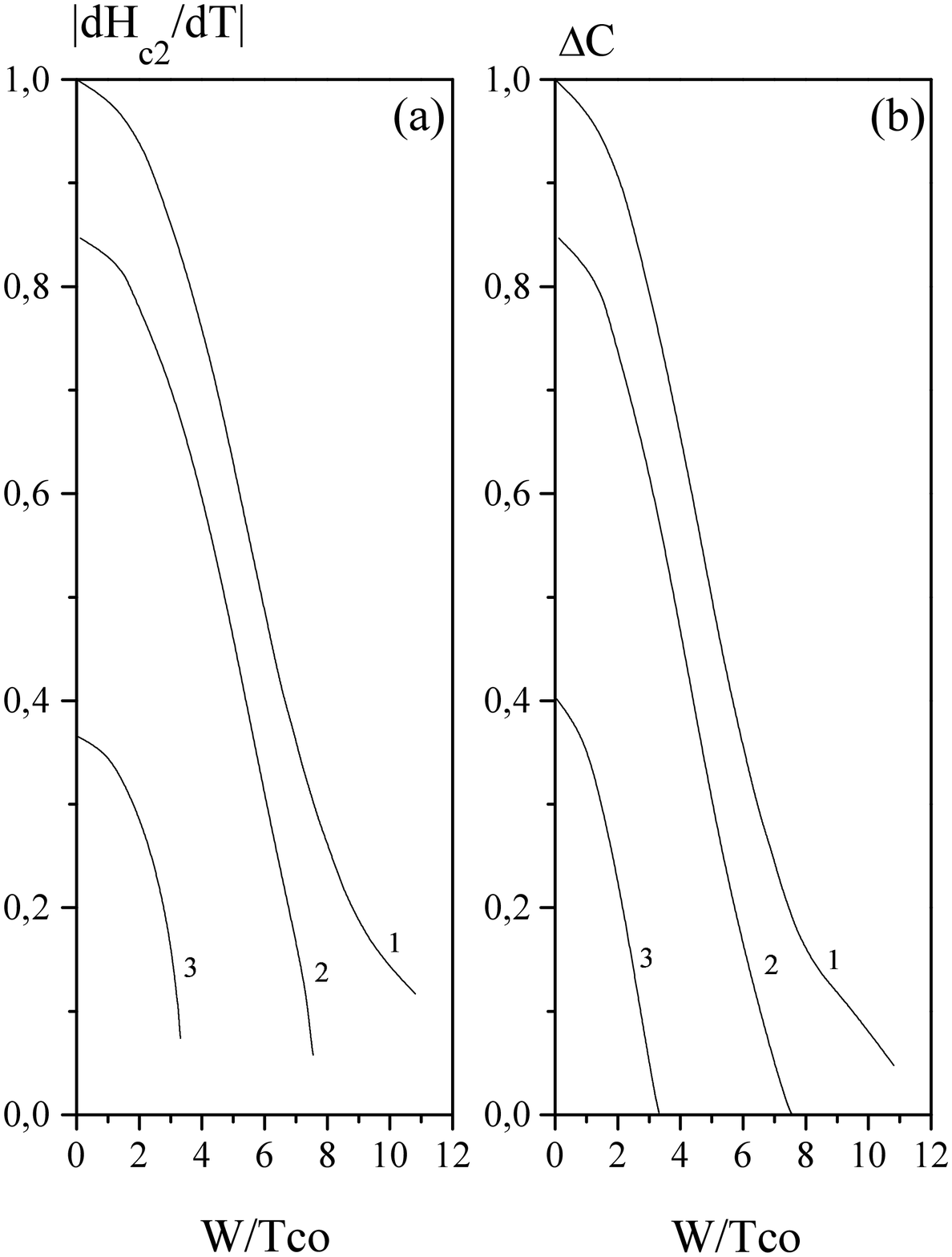}
\caption{Dependence of the slope of the upper critical field and specific heat
discontinuity at the transition on the effective width of the pseudogap
$W$ in case of $d$ -- wave pairing, for several values of impurity scattering
rate $\gamma_0/T_{c0}$: 0 -- 1; 0.18 -- 2; 0.64 -- 3.
} 
\label{sc3} 
\end{figure}

Similar dependences for the slope of the upper critical field and specific
heat discontinuity at the transition point are shown in Fig. \ref{sc3}.
Typically we see fast suppression of these characteristics by pseudogap
fluctuations.
 
Dependence on the value of correlation length of pseudogap fluctuations
is more slow --- in all cases the growth of $\xi$ (drop of $\kappa$) enhances
the effect of pseudogap fluctuations. We drop appropriate results to spare
space.

\begin{figure}
\epsfxsize=7cm
\epsfysize=8cm
\epsfbox{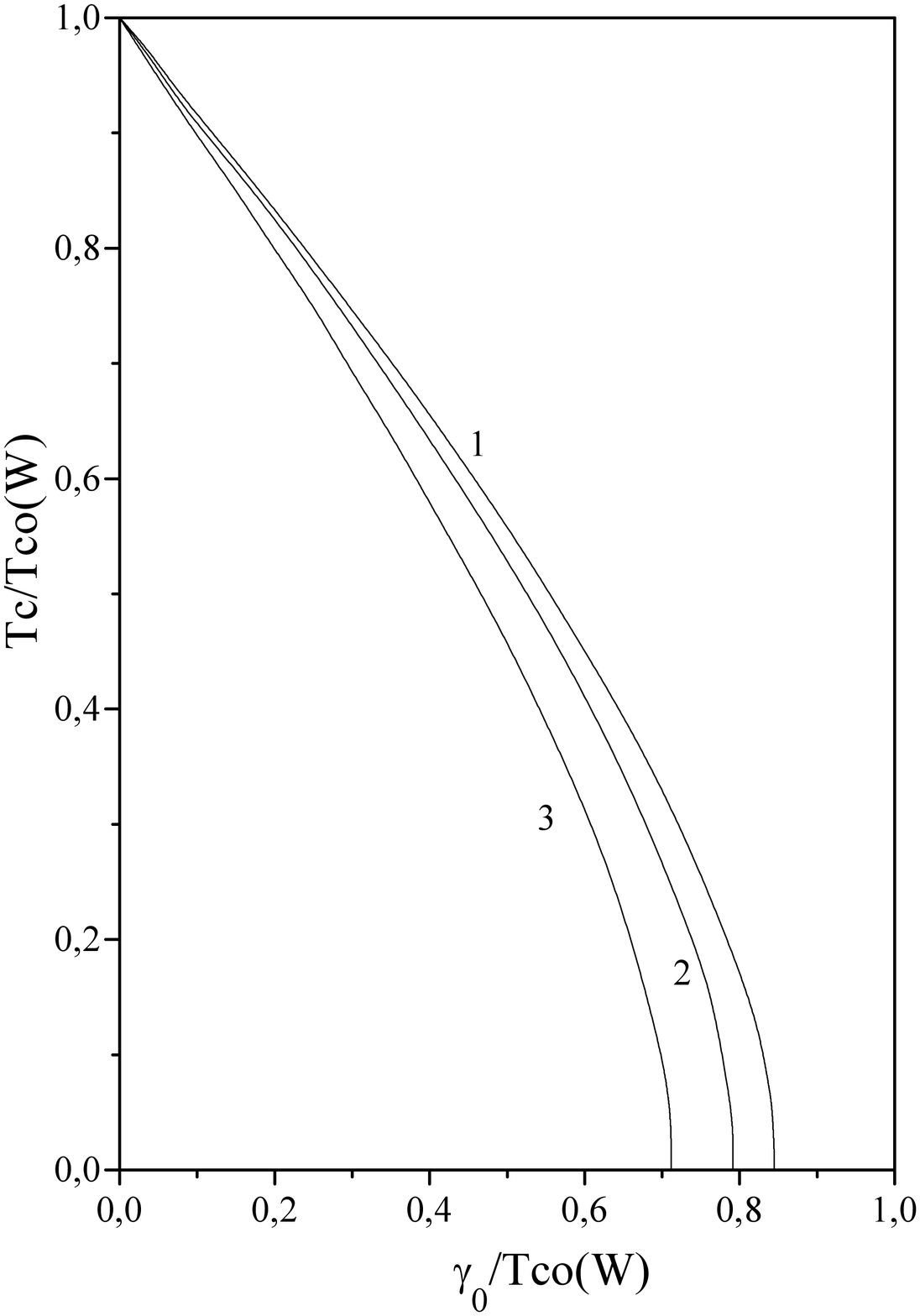}
\caption{Dependence of $T_c$ on impurity scattering rate (disorder) 
$\gamma_0$ for the case of $d$ -- wave pairing, for several values of the
effective pseudogap width:
$W/T_{c0}$: 0 -- 1; 2.8 -- 2; 5.5 -- 3.
} 
\label{sc4} 
\end{figure}

\begin{figure}
\epsfxsize=7cm
\epsfysize=6cm
\epsfbox{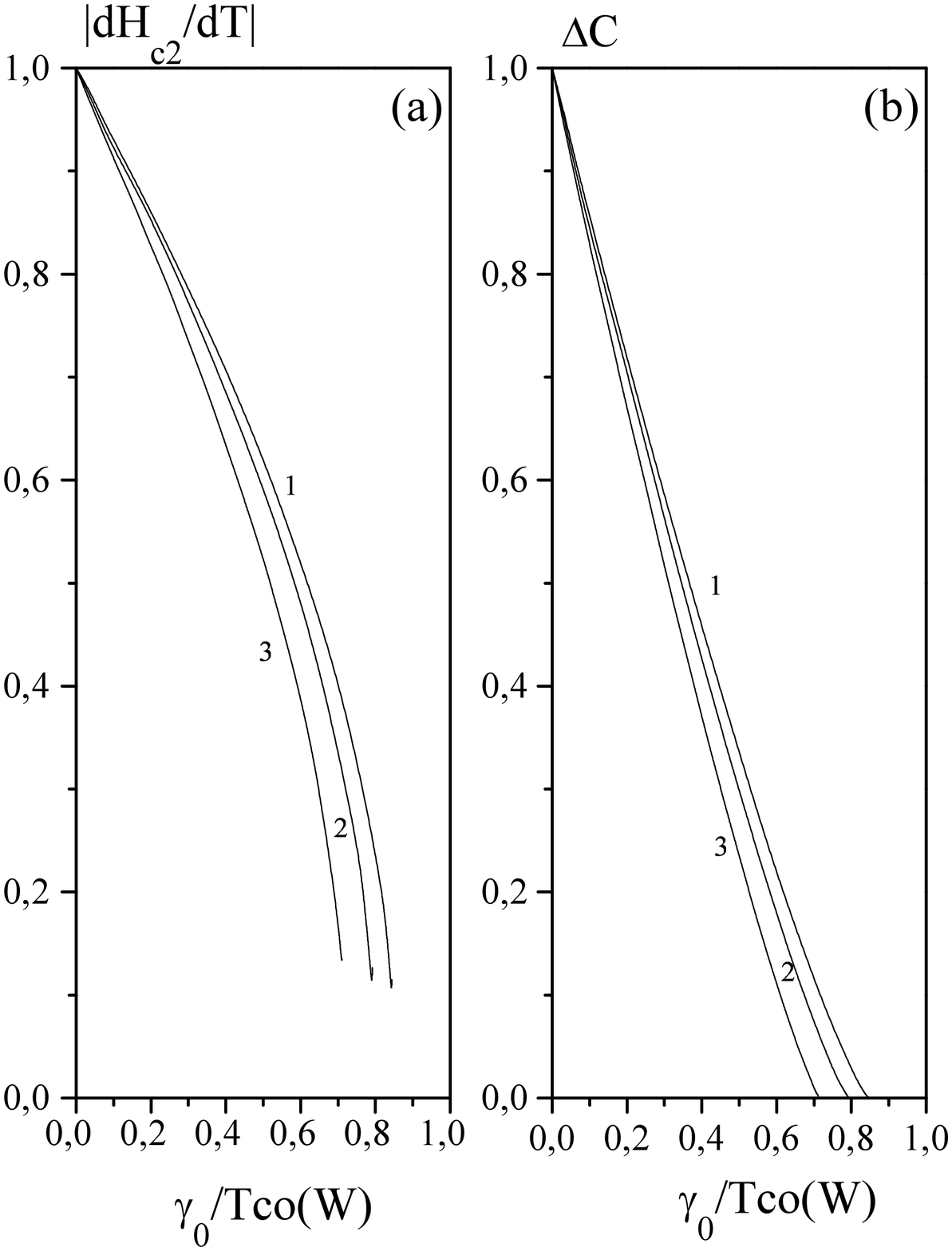}
\caption{Dependence of the slope of the upper critical field and specific
heat discontinuity at the transition on impurity scattering rate (disorder)
$\gamma_0$ in case of $d$ -- wave pairing, for several values of effective
pseudogap width $W/T_{c0}$: 0 -- 1; 2.8 -- 2; 5.5 -- 3.
} 
\label{sc6} 
\end{figure}

In Fig. \ref{sc4} we show the dependences of superconducting transition
temperature $T_c$ on impurity scattering rate $\gamma_0$ for several values
of the effective pseudogap width. We note that in the presence of pseudogap
fluctuations, suppression of $T_c$ by disorder is significantly faster, than
in their absence ($W=0$), when $T_c$ dependence on $\gamma_0$ in case of
$d$ -- wave pairing is described by the standard Abrikosov -- Gorkov curve
(first obtained for $s$ -- wave pairing and scattering by magnetic impurities)
\cite{PS1,Radt}. Similar dependences for the slope of $H_{c2}(T)$ and specific
heat discontinuity are shown in Fig. \ref{sc6}. Again we see, that impurity
scattering (disorder) leads to the fast drop of these two characteristics,
i.e. enhances similar effect of pseudogap fluctuations.

Similar dependences were obtained in Ref. \cite{KSS} for the case of pseudogap
fluctuations of CDW type, when we also obtain recursion procedure for the vertex
part wit alternating signs. At the same time, in this case we observe some 
quantitative difference in the behavior of all characteristics due to another 
combinatorics of diagrams.

Dependences obtained in our model are in qualitative agreement with most of the
experimental data on superconductivity in the pseudogap region (underdoped 
region of the phase diagram of cuprates). Below we shall show that these 
results may be used for direct modelling of a typical phase diagram of a
high -- temperature superconductor.

\subsubsection{$s$ -- wave pairing.}

The case of $s$ -- wave pairing is interesting to us mainly with the aim of
demonstration of major differences from the case of $d$ -- wave case. 
There are no experimental evidence for $s$ -- wave superconductivity in systems
with pseudogap, though such systems may well be discovered in some future.

\begin{figure}
\epsfxsize=7cm
\epsfysize=7cm
\epsfbox{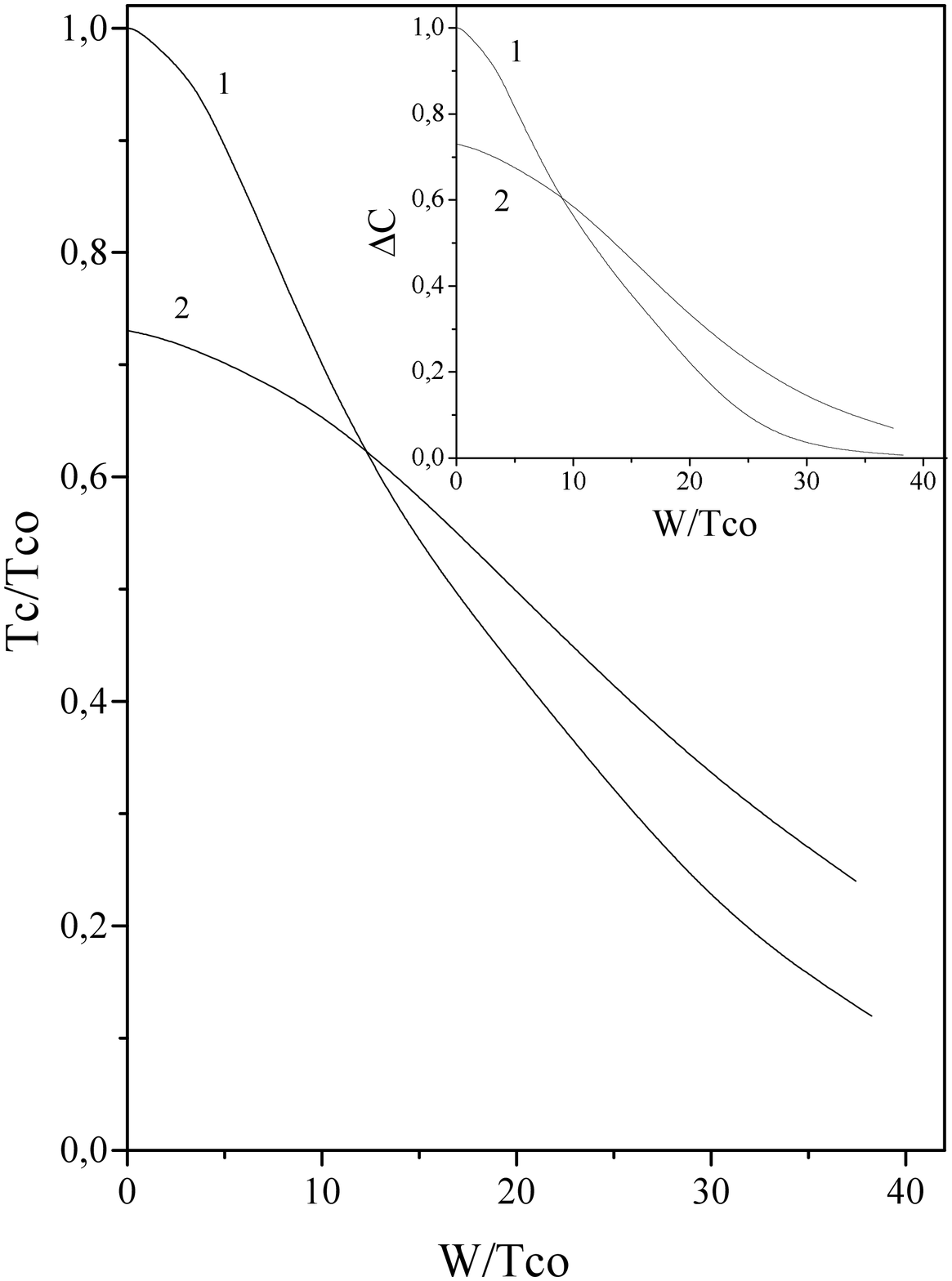}
\caption{Dependence of $T_c$ on effective width of the pseudogap $W$ for the
case of $s$ -- wave pairing, for two values of impurity scattering rate:
$\gamma_0/T_{c0}$: 0 -- 1; 20 -- 2.
Inverse correlation length $\kappa a$=0.2. At the insert -- characteristic
behavior of specific heat discontinuity for the same parameters.
} 
\label{sc7} 
\end{figure}

Our calculation show that pseudogap fluctuations suppress superconducting
transition temperature also in this case (Fig. \ref{sc7}), though 
characteristic scale of these fluctuations, necessary for significant
suppression of superconductivity here is much larger, than in the case of
$d$ -- wave pairing. This result was obtained in Ref. \cite{KSS}, but we 
must note the absence (in the case of Heisenberg (SDW) pseudogap fluctuations)
of characteristic ``plateau'' in the dependence of $T_c$ on $W$, which was 
obtained for the case of scattering by pseudogap fluctuations of CDW type in
Ref. \cite{KSS}. On the same scale of $W$ we also observe here the suppression
of specific heat discontinuity at the transition, which is shown at the insert
in Fig. \ref{sc7}. 

As to dependence of $T_c$ on impurity scattering rate (disorder), we can see
that besides relatively small effect of $T_c$ suppression due to \cite{KK} 
disorder ``smearing'' of the density of states at the Fermi level, even some
weak effect of $T_c$ enhancement with the growth of $\gamma_0$ can be also 
observed, apparently due to the ``filling'' the pseudogap in the density of
states induced by impurity scattering \cite{KKS}.

\begin{figure}
\epsfxsize=7cm
\epsfysize=6cm
\epsfbox{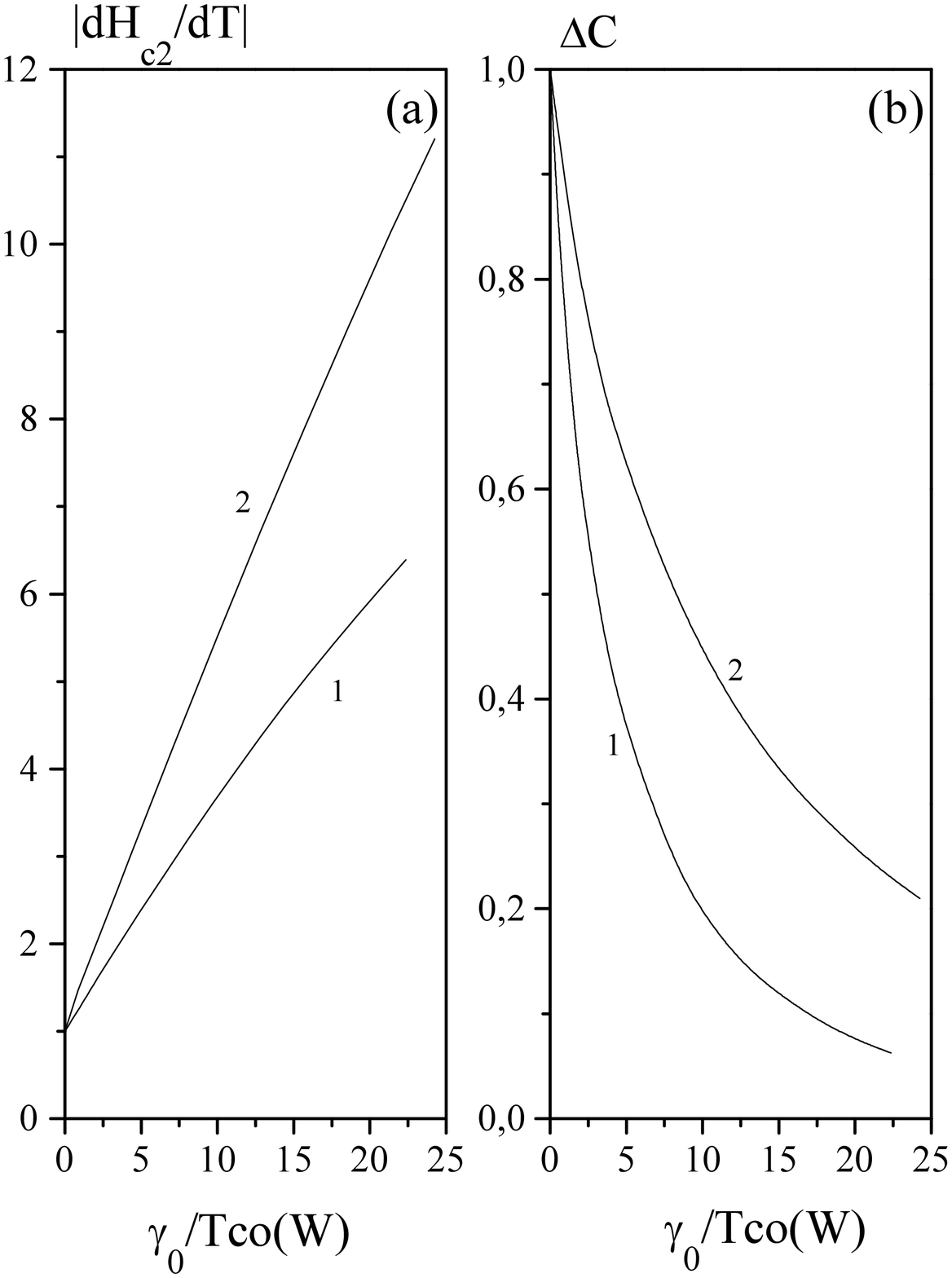}
\caption{Dependence of the slope of the upper critical field and specific heat
discontinuity at the transition on impurity scattering rate (disorder)
$\gamma_0$ in case of $s$ -- wave pairing, for two values of pseudogap width:
$W/T_{c0}$: 0 -- 1; 15 -- 2.
} 
\label{sc11} 
\end{figure}

In Fig. \ref{sc11} we show the influence of impurity scattering (disorder) on
the slope of the upper critical field and specific heat discontinuity.
Specific heat discontinuity is significantly suppressed by disorder, while
the slope of $H_{c2}(T)$ behaves qualitatively different from the case of
$d$ -- wave pairing: the growth of disorder leads to the growth of this
characteristic, as in the case of standard theory of ``dirty'' superconductors
\cite{SC_Loc}, and pseudogap fluctuations enhance the slope of $H_{c2}(T)$.  
In the absence of pseudogap fluctuations, similar differences between $s$ --
wave and $d$ -- wave superconductors in behavior of the slope of $H_{c2}(T)$ 
under disordering were discussed in Ref. \cite{PS1}.

\section{Modeling of the phase diagram of cuprates.}

This model for the influence of pseudogap fluctuations on superconductivity
allows to perform a simple modeling of typical phase diagram of HTSC cuprates
\cite{KKS,KKS1}\footnote{Here we neglect the existence of a narrow region of
antiferromagnetism in Mott insulator state, at low concentrations of doping
impurity, limiting ourselves to a wide region of ``bad'' metallic state.} 
First attempt of such modeling in quite simplified version of our model was
undertaken in Ref. \cite{AC}. Basic idea is to identify our parameter $W$ 
with experimentally observable effective width of the pseudogap  
(temperature of crossover into pseudogap region on the phase diagram)
$E_g\approx T^*$, determined from numerous experiments \cite{Loram,MS}. 
This characteristic, as was already noted above, drops almost linearly
with the growth of the concentration of doping impurity (concentration of
carriers), starting from the values of the order of $10^3$K and vanishing
at some critical concentration $x_c\approx 0.19..0.22$, slightly higher than
``optimal'' value $x_o\approx 0.15..0.17$ \cite{Loram,NT}. Accordingly, we can
assume similar concentration dependence of our pseudogap width $W(x)$\footnote{
Naturally, such identification can be done up to an unknown proportionality
coefficient $\sim 1$.}. In this sense we can say that $W(x)$ is determined
directly from the experiment. Then, the only parameter we need remains
concentration dependence of the ``bare'' temperature of superconducting
transition $T_{c0}(x)$, which would have been existing in the absence of
pseudogap fluctuations. Its knowledge would allow us to determine the
concentration dependence of the observable transition temperature $T_c(x)$, 
solving the equations of our model. Unfortunately, concentration dependence
$T_{c0}(x)$ is, generally, unknown and is not determined by any known
experiment, remaining just free fitting parameter of our theory. 

\begin{figure}
\epsfxsize=7cm
\epsfysize=7cm
\epsfbox{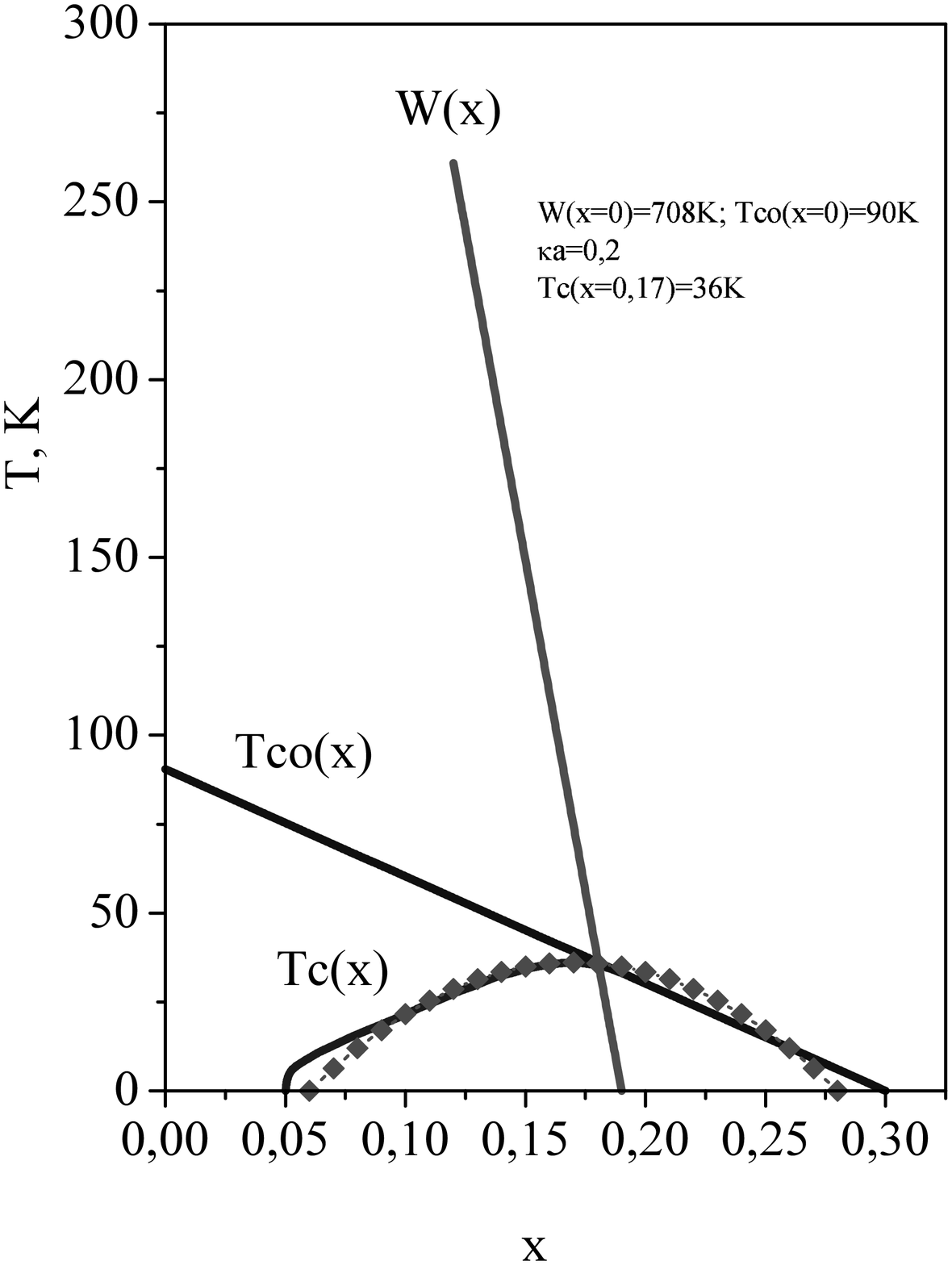}
\caption{Model phase diagram for the case of scattering by pseudogap 
fluctuations of CDW -- type ($d$ -- wave pairing) and ``bare'' superconducting
transition temperature $T_{c0}$ linear in carrier concentration.}  
\label{TcCDW} 
\end{figure}

Following Ref. \cite{AC} we can assume, that $T_{c0}(x)$ can also be described
by linear function of $x$, going to zero at $x=0.3$, and choosing the value of
$T_{c0}(x=0)$, which give us experimentally observed $T_c(x=x_o)$. Then we can
calculate the whole ``observed'' dependence $T_c(x)$. Results of such
calculation for the case of $d$ -- wave pairing and scattering by charge (CDW) 
pseudogap fluctuations \cite{KSS}, using typical $W(x)$ dependence, are shown
in Fig. \ref{TcCDW}.  
We see that even with such arbitrary assumptions the ``hot spots'' model allows
to obtain $T_c(x)$ dependence, which is rather close to experimentally 
observable. Similar calculations for Ising like model of spin fluctuations
(leading to the absence of sign alternation in the recursion procedure for
the vertex part \cite{KSS}) show, that reasonable values for $T_c(x)$ can be
obtained only for unrealistic values of $W(x)$, at least an order of magnitude
larger than observed in the experiments.

\begin{figure}
\epsfxsize=7cm
\epsfysize=7cm
\epsfbox{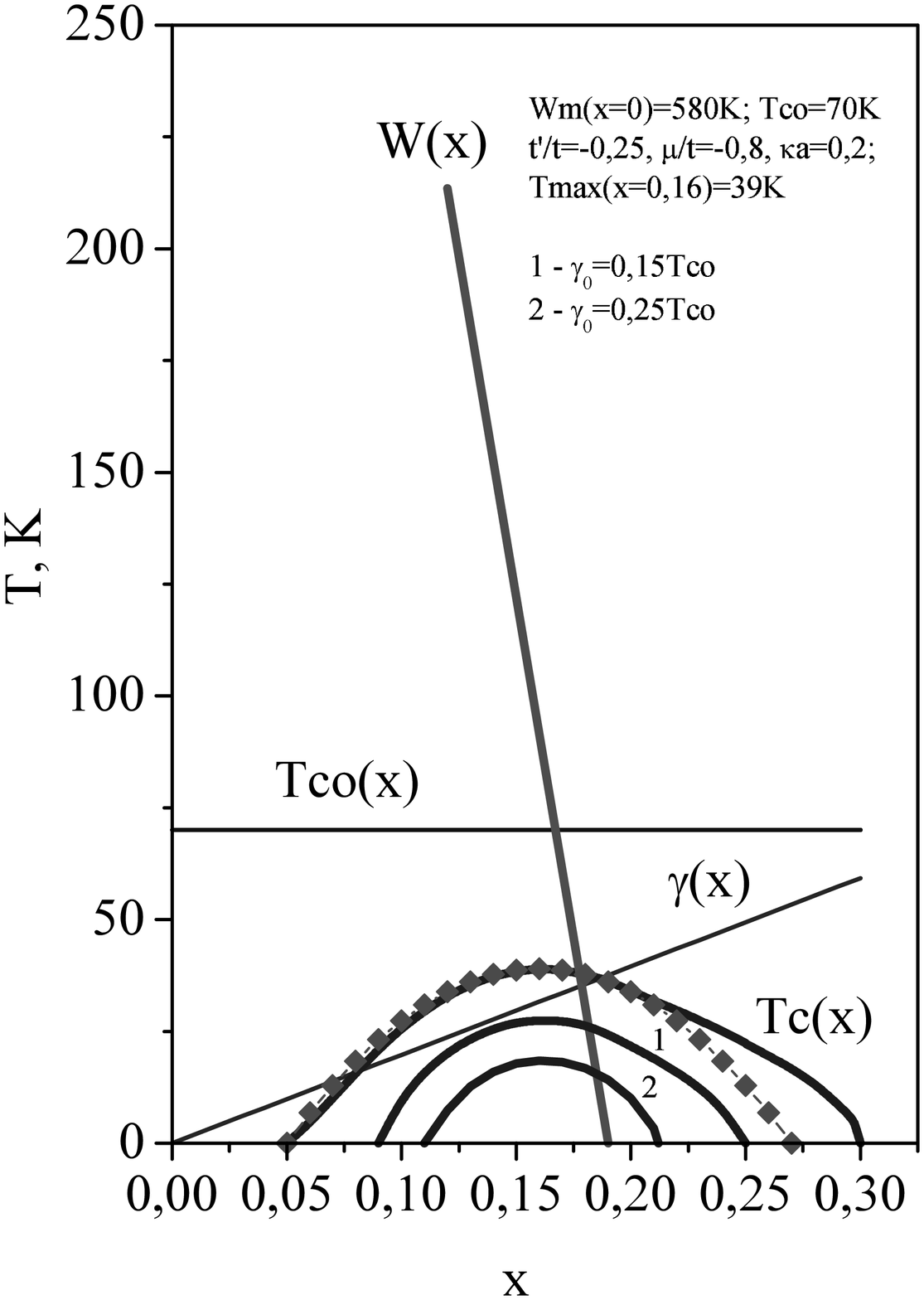}
\caption{Model phase diagram for the case of scattering by Heisenberg
(SDW) pseudogap fluctuations ($d$ -- wave pairing) and ``bare'' temperature
of superconducting transition $T_{c0}$ independent of carrier concentration,
with the account of internal disorder linear in concentration of doping
impurity $\gamma(x)$.}  
\label{TcSDW} 
\end{figure}

In the framework of our BCS -- like model for the ``bare'' $T_{c0}$, an 
assumption of its strong concentration dependence seems rather 
unrealistic\footnote{In this approach any dependence of $T_{c0}$ on $x$ can be
only due to relatively weak changes of density of states at the Fermi
level, as chemical potential moves with $x$.}. Thus, let us assume that the
value of $T_{c0}$ does not depend on carrier concentration $x$ at all, but 
take into account the fact that introduction of doping impurity inevitably
leads to appearance random impurity scattering (internal disorder), which
can be described by appropriate linear dependence of $\gamma(x)$. Let us assume
that it is this growth of disorder, which leads to complete suppression of
$d$ -- wave pairing at $x=0.3$, in accordance with Abrikosov -- Gorkov
dependence \cite{PS1,Radt}. Results of our modeling of the phase diagram for
the system of the type of $La_{2-x}Sr_xCuO_4$, for the case of Heisenberg 
pseudogap fluctuations, with the account of the abovementioned role of 
internal disorder, are shown in Fig. \ref{TcSDW}.
The values of different parameters used in this calculation are also shown at
the same figure.
``Experimental'' values of $T_c(x)$, shown at this figure (as well as at
Fig. \ref{TcCDW}) by ``diamonds'', were obtained from empirical relation 
\cite{NT,PT}:
\begin{equation}
\frac{T_c(x)}{T_c(x=x_o)}=1-82.6(x-x_o)^2
\label{Tcexp}
\end{equation}
which gives rather good fit to experimental data on concentration dependence
of $T_c$ for a number of HTSC cuprates. It is seen, that in the whole
underdoped region our model gives practically ideal description of
``experimental'' data with very reasonable values of $W(x)$. At the end of
overdoped region agreement is less good, but you must take into account, that 
Eq. (\ref{Tcexp}) also usually does not give very good fit here, and the 
obvious crudeness of our assumptions for underdoped region. Also note, that we
have not undertaken any special attempts to improve agreement in this region.

It is interesting to analyze the behavior of superconducting transition
temperature $T_c$ under additional disordering of our system at different
compositions (carrier concentrations). There are many experimental studies
\cite{SC_Loc}, where such an additional disordering was introduced by chemical
substitutions (impurities) \cite{Uch,Tall} or by irradiation with fast
neutrons \cite{Gosch} or electrons \cite{Tolp,Rull}. However, special discussion
of the role of this additional disordering in the context of the studies of
the pseudogap state was made only in Ref. \cite{Tall}.

In our model, such disordering may be simulated by introduction of additional
``impurity'' scattering parameter $\gamma_0$, which is just added to the
parameter of internal disorder $\gamma(x)$. Results of our calculations of
superconducting critical temperature for two values of this parameter are also
shown in Fig. \ref{TcSDW}. We see, that in complete agreement with experiments
\cite{Tall}, introduction of ``impurities'' (disorder) leads to fast narrowing
of the superconducting region on the phase diagram. Also in complete accordance
with our conclusions made above with respect to Fig. \ref{sc4}, and also with
experiments \cite{Gosch,Tall}, superconductivity suppression by disorder in the
underdoped region (pseudogap state) is significantly faster, than at optimal
composition. We could have expected that introduction of ``normal'' disorder.
obviously leading to some suppression of the pseudogap in the density of 
states, could also lead to certain ``slowing'' down of $T_c$ suppression.
However, in case of $d$ -- wave pairing, this effect is apparently absent.

The problem, however, is that in all cases our suppression of $T_c$ 
by disorder is faster than that described by the standard Abrikosov -- Gorkov
dependence for the case of $d$ -- wave pairing \cite{Radt}. At the same time,
attempts to fit the results of majority of experiments on disordering in
HTSC cuprates to this dependence show \cite{Uch,Tolp,Rull}, that such 
suppression actually is significantly slower, than predicted by Abrikosov --
Gorkov dependence. This, still unsolved, problem remains one of the major
problems of the theory of high -- temperature superconductors \cite{SC_Loc}. 
One possible solution may be connected with consistent description of the
role of disorder in superconductors, which belong to the crossover region
from ``large'' pairs of BCS theory to ``compact'' Bosons, appearing in the
limit of very strong coupling \cite{PosSad}. Another interesting possibility
to explain such ``slowing down'' of $T_c$ suppression is connected with the
role of anisotropy of elastic scattering by impurities analyzed in Refs.
\cite{PS1,HN}. This last effect can be rather easily included in our
calculational scheme. It is of particular interest in connection with
established fact of rather strong anisotropy of elastic scattering 
(with $d$ -- wave symmetry), which was observed in ARPES experiments on
$Bi_2Sr_2CaCu_2O_{8+\delta}$ \cite{Kam}. Appropriate scattering rate in these
experiments changed in the interval of  $20 - 60\ meV$ 
(cf. Fig. \ref{gam_anis} (c)) \cite{Kam}, which is nearly an order of magnitude
larger than maximal value of $\gamma(x)$, used in our calculations. This is an
additional evidence of unusual stability of $d$ -- wave pairing in cuprates
towards static disorder. Note that our model for electron self -- energy
in fact describes similar anisotropy of elastic scattering, corresponding
to its growth in the vicinity of ``hot spots'', yet the effect of
``slowing down'' of $T_c$ suppression under disordering is not observed.

Our results show that despite the obvious crudeness of our assumptions, the
``hot spots'' model rather easily leads to a reasonable (sometimes even
semiquantitative) description of superconductivity region on the phase
diagram of HTSC cuprates\footnote{Above we always assumed that we are dealing
with hole doped systems, where concentration dependence of $T^*(x)$ is well
established \cite{Loram,NT}. For electronically doped cuprates similar data
are practically absent.}. The main shortcoming this approach remains
considerable arbitrariness of our ``scenario'' of the origin of concentration
dependence of the ``bare'' superconducting transition temperature.

\section{Conclusion.}

Our discussion demonstrates the variety of results, which can be obtained
in this class of models of the pseudogap state. We must stress once again quite
simplified nature of our approximations, though allowing  to obtain what was
called ``nearly exact'' solution. This solution is interesting, first of all,
from purely theoretical point of view, as rare enough example of the problem,
when it is possible to sum the complete Feynman perturbation theory series
(though, in general, certain classes of higher order diagrams are calculated
approximately). Probably the main theoretical ``lesson'' to learn is, that the
results of such complete summation are rather radically different from the
results, obtained by one or other ``partial'' summation (e.g. of simple
geometrical progression).

Probably the main shortcoming of our approach is the neglect of dynamics of
fluctuations of short -- range order. This is justified, as was shown above,
only for high enough temperatures, which is bad, e.g. from the point of view of
description of superconducting state for $T\ll T_c$ \footnote{Thus, and also
to spare space, we have dropped discussion of Gorkov's equations, which can
also be derived and analyzed \cite{KS3,KK}, taking into account all higher 
orders of perturbation theory over pseudogap fluctuations.} 

Another shortcoming, as was mentioned many times \cite{S79,KS}, is our
limitation to Gaussian fluctuations, which is also justified only for
high enough temperatures. In some sense it is only technical limitation, though
it is quite important for the formal structure of our solution.

We already mentioned that in this class of models we can demonstrate the
breaking of the fundamental property of self -- averaging of superconducting
order parameter \cite{KS1,KS2}. Unfortunately, this can be done only in rather
unrealistic case of infinite correlation length of pseudogap fluctuations
\cite{KS1}, or in very special model with finite correlation length  
\cite{BK,KS2}, and in an oversimplified model with ``hot patches'' on the
fermi surface. Qualitative picture emerging from this analysis demonstrates
the possibility of formation of ``superconducting drops'' at temperatures 
higher than mean -- field $T_c$ \cite{BPS,KS1,KS2}, which may have direct
relation to the existence of the ``low energy'' pseudogap (signatures of
superconducting response at $T>T_c$) and experimentally observed inhomogeneous
superconductivity in cuprates \cite{Pan,Davis}. In the ``hot spots'' model
this kind of analysis is still to be done.

At the same time, we have seen that in some cases we can make even direct
comparison of predictions of these simplified models with the experiments,
and the results of such comparison are rather promising. In this sense the
``hot spots'' model can pretend to give quasi realistic description of the
properties of the pseudogap state in high -- temperature superconductors.

The author is grateful to E.Z.Kuchinskii and N.A.Kuleeva (Strigina) for their
collaboration during our work on the ``hot spots'' model. This work was 
supported in part by RFBR grant 02-02-16031 and programs of the Presidium of
the Russian Academy of Sciences (RAS) ``Quantum macrophysics'' and of the 
Division of Physical Sciences of the RAS ``Strongly correlated electrons in
semiconductors, metals, superconductors and magnetic materials''.

\end{document}